\documentclass[nofootinbib,superscriptaddress,eqsecnum,tightenlines,11pt,longbibliography,notitlepage]{revtex4-1}

\usepackage{hyperref}
\usepackage{graphicx}
\usepackage{amsmath,amssymb,amsfonts,amsthm,stmaryrd,mathtools,bbold,mathtools,xfrac,faktor}
\usepackage{yfonts}
\usepackage{multirow,bigdelim}
\usepackage{dsfont}
\usepackage{color}
\usepackage{graphicx,color,epstopdf,epsfig,psfrag}
\usepackage[dvipsnames]{xcolor}
\usepackage{ mathrsfs }
\usepackage{footmisc}

\usepackage{palatino, eulervm}

\hypersetup{
    colorlinks=true,       
    linkcolor=MidnightBlue,          
    citecolor=BrickRed,        
    filecolor=BrickRed,      
    urlcolor=BrickRed           
}
 
\makeatletter
\newcounter{subsubsubsection}[subsubsection]
\renewcommand\thesubsubsubsection{\thesubsubsection .\@alph\c@subsubsubsection}
\newcommand\subsubsubsection{\@startsection{subsubsubsection}{4}{\z@}%
                                     {-3.25ex\@plus -1ex \@minus -.2ex}%
                                     {1.5ex \@plus .2ex}%
                                     {\centering\normalfont\small\textit}}
\newcommand*\l@subsubsubsection{\@dottedtocline{3}{10.0em}{4.1em}}
\newcommand*{\subsubsubsectionmark}[1]{}
\makeatother

\numberwithin{equation}{section}
\numberwithin{paragraph}{subsection}

\newcommand{\E}{\mathrm{e}}

\newcommand{\SU}{\mathrm{SU}}

\newcommand{\SL}{\mathrm{SL}}
\newcommand{\SO}{\mathrm{SO}}

\renewcommand{\d}{{\mathrm{d}}}

\newcommand{\be}{\begin{equation}}
\newcommand{\ee}{\end{equation}}
\newcommand{\beq}{\begin{eqnarray}}
\newcommand{\eeq}{\end{eqnarray}}
\newcommand{\bes}{\begin{eqnarray}}
\newcommand{\ees}{\end{eqnarray}}

\newcommand{\frakg}{{\mathfrak{g}}}

\newcommand{\la}{\langle}
\newcommand{\ra}{\rangle}

\newcommand{\Tr}{{\mathrm{Tr}}}

\newcommand{\tl}{\widetilde}

\def\pp{\partial}

\newcommand{\G}{{\Gamma}}

\renewcommand{\hat}{\widehat}

\newcommand{\Pexp}{{\mathbb{P}\mathrm{exp}}}
\newcommand{\mQ}{{\mathcal Q}}
\newcommand{\mP}{{\mathcal P}}
\newcommand{\Lie}{{\mathrm{Lie}}}
\newcommand{\ad}{{\mathrm{ad}}}
\newcommand{\Ad}{{\mathrm{Ad}}}
\newcommand{\AD}{{\mathrm{AD}}}
\newcommand{\aD}{{\mathrm{aD}}}
\newcommand{\EM}{{\text{SD}}}
\newcommand{\thetaL}{\theta^L}
\newcommand{\thetaR}{\theta^R}
\newcommand{\fus}{{\circledast}}
\newcommand{\M}{{\mathcal M}^\text{flat}}
\newcommand{\A}{{\mathscr A}}

\DeclareFontFamily{U}{MnSymbolC}{}
\DeclareSymbolFont{MnSyC}{U}{MnSymbolC}{m}{n}
\DeclareFontShape{U}{MnSymbolC}{m}{n}{
    <-6>  MnSymbolC5
   <6-7>  MnSymbolC6
   <7-8>  MnSymbolC7
   <8-9>  MnSymbolC8
   <9-10> MnSymbolC9
  <10-12> MnSymbolC10
  <12->   MnSymbolC12}{}
\DeclareMathSymbol{\contr}{\mathbin}{MnSyC}{'270}

\usepackage{enumerate}

\theoremstyle{definition}

\theoremstyle{remark}

\begin{document}

\title{\fontsize{15pt}{15pt}\selectfont On a self-dual phase space for $3+1$ lattice Yang--Mills theory}

\author{{\bf Aldo Riello}}\email{ariello@perimeterinstitute.ca}
\affiliation{Perimeter Institute, 31 Caroline St North, Waterloo ON, Canada N2L 2Y5}

\date{\today}

\begin{abstract}
I propose a self-dual deformation of the classical phase space of lattice Yang--Mills theory, in which both the electric and magnetic fluxes take value in the compact gauge Lie group. 
A local construction of the deformed phase space requires the machinery of ``quasi-Hamiltonian spaces'' by Alekseev {\it et al.}, which is here reviewed.
The results is a full-fledged finite-dimensional and gauge-invariant phase space, whose self-duality properties are largely enhanced in $(3+1)$ spacetime dimensions.
This enhancement is due to a correspondence with the moduli space of an auxiliary non-commutative flat connection living on a Riemann surface defined from the lattice itself, which in turn equips the duality between electric and magnetic fluxes with a neat geometrical interpretation in terms of a Heegaard splitting of the space manifold.
Finally, I discuss the consequences of the proposed deformation on the quantization of the phase space, its quantum gravitational interpretation, as well as its relevance for the construction of $(3+1)$ dimensional topological field theories with defects.
\end{abstract}

\maketitle

\section{Introduction}

The non-perturbative study of Yang--Mills theories and quantum gravity often appeals to a lattice-like regularization.
One feature of this type of regularization is that of spoiling the symmetry between the electric and magnetic components of the field strength tensor.
In fact, on the lattice, magnetic fluxes are encoded in Lie group variables, whereas electric fluxes are encoded in Lie algebra variables. This distinction is particularly relevant since the first live in a curved compact space, while the letter in a non-compact linear space. 
Quantum mechanically, this leads to non-commutative electric flux operators---even at the gauge-covariant level---with a discrete spectrum, while magnetic flux operators keep enjoying a continuum spectrum and commute among themselves.

In this paper, I show how to construct a self-dual deformation of the (classical) Yang--Mills lattice phase space, with both electric and magnetic degrees of freedom encoded in {\it compact} Lie group variables. 
Now, while $\mathrm T^\ast G \cong G\times \Lie(G)^\ast$, $G$ a (compact semisimple) Lie group, posses a canonical symplectic structure coinciding with that of Yang--Mills theory restricted to an edge of the lattice, the space $G\times G$ generally does not admit {\it any} symplectic structure at all. 
This is the main difficulty in the path to a self-dual phase space. To solve it, I turn to the mathematical framework of ``quasi-Hamiltonian spaces''.

This framework was developed in the late nineties mostly by Alekseev, Kosmann-Schwarzbach, Malkin, and Meinrenken \cite{Alekseev1998,Alekseev2000,Alekseev2000b,Alekseev2002}. 
Their principal motivation was to provide a finite dimensional construction of the symplectic structure on the moduli space of flat connection on a Riemann surface \cite{Atiyah1983,Goldman1984,Jeffrey1994,Alekseev1994,Alekseev1998}, hence generalizing the work of Fock and Rosly \cite{Fock1999} to the compact group case. In the process, they solved the more abstract problem of constructing a complete and satisfactory theory of group-valued momentum maps. 
Their work furthermore connects to a line of research seeking the ``classical'' analogue of the quantum group structures of Drinfeld's \cite{Kosmann-Schwarzbach1991}.

While these problems are tied to the rich and fruitful interplay between low-dimensional topology and gauge theories, some of the results can be put to fruition in higher dimensions too. 
Indeed, one of the auxiliary constructions of Alekseev {\it et al.} is that of a quasi-Hamiltonian structure on the ``double'' $D(G)\cong G\times G$, which generalizes the canonial symplectic structure on $\mathrm T^\ast G$.
The starting point of this paper is simply that of employing such a structure to build the self-dual deformation of the phase space of lattice Yang--Mills theory advocated above. 
One important feature of this construction is its non-locality: the building blocks associated to the edges of $\G$ are quasi-Hamiltonian spaces which do not carry an actual symplectic structure. The final gauge-invariant phase space---obtained via a reduction by the deformed Gau{\ss} constraints---is, instead, a full-fledged symplectic space.

In $(3+1)$ spacetime dimensions, the self-dual structures of the deformed phase space are enhanced by an intimate interplay between symplectic geomtry and topology. This enhancement spurs from the isomorphism between the deformed phase space on a lattice $\G$ and the moduli space of an auxiliary flat connection on the Riemann surface $S_\G$ bounding a tubular neighborhood $H_\G$ of the lattice.%
\footnote{
{\it Note added} A deformation of Hamiltonian Yang--Mills theory similar in spirit to the present one, as well as its relation to the moduli space of flat connection on a Riemann surface, was proposed and studied by Frolov in the 1990s \cite{Frolov1995a,Frolov1995b,Frolov1995c}. 
For more on the analogies and differences between the two approaches see the {\it Note added} at the end of this paper. 
}

Furthermore, when interpreted in the light of loop quantum gravity---it is in this context that many of the ideas presented in this paper first appeared \cite{HHKR,HHKRplb,Haggard2016}---another layer is added to the interplay between dual structures. A four-valent lattice $\G$ dual to a triangulation $\Delta$ of a three-dimensional Cauchy surface $\Sigma$ turns out to encode the geometry of a homogeneously curved tetrahedra, in which dihedral angles and edge lengths plays dual roles. The appearance of homogeneously curved (twisted) geometries can be explicitly related to a non-vanishing cosmological constant \cite{HHKR,HHKRplb,Haggard2016}. Moreover, another appealing feature of the loop gravitational interpretation is the fact that the quite natural Fenchel--Nielsen coordinates on the reduced phase space have a clear geometrical significance too \cite{Haggard2015,Han2017} (also \cite{Delcamp2016,Dittrich2017}).
Notice, also, that similar ideas, mostly relying on the Fock--Rosly construction, have been largely studied in $(2+1)$ dimensional loop quantum gravity \cite{Noui2006,Dupuis2013,Dupuis2014,Bonzom2014a,Bonzom2014,Charles2015,Charles2017} (see also \cite{Meusburger2016,Meusburger2016a}).

For what concerns the quantization of the deformed phase space, from its compactness one can immediately deduce that the ensuing Hilbert space is finite dimensional and all the operators must hence be bounded---a fact particularly appealing from the quantum gravitational perspective \cite{HHKR,HHKRplb,Haggard2016} (also \cite{Noui2003a,Han2011b,Fairbairn2012}).
Although I do not attempt a rigorous quantization of the deformed phase space in this paper, it is clear from the previous discussion that a tight relationship with quantum groups is present. In particular, in $(3+1)$ dimensions, the connection with Chern--Simons theory and the moduli space of flat connection on $S_\G$ proves that a quantization of this phase space has already been constructed by Dittrich \cite{Dittrich2017}, whose motivation were rooted in the study of four-dimensional topological quantum field theories with defects \cite{Walker2012,Wan2015, Lan2017,Barenz2016}.%
\footnote{For further connections with the subject of  topological phases of matter, see \cite{Dittrich2017} and references therein.}
The present work sheds further light on her construction, providing its classical limit and more evidence for its connection with the Crane--Yetter model \cite{Crane:1993if,Barrett2007} (see also \cite{HHKR,Kodama1988} in relationship with this connection). It also suggests new and more geometrical ways to couple her model to lower dimensional defects, a compelling fact from the viewpoint of extended topological field theories.\\

The paper is organized as follows. In section \ref{sec_overview}, I give an overview of the technical content of the paper. This is meant to serve as a roadmap (and summary) of the many technical issues touched upon in the following sections. This section is not required for following the rest of the paper, since all notation will be reintroduced at due time. 
Section \ref{sec_classical} is where the paper actually starts. It reviews the formal construction of the phase space of lattice Yang--Mills theory, hence preparing the terrain for the subsequent deformation presented in section \ref{sec_selfdual}. Reviewing the work of \cite{Alekseev1998} this section is mostly technical. It contains, however, some more general comments about the structure of the double and its decomposition on holonomy and flux space which I believe of more general interest (subsection \ref{sec_GtimesG}). Then, in section \ref{sec_remarks}, I briefly collect some remarks on the notion of self-duality which emerged so far. This serves as a motivation for section \ref{sec_2surface}, which discusses the enhanced duality found in $(3+1)$ dimensions. This section is divided in three parts: in the first part, I present the main ideas and features, mostly in a discursive way; in the second part, I discuss two basic examples quite explicitly; and, finally, in the third part I give a proof---adapted from \cite{Alekseev1998}---of the statement that the deformed (reduced) phase space of lattice Yang--Mills theory on $\G$ is naturally isomorphic to the moduli space of an auxiliary flat $G$-connections on $S_\G$. This third part contains many technical details and is not necessary to follow the rest of the discourse. The remaining three section are again largely discursive and explore in a preliminary way the consequences of the duality structure of the deformed $(3+1)$ dimensional phase space for the concepts of polarization, excitations, and defects (section \ref{sec_polarization}), for quantization (section \ref{sec_quantum}), and for quantum gravity with a cosmological constant (section \ref{sec_gravity}). Section \ref{sec_conclusions} contains some closing remarks. 
Finally, there are three short appendices. The first two regard the relation between the (quasi-)symplectic and (quasi-)Poisson frameworks, appendices \ref{app_A1} and \ref{app_A2}, while the last one discusses the relation between the Poisson (non-)commutativity of the holonomies in the quasi-Poisson framework with the  failure of the Jacobi identity.

\section{Overview \label{sec_overview}}

In this appendix, I briefly and schematically overview the construction advocated in the paper. 

The first step is to regularize the phase space of Yang--Mills theory through the introduction of a graph $\G\subset\Sigma$, dual to a cellular discretization of $\Sigma$, so that its edges carry holonomies and (electric) flux variables.
This construction, however, leads to an asymmetric treatments of the magnetic and electric fields: electric degrees of freedom are encoded into Lie-algebra elements, while magnetic ones into Lie-group elements.
To restore the symmetry at the discrete level, I advocated a deformation of the phase space by a replacement of the Lie-algebra-valued electric fluxes, with Lie-group-valued electric fluxes: 
\be
(h_e,X_e)\in (G\ltimes_\Ad \frakg)_e \cong (\mathrm T^\ast G)_e 
\quad\leadsto\quad
(h_e,g_e)\in (G_\text{h}\ltimes_\AD G_\text{f})_e \cong D_e(G),
\ee
where $D(G)$ is the {\it double} of $G$, and $\AD$ is the action of $G$ on itself by conjugation. I referred to this procedure as {\it flux exponentiation}, see figure \ref{fig_DG}.

Aiming for a local construction of the phase space of Yang--Mills theory on $\G$, this idea soon encounters a stumbling block. The issue is that, while $\mathrm T^\ast G$ carries a canonical symplectic structure, $D(G)$ generally cannot carry any symplectic structure at all.
Following \cite{Alekseev1998}, I temporarily put this issue aside and rather focused on the symmetry properties of the phase space.
 
Yang--Mills theory, regularized on $\G$ in terms of holonomy-flux variables as above, still supports residual gauge symmetries at the vertices $v\in\G$.
These symmetries are
\be
(h, X) \mapsto (g_t^{-1} h g_s, \Ad_{g_s} X)
\quad\leadsto\quad
(h,g) \mapsto (g_t^{-1} h g_s, \AD_{g_s} g)
\ee
At the level of a single edge $e\in\G$, these symmetries are generated by the electric fluxes
\be
X_e \quad\text{and}\quad \tl X_e = - \Ad_{h_e} X_e
\quad\leadsto\quad
g_e \quad\text{and}\quad \tl g_e = \AD_{h_e} g^{-1}_e,
\ee 
at the source and target vertices of $e$, respectively.

In the undeformed case, this means that an infinitesimal gauge transformation $Y\in \frakg$ at the source vertex $s(e)$ is generated by the Hamiltonian function
\be
H_Y = Y_i X^i = \la \mu, Y \ra
\qquad\text{with}\qquad
\mu(h,X) = X^i\tau_i,
\ee 
where the undeformed {\it momentum map} was introduced,
\be
\mu : (\mathrm T^\ast G) \to \frakg, \qquad (h,X) \mapsto X
\ee
as a projection on the flux component of the phase space. (Similarly, there is a momentum map $\tl \mu(h,X) = \tl X$ assocaited to gauge transformations at the target vertex of $e$.)
 
Denoting ${Y}^\sharp$ the flow of the gauge transformation $Y$ in $\mathrm T^\ast G$, one has
\be
Y^\sharp = \{ H_Y , \; \cdot \;\} 
\qquad\text{iff}\qquad
Y^\sharp \contr \omega = \d H_Y = \la \d\mu, Y \ra,
\ee
where the symplecitc form on $\mathrm T^\ast G\cong G\times \frakg$ is
\be
\omega(h,X) = -\d \la X, \theta_h^L \ra = \la \theta^L_h \stackrel{\wedge}{,} \d X\ra+ \tfrac12\la \ad_X\theta^L_h,\stackrel{\wedge}{,}\theta^L_h\ra
\ee
where $\theta^L_h = h^{-1}\d h $ and $\theta^R_h = \d h h^{-1}$.

Observing that at lowest order in the expansion $g\approx \mathbb 1 + X + \dots$, 
\be
\theta^R_g \approx \d X  \approx \theta^L_g ,
\ee
the following generalizations were proposed \cite{Alekseev1998},
\begin{align}
\omega(h,X)= - \d \la X, \theta^L_h\ra & \leadsto \omega(h,g)  = \la \theta^L_h \stackrel{\wedge}{,} \tfrac12 (\theta^L_g + \theta^R_g)\ra+ \tfrac12\la \Ad_g\theta^L_h,\stackrel{\wedge}{,}\theta^L_h\ra\\
Y^\sharp \contr \omega = \la \d\mu , Y\ra& \leadsto Y^\sharp \contr \omega =  \la \tfrac12(\theta^L_\mu + \theta^R_\mu), Y \ra ,\label{eq_qflow}
\end{align}
The two-form $\omega$ on $D(G)$ is however not symplecitc, since it is neither non-degenerate, nor closed:
\begin{align}
\ker \omega_{(h,g)} & = \big\{ \mathfrak{v} = \mathcal Y^\sharp \; \text{for some} \;  \mathcal Y \in \ker( \Ad_{\mu(h,g)} +1 ) \big\} \subset \mathfrak{X}^1(D) \label{eq_S1}\\
\d \omega & = \chi_\mu.\label{eq_S2}
\end{align}
For these reasons, I referred to $\omega$ in the deformed case as a ``quasi-symplectic'' two-form.
The fact that $\omega$ is not closed can be interpreted as a violation of the Jacobi identity. As discussed at the end of Appendix \ref{app_A2}, one of the physical information subsumed by the violation of the Jacobi identity is that the holonomy variables do not Poisson-commute with one-another.
This fact is a consequence of the curvature of the flux space, which---after the deformation $\frakg \leadsto G$---fails to be a linear space.

To build the phase space associated to the {\it whole } graph $\G$, one needs to stitch together multiple copies of $\mathrm T^\ast G \leadsto D(G)$. 
On the deformed side, difficulties arise when one is required to make sense of the ``total'' generator of the gauge transformation at each vertex. ``Total'' here means that the gauge transformation so generated acts ``in the same way'' (diagonally) on the end-points of all the edges that end (or start) at the given vertex.
The starting point is the requirement that the total momentum map---representing the (deformed) Gau{\ss} constraint at $v\in\G$---is deformed by
\be
\mu_v = \sum_{e:v\in\pp e} \mu_e 
\qquad\leadsto\qquad
\mu_v = \overleftarrow{\prod_{e:v\in\pp e}}\mu_e
\ee
(here the assumption was made that the vertex $v$ is the source of for all $e$ such that $v\in\pp e$).
It is clear, however, that the (cyclical) order of the factors is crucial on the deformed side. 
Moreover, since $\theta^L_{g_2g_1} \neq \theta^L_{g_2} + \theta^L_{g_1}$, the total quasi-symplectic two forms cannot be simply $\sum_e \omega_e$ as in the underformed case, otherwise equation \eqref{eq_qflow} would not be satisfied for the total gauge transformation described by the above momentum map. 

Consequently, the total phase space associated to $\G$ has to be built step by step by {\it fusing} together one edge after the other taking care of the ordering of this fusion product and appropriately twisting the corresponding total quasi-symplectic form.

Finally, once the total, gauge-{\it co}variant phase space has been constructed, one can ask how to reduce it to obtain a gauge-{\it in}variant phase space:
\be
\mP_\G = (\mathrm T^\ast G)^{\times E}// G^{\times V}
\qquad\leadsto\qquad
\mP^\EM_\G = D(G)^{\times E}// G^{\times V}.
\ee
The reduction procedure is completely analogous in the two cases, and interestingly leads in both cases to an actual {\it symplectic} phase space structure on the space of gauge orbits withing the constraint surface $\mathrm{ker}(\mu)$. Intuitively, this can be traced back to the fact that all the anomalies related to the deformed case depend on the momentum map $\mu$ (see the rhs of the equations \eqref{eq_S1} and \eqref{eq_S2}), whereas at the reduced level all symmetries have been ``taken care of'' and no momentum map is left to source such anomalies (in other words, a quasi-symplectic space with trivial group-valued momentum map, is symplectic). 

The above construction takes a special meaning in $(3+1)$ spacetime dimensions.
In this case, the group elements $(h,g)$ can be interpreted as the longitudinal and transverse parallel transports associated to an auxiliary (Poisson non-commutative) $G$-connection $\A$ along the boundary of the thickened edges $e\in\G$, see figure \ref{fig_cyl}. 
More specifically, the above phase space is isomorphic to the (Atyiah--Bott) moduli space of flat connection on the surface $S_\G$ obtained by thickening $\G$ itself:
\be
\mP^\EM_\G \cong \M(S_\G , G).
\ee
 Since $S_\G$ is the Heegaard surface associated to the Heegaard splitting induced by the cellular decomposition  $\Delta$, it  turns out that the face holonomies, $h_f = \overleftarrow{\prod}_{e:e\in\pp f} h_e^\epsilon$, and the exponentiated fluxes $g_e$ play perfectly symmetric roles. 
This symmetry originates in the symmetric roles played by the two handlebodies $H_\G$ and $H_{\G^\ast}$ constituting the Heegaard decomposition of $\Sigma$,
\be
\Sigma \cong H_{\G} \cup_{S_\G} H_{\G^\ast}
\qquad \text{where} \qquad 
\pp H_{\G} \cong S_\G \cong \pp H_{\G^\ast}^\text{op},
\ee
induced by the cellular decomposition $\Delta$, see figure \ref{fig_HG_HGstar}.
In particular, the Gau{\ss} constraint and the discrete Bianchi identities (i.e. the divergenceless of the electric and magnetic fields in the absence of electric sources) acquire perfectly dual geometrical interpretations, since both are related to the topological contractibility of some paths on $S_\G$ around the vertices of $\Delta$ or $\Delta^\ast \cong \G$, respectively (see figure \ref{fig_vertexundone}).

Therefore, in $(3+1)$ dimensions, three notions of symmetry and duality mirror each other in a mutual interplay: the symmetry between the Yang--Mills electric and magnetic fields, the symplecto-geometric symmetry between position and momentum space (Born reciprocity \cite{Born}), and the symmetry between the graph $\G$ and its dual enhanced to the symmetry between the two handlebodies in a Heegaard decomposition of the Caucy surface $\Sigma \cong H_{\G} \cup_{S_\G} H_{\G^\ast}$.

If one further specialized to $(3+1)$ loop quantum gravity and curved (twisted) geometries, another element is integrated in the above picture: the reduced phase space attached to a four-valent vertex of $\G$ can be interpreted as the space of shapes of homogeneously curved tetrahedra, for which the above symmetries take the form of a duality between the edge length and the dihedral angles.

\section{The classical spin-network phase space \label{sec_classical}}

In this section, I briefly review the phase space structure underlying spin networks states.

Spin networks constitute a specific basis of the gauge invariant Hilbert space of a Yang--Mills (YM) theory on a graph $\Gamma$, which can be---but does not have to be---a regular lattice.
They are characterized by the property of diagonalizing the electric flux operators on the edges of the graph.
I will use the term more loosely, essentially indicating a gauge-field configuration on $\Gamma$. 

More specifically, a configuration of the magnetic-potential $A(x)\in\Omega^1(\Sigma)\otimes\Lie(G)$ on the $d$-dimensional space-like manifold $\Sigma$ is regularized via the introduction of an embedded graph $\Gamma\subset \Sigma$. $\Sigma$ is here understood to be a Cauchy surface in a globally hyperbolic spacetime $M\cong\Sigma\times\mathbb R$. Moreover, it will be useful to assume that the graph $\Gamma$ is the connected 1-skeleton of $\Delta^\ast$, the dual of a cellular discretization $\Delta$ of $\Sigma$. 
The graph $\Gamma$ has, say, $E=\#\{e\}$ oriented edges and $V=\#\{v\}$ vertices, and is closed, i.e. all its vertices have valencies larger than 2, which corresponds to $\pp \Sigma = \emptyset$.
On $\Gamma$, $A(x)$ is replaced by parallel transports (hereafter referred to as holonomies) associated to the oriented edges of the graph,
\be
h_e=\overleftarrow{\Pexp} \int_e A \in G,
\label{eq_he}
\ee
and the group of residual gauge transformations is left to act only at vertices, 
\be
h_e\mapsto g_{t(e)}^{-1} h_e g_{s(e)}
\label{eq_htoghg}
\ee
 with $(s,t)$ representing the source- and target-vertex maps.
Therefore, the classical configuration space $\mQ_\Gamma$ associated to a spin network $\Gamma$ is given by one copy of the compact and semisimple gauge group $G$ per link of the graph, modulo the residual gauge transformations. Schematically,
\be
\mQ_\Gamma\cong G^{\times E}// G^{\times V}.
\ee

The conjugate variable to the gauge-potential $A(x)$ is the Lie-algebra valued electric field%
\footnote{To be precise, in a $(d+1)$-dimensional spacetime, $A\in\Omega^{1}(\Sigma)\otimes\mathrm{Lie}(G)$ and $E\in\Omega^{d-1}(\Sigma)\otimes{Lie}(G)$. Under gauge transformations $A$ transforms as a connection, and $E$ in the adjoint representation.  Symbolically, $\{A,A\}=0=\{E,E\}$ and $\{ \ast E, A \} = \delta$ where $\ast$ is the Hodge dual in $\Omega^\bullet(\Sigma)$ and, in this deWitt-like notation, $\delta$ stands for a Kronecker delta between internal and tangent-space indices, as well as for a Dirac delta between the space locations.}
 $E(x)$.
 Upon regularization, $E(x)$ is replaced by Lie-algebra-valued electric fluxes $X_e$ associated to the codimension-1 cells of $\Delta$, dual to graph edges,
\be
X_e = X_e^i\tau_i \in \frakg,
\ee
where $\{\tau_i\}$ is an orthonormal basis of $\frakg=\Lie(G)$ with respect to the Killing form $\la\cdot,\cdot\ra$, which is used to
raise and lower Lie algebra indices.

This flux has to be understood as residing at the source vertex of $e$.
Then, the action of a gauge transformation is 
\be
X_e \mapsto \Ad^{-1}_{g_{s(e)}} X_e.
\label{eq_XtoAdgX}
\ee
Fluxes $\tl X_e$ residing at the target vertices of their edges can be readily defined by parallel transport,
\be
\tl X_e = -\Ad_{h_e} X_e.
\label{eq_AdhX}
\ee

The phase space $\mP_\Gamma$ associated to $\Gamma$ is therefore
\be
\mP_\G \cong \big(G\times \frakg\big)^{\times E}//G^{\times V}.
\label{eq_Gg//G}
\ee

The Yang--Mills action, together with a proper understanding of the flux regularization leading to the variables $X_e$ (see e.g. \cite{Freidel2011}), equips $\mP_\Gamma$ with the following symplectic structure:
\be
\{h_e,h_{e'}\}=0,
\qquad
\{X_e^i,h_{e'}\} = \delta_{e,e'}  h_e \tau^i,
\qquad\text{and}\qquad
\{X_e^i, X_{e'}^j \} = \delta_{e,e'} f^{ij}{}_k X^k_e ,
\label{eq_T*G}
\ee
where $f_{ijk}= \la[\tau_i,\tau_j], \tau_k\ra$ are the structure constants of $\frakg$.
The need to lower and raise Lie algebra indices, in otherwise very natural expressions, betrays an identification of $\frakg$ and its dual $\frakg^\ast$ via the Lie algebra Killing form and the consequent translation of the natural symplectic structure on $G\times\frakg^\ast$. 

Indeed, the symplectic structure of equation \eqref{eq_T*G} is nothing but the canonical symplectic structure on the cotangent bundle $\mathrm T^\ast G$ written in coordinates associated to the left trivialization of $\mathrm T^\ast G \cong G\times\frakg^\ast$.  This trivialization is, however, most intuitively expressed in $\mathrm TG$, where it identifies $T_{h=e}G=\frakg$ and $\mathrm T_h G$ via the one-to-one map which identifies a Lie algebra element $X\in\frakg$ with the value of the right-invariant vector-field $\hat X^R\in\mathfrak X^1(G)$ at $h\in G$. In formulas,
\be
X\in\frakg \mapsto \hat X^R_h \in \mathrm T_h G
\qquad\text{where}\qquad
\hat X^\text{R}_h f(h) = \frac{\d}{\d t}_{|t=0} f(\E^{-tX}h) 
\qquad \forall  f\in\mathcal C^1(G).
\ee
Similarly, using $\tl X$ as coordinates on the phase space corresponds to  choosing the opposite trivialization through the left-invariant vector-field: $ \big(\hat{\tl X}\big)^R_h f(h) = \frac{\d}{\d t}_{|t=0} f(\E^{-t\tl X}h) = \hat X^L_h f(h)$. 
Henceforth, I will leave the identification between $\frakg$ and its dual via the Killing form implicit.

Thus, the classical spin-network phase space is simply
\be
\mP_\Gamma \cong \big( \mathrm T^\ast G\big)^{\times E}//G^{\times V},
\label{eq_P=T*G}
\ee
where $\mathrm T^\ast G$ is equipped with the canonical symplectic structure associated to cotangent bundles.

In order to be able to generalize this construction appropriately in the following sections, a more formal understanding of the residual gauge symmetry is needed. 
As can be read directly from equation \eqref{eq_T*G}, gauge symmetries at the source vertex of an edge (see equations \eqref{eq_XtoAdgX} and \eqref{eq_htoghg}) are generated  by the flux $X_e^i$ (in turn, at the target vertex, the symmetries are generated by $\tl X^i_e$). 
The Hamiltonian generator of an infinitesimal symmetry parametrized by $Y\in\Lie(G)$ is hence
\be
H^e_Y = X^i_e Y_i = \la X, Y\ra,
\ee
since
\be
\{ H^e_Y, h_{e'}\} = \delta_{e,e'} h_e Y
\qquad\text{and}\qquad
\{H^e_Y, X_{e'} \} = \delta_{e,e'} Y_i f^{ij}{}_k X^k_e \tau_j =\delta_{e,e'} [X,Y].
\ee

Although the previous discussion is tailored to the residual $G$ symmetry at a vertex of $\G$, this can be readily generalized to an arbitrary phase space $\mP$ acted upon by a group $\mathcal G$.
It is a general fact that the Hamiltonian generator of an infinitesimal $\mathcal G$-symmetry parametrized by a Lie algebra element $\mathcal Y\in\Lie(\mathcal G)$ is a function on phase space which is linear in $\mathcal Y$.
For this reason, on very general grounds, one introduces for each symmetry a momentum map $\mu$ from the phase space $\mP$ into the dual of the Lie algebra
\be
\mu : \mP \to \Lie(\mathcal{G})^\ast,
\ee
so that the Hamiltonian generator takes the form
\be
H_\mathcal{Y} = \mu(\mathcal{Y}) = \la \mu, \mathcal Y\ra
\qquad\text{and}\qquad
\{ H_{\mathcal Y}, \cdot \} = \pounds_{\mathcal Y^\sharp},
\label{eq_PoissonHam}
\ee
where $\mathcal Y^\sharp$ stands for the phase space flow (vector field)  associated to the infinitesimal symmetry $\mathcal Y\in\Lie(\mathcal G)$ and $\pounds$ for the Lie derivative.
If the Poisson algebra of the Hamiltonian generators is a representation of the symmetry algebra $\Lie(\mathcal G)$, then this property is translated into the equivariance of the momentum map,
\be
\pounds_{\mathcal Y^\sharp} \mu \equiv \mathcal Y^\sharp \contr \d \mu = \ad^\ast_\mathcal{Y}\mu = - \ad_{\mathcal Y} \mu,
\ee
where in the last step we identified $\Lie(\mathcal G)^\ast$ and $\Lie(\mathcal G)$ via the Killing form.

In the case considered above, the momentum map for the gauge symmetry at the source of $e$---when written in the left trivialization $T^\ast G \cong G\times \frakg$---is a projection on the flux factor $\frakg$.

The quotienting procedure of equation \eqref{eq_P=T*G} is a prototypical example of a Marsden--Weinstein symplectic reduction, which can be described as follows.
Consider a phase space $\mP'$ invariant under the action of some Lie group $\mathcal G$, with equivariant momentum map $\mu:\mP'\to\mathbb \Lie(\mathcal G)^\ast$.  
Then a canonical phase space structure can be produced on the ``constraint surface'' $\{\mu =0\}\subset\mP'$ once symmetry-related configurations are identified, i.e. on $\mP = \{\mu=0\}/\sim_\mathcal{G}$. 
This is often denoted as
\be
\mP = \mP'//\mathcal G.
\ee
The double quotient reminds us that both restriction to the constraint surface {\it and} identification among symmetry-related configurations are needed.
In physicists' parlance, $\dim(\mathcal G)$ first-class constraints $\mu_i=0$ have been imposed in the passage from $\mP'$ to $\mP$.

Let me go back to equations \eqref{eq_Gg//G} and \eqref{eq_T*G}, and reformulate the symmetry properties of $\mP'$ in the momentum map language. 
In this context, $\mP' \cong (G\times \frakg)^{\times E}$ and $\mathcal G = G^{\times V}$. Restricting the attention to a single edge $e\in\Gamma$, the gauge group reduces to two copies of $\mathcal G_e=G_s\times G_t$, associated to the source and target of $e$ respectively. Moreover, the flow $\mathcal Y_e^\sharp$ of a gauge transformation $\mathcal Y_e=(Y_s, Y_t)$ has the following form on the holonomy and flux spaces:
\be
\mathcal Y_e^\sharp = \mathcal Y_e^\sharp|_\text{h} + \mathcal Y_e^\sharp|_\text{f} ,
\qquad\text{with}\qquad
\mathcal Y_e^\sharp|_\text{h} =   \hat Y_s^L + \hat Y^R_t
\qquad\text{and}\qquad
\mathcal Y_e^\sharp|_\text{f}= - \ad_{Y_s} .
\label{eq_Ysharpflat}
\ee
Finally, an inspection of the symplectic structure associated to $e$ shows that the momentum map generating the gauge transformation $\mathcal Y$ above is given by the source and target fluxes:%
\be
\mu_e=(\mu_s,\mu_t):  G\times\Lie(G) \to \Lie(\mathcal G),
\qquad
 (h,X) \mapsto (X, \tl X),
\label{eq_mue=X}
\ee
with the identification between $\Lie(\mathcal G)$ and its dual left implicit. 

Hence, going back to the phase space associated to the full graph $\G$, the momentum map of residual gauge transformations on $\Gamma$ is
\be
\mu: \mP'= (G\times \frakg)^{\times E} \to \frakg^{\times V},
\qquad
\mu = \Big( \mu_v = \sum_{e:v=s(e)} \mu_{s(e)} + \sum_{e:v=t(e)} \mu_{t(e)}\Big)_v.
\label{eq_mu=sumX}
\ee

Physically, constructing the reduced phase space $\mP_\Gamma$ as the symplectic reduction of $( G \times \frakg)^{\times E}$ with respect to the momentum maps $\mu_v =0$, means restricting to those points in phase space which respect the vacuum Gauss constraint (vanishing of the total electric flux) at each vertex, while at the same time identifying gauge-related configurations. 

All the above statements can be translated from a Poisson-theoretic to a symplectic-theoretic language.
Consider a phase space $\mP$ equipped with Poisson brackets $\{\cdot,\cdot\}$. 
Introduce the bivector $P$, i.e.
\be
P \in \mathfrak{X^1}(\mP)\otimes_A \mathfrak{X^1}(\mP)
\ee
with $\otimes_A$ standing for the antisymmetric part of the tensor product, defined by
\be
P(\d f_1 \otimes \d f_2) = \{ f_1, f_2 \}
\qquad\forall f_1,f_2\in\mathcal C^1(\mP).
\ee
Its inverse%
\footnote{
The bivector $P$ can be seen as a map $P^\sharp$ from 1-forms to vector fields:
\be
P^\sharp:\Omega^1(\mP) \to \mathfrak{X}^1(\mP), \quad \d f \mapsto P^\sharp(\d f) = P(\d f, \cdot) = \{f, \cdot \}.\notag
\ee
The standard definition of a phase space requires this map to be invertible. Denote its inverse by $ \omega_\flat : \mathfrak{X}^1(\mP) \to \Omega^1(\mP)$,
\be
  (P^\sharp) \circ \omega_\flat = \mathrm{id}.\notag
\ee
This can similarly be understood as descending from a 2-form, $\omega\in\Omega^2(\mP)$, via $\omega_\flat({\frak v}) = \frak v\contr\omega$, for all ${\frak v}\in\mathfrak X^1(\mP)$. 
\label{fn_Pomega=1}
}
is a two-form $\omega\in\Omega^2(\mP)$ known as the symplectic form on $\mP$. The existence of such an inverse corresponds to a non-degeneracy condition on either $P$ or $\omega$, while the Jacobi identity satisfied by the Poisson brackets%
\footnote{In terms of $P$, the Jacobi identity is encoded in the vanishing of the so-called Schouten bracket of $P$ with itself, $[\![ P,P ]\!]=0$. See \cite{Kosmann-Schwarzbach?}.}
becomes a simple closure requirement,
\be
\d \omega = 0.
\ee

From the symplectic perspective, the Hamiltonian flow equation \eqref{eq_PoissonHam} reads
\be
\mathcal Y^\sharp\contr\omega =   \d H_\mathcal{Y}  \equiv \la \d\mu, \mathcal Y\ra .
\label{eq_symplHam}
\ee

Finally, ``dual'' to the left (right) invariant vector fields $\hat X^L$ ($\hat X^R$, resp.) on $G$ are the Lie-algebra valued left (right) Maurer--Cartan one-forms $\thetaL\in\Omega^1(G)\otimes \frakg$ ($\thetaR$, resp.), fully determined by the conditions
\be
\hat Y^L \contr\thetaL = Y
\qquad (\hat Y^R \contr\thetaR = -Y , \text{ resp.}).
\ee
They satisfy the Maurer--Cartan equations%
\footnote{
Recalling that $\theta^{L,R}\in\Omega^1(G)\otimes\Lie(G)$, the `commutator-wedge' notation means that the wedge-product is taken among the one-form parts of $\theta^{L,R}$ and the commutator is taken among its Lie-algebra parts. Take $\{z^a\}$ to be coordinates on $G$, then the (left) Maurer--Cartan form reads
$$
\theta^L = \theta^{L}{}^i_\mu \d z^a \tau_i  = h^{-1}(z)\frac{\pp h(z)}{\pp z^a} \d z^a,
$$
where in the last equality $G$ is assumed to be a matrix group, and the Maurer--Cartan equation \eqref{eq_MCeq} is
$$
\d \theta^L{}^i = -\tfrac12 f_{jk}{}^j \theta^L{}^j \wedge \theta^L{}^k.
$$
}
\be
\d \thetaL = -\tfrac12 [\thetaL\stackrel{\wedge}{,}\thetaL]
\qquad (\d \thetaR = -\tfrac12 [\thetaR\stackrel{\wedge}{,}\thetaR] , \text{ resp.}).
\label{eq_MCeq}
\ee
It is useful to recall that in coordinates, at $h\in G$, they read
\be
\thetaL_h = h^{-1} \d h 
\qquad (\thetaR_h = \d h h^{-1}, \text{ resp.}).
\ee

For future reference, let me remark that if $h$ is close to the identity, $h=\E^{X}$, at first order in $X$ one has
\be
\theta^L_h \approx \d X \approx \theta^R_h.
\label{eq_approx}
\ee

Hence, it is easy to verify that the canonical symplectic structure on $(T^\ast G)^{\times E}$ derived from equation \eqref{eq_T*G} takes the following simple form (see Appendix \ref{app_A1}): 
\be
\omega = \sum_e \omega_e
\qquad\text{where}\qquad
(\omega_e)_{(h,X)} = - \d \la X, \thetaL_h \ra = \la \thetaL_h\stackrel{\wedge}{,}\d X\ra + \tfrac12 \la \ad_X \thetaL_h\stackrel{\wedge}{,}\thetaL_h\ra,
\label{eq_omegaTG}
\ee
where in the last equality the $\ad$-invariance of the Killing form was used.
Notice that $\omega$ is invariant under the action of $\mathcal G_e$, and that---$\omega$ being exact---the Jacobi identity is manifestly satisfied. 
Notice also that the curvature of the compactified configuration space $G$ ($\d\theta\neq0$) induces a Poisson non-commutativity in the conjugate flux variables.

\section{Self-dual phase space on $\Gamma$ \label{sec_selfdual}}

In the description above, the introduction of $\Gamma$ results in the compactification of the configuration variable $A(x)$ to the set of group-valued variables $\{h_e \in G\}$.

While in the continuum the magnetic field is encoded in the curvature of $A$, i.e. $B= F[A] =\d A + \tfrac12[A\stackrel{\wedge}{,}A]$, on $\Gamma$ magnetic fluxes are encoded in face- (or plaquette-)holonomies and thus got ``compactified'' too:
\be
h_f = \overleftarrow{\prod_{e:e\in\pp f}} h_e^{\epsilon_{ef}} \in G 
\label{eq_hf}
\ee
with $\epsilon_{ef} = \pm1$ accounting for the relative orientation of $e$ and $f$.

In turn, electric fluxes are valued in the Lie algebra in the continuum as well as on $\Gamma$. 
Self-duality (SD) on $\Gamma$ hence demands a further compactification of the electric fluxes. 
Requiring a passage from $\frakg$  to $G$, this procedure is sometimes called ``flux exponentiation'' (more on this at the end of the section).
Here, I won't attempt its justification from an action principles, although further comments on this are provided at the end of sections \ref{sec_polarization}, \ref{sec_quantum}, and especially in the section about the gravitational interpretation, i.e. section \ref{sec_gravity}. 
Also, at this level, the ``flux exponentiation'' works independently of the dimension of $\Sigma$.
In $3+1$ spacetime dimensions, however, a completely self-dual dual description can be given.
I will discuss specifically this situation in section \ref{sec_2surface}.

The goal is to build a SD phase space which reduces to the standard one in the appropriate small-electric-flux limit.  Schematically, it will have the form
\be
\mP_\Gamma^\EM\cong \text{``}\big( G\times G \big)^{\times E}// G^{\times V}\text{''} .
\ee
The difficulty is that, for $G$ compact and simply connected, $G\times G$ cannot carry {\it any} symplectic structure (i.e. a non-degenerate closed two-form), since its second cohomology is trivial. Let alone a canonical one.
Therefore, it might be surprising that something like $\mP^\EM_\Gamma$ can be given a symplectic structure at all. 
This can be achieved by introducing the notions of {\it quasi-Hamiltonian $G$-spaces}, which allows one to equip $G\times G$ with a ``quasi-symplectic'' structure, and that of {\it fusion}, which allows one to assemble quasi-Hamiltonian $G$-spaces together. Then, a generalized version of a Marsden--Weinstein symplectic reduction, can be used  to turn the quasi-Hamiltonian space associated to $\Gamma$, as obtained by fusion, into an actual symplectic space. 

All these technologies have been developed by Alekseev, Kosmann-Schwarzbach, Malkin, and Meinrenken \cite{AKMM}.
In this paper, I will review them and put them into use. 
I will proceed in two steps: ({\it i}) first I introduce the quasi-symplectic analogue of the edge phase space; then ({\it ii}) I describe the fusion procedure to obtain $\mP^\EM_\Gamma$. I will work in the (quasi-)symplectic framework, although a (quasi-)Poisson framework is also available and is presented in the appendix.

\subsection{The quasi-Hamiltonian $G$-space $D(G)=G\times G$\label{sec_D}}

Since many instances of $G$ will appear playing different roles, I introduce subscripts as mnemonic labels. E.g., I denote the {\it double} $D(G) \cong G\times G$ as
\be
\boxed{\;\;\phantom{\int}
D \cong G_\text{h} \times G_\text{f} \ni (h,g)
\quad}
\ee
whenever its decomposition is thought of in terms of holonomies $h\in G_\text{h}$ and exponentiated fluxes $g\in G_\text{f}$. Notice, I am {\it not} demanding $D$ to inherit the product group structure. I will come back on this point at the end of the section.

For convenience, I also introduce the notation $\AD$ for the action of $G$ on itself by conjugation. This is inspired by the ``exponentiation ladder''
\be
\ad_Y X = [Y,X]
\quad\leadsto\quad
\Ad_g X = g X g^{-1}
\quad\leadsto\quad
\AD_g h = g h g^{-1}.
\ee

On a single edge, the fundamental properties that we want to retain from the standard construction are:
\begin{enumerate}
\item the parallel transport of the flux from one end of the edge to the other according to
\be
\boxed{\;\;\phantom{\int}
\tl g = \AD_h g^{-1};
\quad}
\label{eq_ADhg}
\ee
\item gauge transformations $(g_s, g_t)\in \mathcal G_e$ acting on $D$ as
\begin{subequations}
\be
h \mapsto g_t^{-1} h g_s
\qquad\text{and}\qquad
g\mapsto \AD^{-1}_{g_s} g ,
\ee
or, infinitesimally, to $\mathcal Y_e = (Y_s,Y_t)\in \Lie(\mathcal G_e)$ correspond flows
\be
\boxed{\;\;\phantom{\int}
\mathcal Y_e^\sharp|_{\text{h}} = \hat Y_s^L  + \hat Y^R_t 
\qquad\text{and}\qquad
\mathcal Y_e^\sharp|_{\text{f}}=  \hat Y_s^L + \hat Y_s^R  ; 
\quad}
\ee
\label{eq_gaugetr}
\end{subequations}
\item fluxes $g$ ($\tl g$) ``generate'' gauge transformations at the source (target) end of the edge, via an  equivariant ''momuntum map'';
\item the invariance of the ``quasi-symplectic two-form'' $\omega_e\in\Omega^2(D)$ under the action of $\mathcal G_e$:
\be
\pounds_{\mathcal Y_e^\sharp} \omega_e = 0.
\ee
\end{enumerate}
The terms that need clarification in the list above are put in inverted commas.
The analysis being restricted to a single edge, the label $e$ will be omitted in the rest of this section.

Point ({\it iii}) suggests that the proper generalization of the momentum map generating gauge transformation at the source end of an edge $e$ (equation \eqref{eq_mue=X}) is a {\it group-valued} momentum map consisting in the source and target fluxes:
\be
\boxed{\;\;\phantom{\int}
\mu =(\mu_s,\mu_t): D \to \mathcal G,
\qquad 
\mu(h,g) = \big(g, \tl g\big).
\label{eq_gmu}
\quad}
\ee
Notice, the proper generalization of equivariance  is guaranteed by construction, $\pounds_{\mathcal Y^\sharp} \mu =  -\aD_{\mathcal Y }\mu$, thanks to point ({\it ii}).

So far, the meaning of ``generating'' is however still vague. 
Notice, that a notion of Hamiltonian function $H_\mathcal{Y}$ for the infinitesimal symmetry $\mathcal{Y}\in \Lie(\mathcal{G})$ is problematic: no natural pairing between group and Lie algebra elements exists which is linear in the latter (cf. equation \eqref{eq_PoissonHam}).

Indeed, the solution to the above conundrum consists in bypassing the definition of an Hamiltonian function altogether, and finding directly a replacement for equation \eqref{eq_symplHam}, that is $\mathcal Y^\sharp \contr \omega = \la \d\mu, \mathcal Y\ra $, and in particular for its right-most term.
In this term, the only thing that is needed is a pairing between a {\it one-form} on the image space of the momentum map---that is the group in the deformed case---and $\mathcal Y\in\Lie(\mathcal G)$. 
In the deformed case, this can be achieved by resorting to the $\Lie(\mathcal G)$-valued Maurer--Cartan (MC) form on $\mathcal G$ (as well as to the Killing form, as above).
The first obvious issue is that there is not {\it one} MC form, but two. 
Remarkably, only the symmetric combination is compatible with the antisymmetry of the quasi-symplectic form and with the gauge transformation of fluxes ({\it ii}):%
\footnote{
Suppose , $\mathcal Y=(Y,0)$, then $\mathcal Y^\sharp\contr\omega = \la a \thetaL_{\mu_s}+ b \thetaR_{\mu_s}) , Y\ra$. Now, contracting equation \eqref{eq_qsymplHam} with $\mathcal Y^\sharp$ gives on the lhs $(Y^\sharp\otimes Y^\sharp)\contr\omega\equiv0$, and on the rhs $\la (\hat Y^L + \hat Y^R) \contr(a \thetaL_{\mu_s}+b\thetaR_{\mu_s}) , Y\ra = \la (a-b) Y + b \Ad_{\mu_s} Y - a\Ad_{{\mu_s}}^{-1} Y, Y\ra$. Using the symmetry and $\Ad$-invariance of the Killing form this gives $ (a-b)\big[\la Y,Y\ra -\la\Ad_{\mu_s} Y,Y\ra\big]$, which vanishes identically for all $Y$ iff $a=b$. A similar computation holds for $\mathcal Y = (0,Y)$, while the general case is recovered by linearity. 
}
\be
\boxed{\;\;\phantom{\int}
\mathcal Y^\sharp\contr\omega = \la \tfrac12(\thetaL_\mu+\thetaR_\mu) ,\mathcal Y\ra,
\label{eq_qsymplHam}
\quad}
\ee
the factor $\tfrac12$ guarantees the correct small-flux limit via equation \eqref{eq_approx}. 
For clarity, let me spell out my notation: $\theta^{L,R}_\mu$ is the MC form on $\mathcal G$ evaluated at the image of the momentum map $\mu$;%
\footnote{More precisely, $\theta^L_\mu \equiv \mu^\ast\theta^L$ is the pullback to $D$ of the MC form on $\mathcal G$ via the momentum map $\mu$.} also, if $\mathcal G$ is the Cartesian product of multiple groups, say $\mathcal G = G_1\times G_2$, the above has to be read as follows $\theta^L_\mu = (\theta^L_{\mu_1}, \theta^L_{\mu_2})$, $\mathcal Y = (Y_1, Y_2)$, and $ \la \thetaL_\mu,\mathcal Y\ra = \la \thetaL_{\mu_1},\mathcal Y_1\ra + \la \thetaL_{\mu_2},\mathcal Y_2\ra$. 

Now that a replacement for the Hamiltonian flow equation has been found, I can discuss its interplay with the other desiderata.

Condition ({\it iv}) requires $\omega$ to be $\mathcal G$-invariant, i.e. $\pounds_{\mathcal Y^\sharp}\omega \stackrel{!}{=} 0$. Using the equation above, this is%
\footnote{Recall Cartan's formula: $\pounds_{Y_e^\sharp}\nu= \d (Y^\sharp \contr \nu) + Y^\sharp\contr \d \nu$, for any $\nu\in\Omega^\bullet(D)$.}
\be
0 \stackrel{!}{=}  \mathcal Y^\sharp \d \omega + \la \tfrac12\d(\thetaL_\mu+\thetaR_\mu) , \mathcal Y\ra,
\label{eq_Lieomega}
\ee
which readily implies
\be
d\omega \neq 0.
\ee
Therefore, $\omega$ cannot be symplectic.
In the (quasi-)Poisson framework, this translates into the failure of the Jacobi identity.
To minimize damages, $\omega$'s violation of closedness is required to take a controllable form---hence the ``quasi'' nomenclature---compatible with equation \eqref{eq_Lieomega}, i.e. 
\be
\boxed{\;\;\phantom{\int}
\d \omega = -\chi_\mu
\qquad\text{with}\qquad
\chi = \tfrac{1}{12}\la \thetaL  \stackrel{\wedge}{,} [\thetaL \stackrel{\wedge}{,}\thetaL ]\ra  = \tfrac{1}{12}\la \thetaR  \stackrel{\wedge}{,} [\thetaR   \stackrel{\wedge}{,}\thetaR   ]\ra,
\quad}
\label{eq_chi}
\ee
where the $\Ad$-invariance of the Killing form was used, and $\chi_\mu\equiv\mu^\ast \chi$ is the pullback of the 3-cocyle $\chi\in\Omega^3(G_\text{f})$ to $D(G)$ by the momentum map $\mu:D(G)\to G_\text{f}$.

Another consequence of condition \eqref{eq_qsymplHam} is the failure of $\omega$ being non-degenerate everywhere on $D$. Indeed, using the $\Ad$-invariance of the Killing form, its rhs can be written as $\la\thetaR_\mu, \tfrac12(1 + \Ad_\mu) \mathcal Y \ra $, which at $(h,g)\in D$ vanishes for those $\mathcal Y$ such that $\Ad_h \mathcal Y = -\mathcal Y$. Thus, a minimal relaxation of the non-degeneracy condition means requiring these are the only vector fields in the kernel of $\omega$, i.e.
\be
\boxed{\;\;\phantom{\int}
\ker \omega_{(h,g)} = \big\{ \mathfrak{v} = \mathcal Y^\sharp \; \text{for some} \;  \mathcal Y \in \ker( \Ad_{\mu(h,g)} +1 ) \big\} \subset \mathfrak{X}^1(D). 
\label{eq_Ad+1}
\quad}
\ee

Notice how the whole quasi-symplectic structure of $(D,\omega)$ revolves around the chosen symmetry properties: even such fundamental features as the closedness and non-degeneracy violations depend on the momentum map.
If $G=\{\E\}$ is trivial, the definition reduces to that of a (standard) symplectic space.

Let me summarize what has been done so far. 
Starting from the idea of implementing a group-valued momentum map and from the only sensible ensuing generalization of the Hamiltonian flow equation \eqref{eq_qsymplHam}, it was noticed that both the closedness and non-degenetacy of $\omega$ had to be relaxed. Doing this in a minimal way led to the requirements \eqref{eq_chi} and \eqref{eq_Ad+1}, respectively. 
These three equations can be abstracted from the context in which I have discussed them, and indeed {\it define} quasi-Hamiltonian $G$-spaces \cite{Alekseev1998}. 

Now, comparison with equation \eqref{eq_omegaTG}, suggests the following generalization of $\omega$ from $\mathrm T^\ast  G$ to $D(G)$:
\be
\boxed{\;\;\phantom{\int}
\omega_{(h,g)} =  \la \thetaL_h \stackrel{\wedge}{ ,} \tfrac12( \thetaL_g + \thetaR_g) \ra + \tfrac12 \la \Ad_g \thetaL_h \stackrel{\wedge}{ ,} \thetaL_h\ra,
\quad}
\ee
where, again, $\d X \leadsto \tfrac12( \thetaL_g + \thetaR_g)$ and $\ad_X \leadsto \Ad_g$.
In fact, as it turns out, this $\omega$ together with the momentum map of equation \eqref{eq_gmu} satisfies the three conditions listed above.%
\footnote{This is much easier to verify in the $(a,b)$ variables introduced in the next section.}

This concludes the discussion of the quasi-symplectic structure on the deformed edge ``phase space'', $D=G_\text{h}\times G_\text{f}$. 
Before moving on to the assemblage of many $D$'s, one more comment is in order. 

\subsection{Too easy to say ``$G\times G$'' \label{sec_GtimesG}}

\begin{figure}[t!]
\includegraphics[width=.65\textwidth]{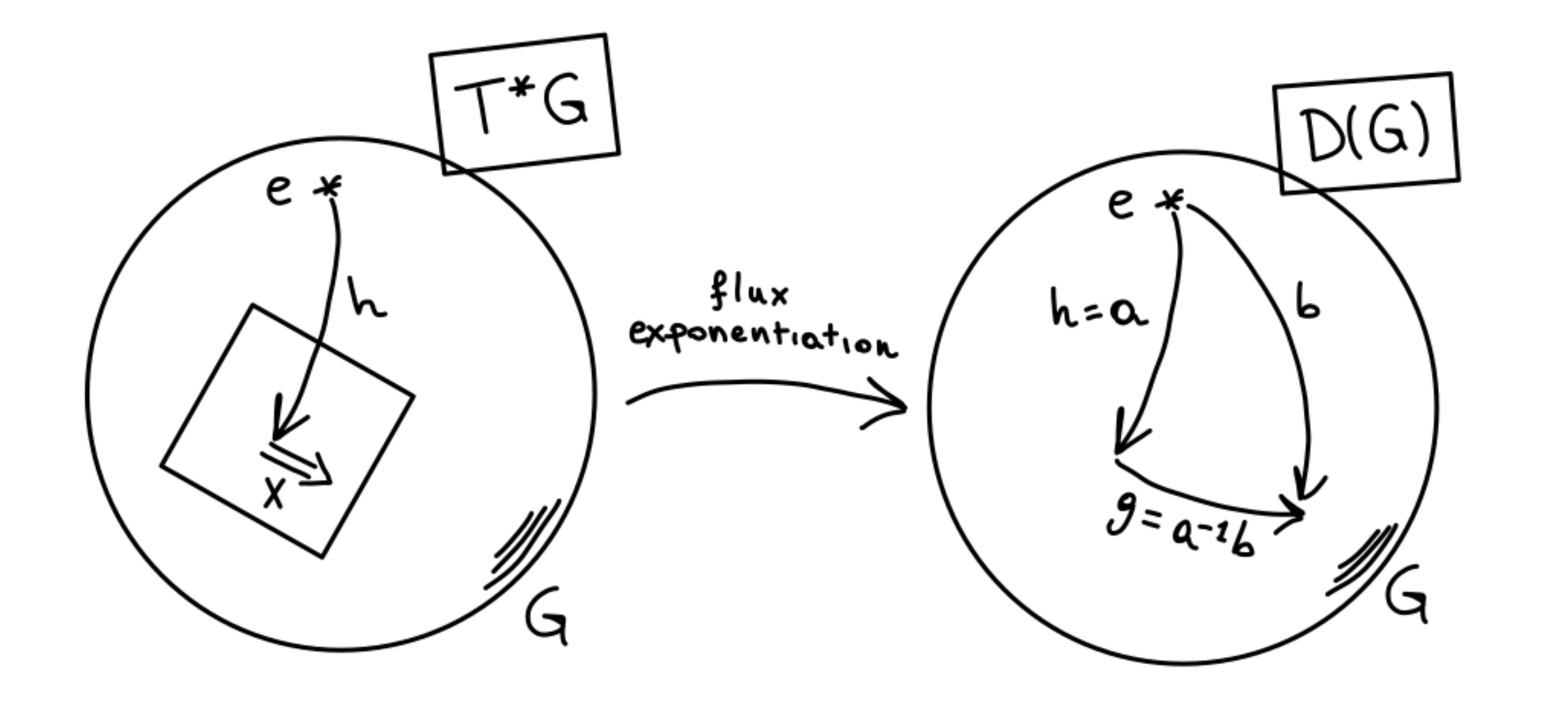}
\caption{A representation---within a single copy of $G$---of the flux exponentiation and the ensuing relation between the a-b and holonomy-flux decompositions of the double $D(G)$, i.e. $D(G)=G_\text{a}\times G_\text{b} \cong G_\text{h}\ltimes G_\text{f}$.}
\label{fig_DG}
\end{figure}

So far, we have chosen as coordinates on $D$ the generalization of the holonomy and flux variables on $\mathrm TG$ via a left trivialization. 
In particular, we notice the natural semi-direct product structure that this decomposition carries,
\be
D = G_\text{h} \ltimes_\AD G_\text{f};
\ee
the holonomies parallel transport the fluxes via an $\AD$-action, in strict analogy with $\mathrm TG = G\ltimes_\Ad \frakg$ (see equations \eqref{eq_ADhg} and \eqref{eq_AdhX}).
It is, however, enlightening to introduce a more symmetric decomposition of $D$, where both copies of $G$ play the same role (see figure \ref{fig_DG}).
Let me denote it by
\be
D = G_\text{a} \times G_\text{b} \ni (a,b),
\ee
and define it via the following change of variables:
\be
a = h 
\qquad\text{and}\qquad
b = hg.
\label{eq_abhg}
\ee
This decomposition allows to turn $D$ into a group, by letting it inherit the natural product group structure from $G_\text{a} \times G_\text{b}$.
In the new variables, whose naming will be clarified in section \eqref{sec_2surface}, the gauge transformation, the momentum map, and the quasi-symplectic form take a more symmetric form.
Indeed, gauge transformations become
\be
a \mapsto g_t^{-1} a g_s 
\qquad\text{and}\qquad
b \mapsto g_t^{-1} b g_s,
\ee
the momentum map
\be
\mu=\big( a^{-1}b, ab^{-1} \big),
\ee
and the quasi-symplectic two-form
\be
\omega_{(a,b)} = \tfrac12 \la \thetaL_a \stackrel{\wedge}{,} \thetaL_b \ra + \tfrac12 \la \thetaR_a \stackrel{\wedge}{,} \thetaR_b \ra.
\ee
From this perspective, we see that the $G_\text{a} \times G_\text{b}$ is the most natural decomposition, while the $G_\text{h} \times G_\text{f}$ arises as a ``left trivialization'' in which holonomies constitute the diagonal subgroup of $D$,
\be
G_\text{h}\subset_\d  G_\text{a} \times G_\text{b},
\ee
and the fluxes are the ``rest''
\be
G_\text{f} \cong \faktor{G_\text{a} \times G_\text{b}}{ G_\text{h}}.
\ee
This discussion highlights the fact that the fluxes do not carry any natural group structure in $D$ (they are a co-set).
The latter is in a sense only inherited from the group composition law of the momenta---see the next section---which in turn provides a preferred parametrization of the flux space such that the (source) momentum map looks like a projection.%

Then, the $\AD$-action of the holonomies on the fluxes is just a consequence of the group multiplication in  $D$ and of the choice of coordinates \eqref{eq_abhg} on the quotient:
\begin{align}
\big(h_3,h_3g_3\big) &= \big(a_3,b_3\big) = \big(a_2,b_2\big)\cdot_D \big(a_1,b_1\big) = \big(a_2 a_1 , b_2 b_1\big)  =\big(h_2h_1, h_2 g_2 h_1 g_1\big)\notag\\
& = \big( h_2h_1 , (h_2 h_1) (\AD^{-1}_{h_1}g_2) g_1\big).
\end{align}

I conclude this section with a side remark: let me compare this situation with another possible deformation of $\mathrm T^\ast G$, which is $G^\mathbb{C}$. In this case, $G^\mathbb{C}$ can be equipped with a natural symplectic structure via the Iwasawa decomposition of $G^\mathbb{C}$. For $G=\SU(2)$ reads $\SL(2,\mathbb{C}) = \SU(2)\times\mathrm{SB}(2,\mathbb C)$ (see \cite{Dupuis2013,Dupuis2014,Bonzom2014a,Bonzom2014,Fock1999,Charles2015,Charles2017}). The present issue with this decomposition, which has the nicest symplecto-geometric properties, is that although the $\SU(2)$ is readily interpreted as the rotation subgroup, the momentum space does not transform covariantly under rotations. However, boosts would. These, on the other hand carry no group structure, they are just a co-set, $K=\SL(2,\mathbb{C})/\SU(2)$. The construction described in the main text, when applied to $G=\SU(2)$, gives just the Euclidean version of the rotation-boost construction, the $\SU(2)$ fluxes corresponding to ``Euclidean boosts''. In the Euclidean, no Iwasawa-like decomposition is available. On this topic, see also the concluding paragraph of the appendix.

\subsection{Fusion and reduction}

In the previous section, I reviewed the construction of the quasi-Hamiltonian $G$ space $D_e \cong D(G)$ associated to a single edge $e\in\Gamma$.
The direct product $\bigtimes_{e} D_e$ is of course the space in which to perform the analogue of the symplectic reduction. 
The question is with respect to which momentum map and which symplectic structure. 
For simplicity, I start from the case of a single vertex being the source of various edges.

In the flat case, momenta are valued in the linear space $\frakg^\ast$ and therefore they can be summed to one-another to obtain the ``total momentum map''. For example, to a given vertex of $\Gamma$, where multiple edges convene, one associates the total momentum of equation \eqref{eq_mu=sumX}, $\mu_v = \sum \mu_{s(e)}$.
Because of the linearity of this procedure, the symplecitc form is consistently obtained simply as the sum $\omega=\sum\omega_e$.
The Hamiltonian flow equation for the gauge symmetry at $v$, embodied by the diagonal action of $G\subset_d G^{\times\#\{e\}}$, is then automatically satisfied.

Consider now the deformed case.
Here, momenta are valued in the non-linear space $\mathcal G$, and their natural composition is the group product. Schematically,
\be
\boxed{\;\;\phantom{\int}
\mu_v = \overleftarrow\prod \mu_{s(e)}.
\quad}
\ee
Not surprisingly, a total quasi-symplecitc form defined as the linear composition $\sum\omega_e$ turns out {\it not} to be compatible with the quasi-Hamiltomian flow equation \eqref{eq_qsymplHam} for the total momentum.
Thus the simple sum formula has to be twisted, and, due to the non-commutativity of $G$, this twist shall depend on the order of the factors in $\mu_v$.

Let me formalize the situation a little more. Consider two edge spaces, $\big(D_1,\omega_1, \mathcal G_1, \mu_1=(\mu_{s_1},\mu_{t_1}) \big)$ and $\big(D_2,\omega_2, \mathcal G_2, \mu_2=(\mu_{s_2},\mu_{t_2}) \big)$.
Both spaces are $\mathcal G \cong G^2$ quasi-Hamiltonian spaces, whose (deformed) symmetry group is $\mathcal G=G_s\times G_t$.
The space $\big(D_2\times D_1, \omega_2 + \omega_1, \mathcal G_2 \times \mathcal G_1, (\mu_1, \mu_2) \big)$ is then naturally equipped with a $G^4 = \mathcal G_2 \times \mathcal G_1$ quasi-Hamiltonian structure. 
The {\it fusion} procedure aims at turning this space into a $G^3$ quasi-Hamiltonian space: 
assuming for definiteness that $v$ is the source of both $e_1$ and $e_2$, after fusion the symmetry group and momentum map shall be $G^3=G_v\times G_{t_2}\times G_{t_1}$---with $G_v$ the diagonal subgroup of $G_{s_2} \times G_{s_1}$---and $(\mu_v = \mu_2 \mu_1, \mu_{t_2}, \mu_{t_1})$, respectively.
Notice, that the group $H = G_{t_2}\times G_{t_1}$ shall be a complete spectator in the procedure above, hence playing no role at all.
This observation allows to formulate fusion in its most general form, without any further complication.

Given two quasi-Hamiltonian spaces%
\footnote{With apologies for the slight abuse of notation: in the parenthesis, $\mP$ stands for the manifold underlying the quasi-Hamiltonian space, stripped of the rest of its structure.}
$\mP_\alpha=\big(\mP_\alpha, \omega_\alpha, \mathcal G_\alpha = G\times H_\alpha, \mu_\alpha = (\mu_{G_\alpha},  \mu_{H_\alpha})\big)$, define $ \mP_2 \fus \mP_1 = \mP_{2\fus1} = (\mP_{2\fus1}, \omega_{2\fus1},  \mathcal G_{2\fus1}, \mu_{2\fus1})$ as follows:
\begin{enumerate}
\item  $\mP_{2\fus1} = \mP_2\times \mP_1$ as a manifold;
\item symmetry group
\be
\mathcal G_{2\fus1} = G \times H_2 \times H_1,
\ee
with $G$ the diagonal subgroup of $G\subset_\d G\times G \subset \mathcal G_2 \times \mathcal G_1$;
\item  total momentum map
\be
\boxed{\;\;\phantom{\int}
\mu_{2\fus1} = \big(\mu_G=\mu_{G_2}\mu_{G_1} \, , \, \mu_{H_2} \, , \, \mu_{H_1}\big);
\quad}
\ee
\item and quasi-symplectic structure
\be
\boxed{\;\;\phantom{\int}
\omega_{2\fus1} = \omega_2 + \omega_1 + \frac12 \la \thetaL_{\mu_2} \stackrel{\wedge}{,} \thetaR_{\mu_1} \ra
\quad}
\label{eq_omegafus}
\ee
(in the limit of small momenta---read, fluxes---the additional term in this equation is of higher order with respect to the standard symplectic structure, and can be dropped).
\end{enumerate}

It is not hard to check that $\mP_{2\fus1}$ satisfies the axioms of quasi-Hamiltonian spaces, and I will not do it explicitly.
However, I want to highlight the role of the last term of equation \eqref{eq_omegafus}. For this purpose, it is enough to compute the rhs of equation \eqref{eq_qsymplHam}, with $\mu= \mu_2\mu_1$:
\begin{align}
\la\thetaL_{\mu} + \thetaR_{\mu}, \mathcal Y\ra  
& = \la\thetaL_{\mu_2\mu_1} + \thetaR_{\mu_2\mu_1}, \mathcal Y\ra
= \la \Ad_1^{-1} \thetaL_2 + \thetaL_1 + \Ad_2 \thetaR_1 + \thetaR_2, \mathcal Y\ra \notag\\
& =    \la  \thetaL_2 +  \thetaR_2, \mathcal Y\ra + \la  \thetaL_1 +  \thetaR_1, \mathcal Y\ra 
+ \la \thetaL_2, (\Ad_1 - 1) \mathcal Y\ra + \la \thetaR_1, (\Ad^{-1}_2 - 1) \mathcal Y\ra,
\end{align}
with obvious short-hand notations.
The last two terms arise as a consequence of non-commutativity of $G$, and are precisely compensated by the extra term in $\omega_{2\fus1}$. Indeed, (neglecting the $H_\alpha$ factors)
\be
\mathcal Y_{2\fus1}^\sharp \contr \la \thetaL_{\mu_2} \stackrel{\wedge}{,} \thetaR_{\mu_1} \ra = 
\la \mathcal Y^\sharp\contr \thetaL_{2}, \thetaR_{1}\ra 
- \la  \thetaL_{2}, \mathcal Y^\sharp \contr \thetaR_{1}\ra\notag\\
= \la (1- \Ad_{\mu_2}^{-1})\mathcal Y , \thetaR_{1}\ra
 -  \la \thetaL_{2} , (\Ad_{1} -1 )\mathcal Y\ra,
\ee
where the equivariance of $\mu_\alpha$ was used to compute $\mathcal Y^\sharp\contr \thetaL_{\mu_\alpha} = (1- \Ad_{\mu_\alpha}^{-1})\mathcal Y $.

Another fact that can be easily checked is the associativity of the fusion procedure: 
\be
\mP_{3 \fus (2 \fus 1 )}= \mP_{(3 \fus 2) \fus 1}.
\ee

Subtler is the question about commutativity. 
Indeed, although the fusion product is manifestly non-commutative, $\mP_{2\fus1}$ and $\mP_{1\fus2}$ are nontheless {\it isomorphic}.
The relevant isomorphism is called the {\it braid isomorphism} and is denoted $R_{12}$. 
It essentially consists in acting on the space $\mP_2$ via $\mu_1$ (note the reverse labels) before proceeding to the ``inverted'' fusion.
I will comment more about it later.
See \cite{Alekseev1998}[Theorem 6.2] for the precise definition.

Hence, the fusion of the group actions associated to every single vertex in the graph leads to the quasi-Hamiltonian space
\be
\boxed{\;\;\phantom{\int}
(\mP')^\EM_\Gamma = \big( D^{\times e} , \omega_\Gamma, G^{\times V}, \mu = ( \mu_v ) \big).
\quad}
\ee
To build the gauge-invariant {\it symplectic} space $\mP^\EM_\Gamma$, a generalization of the Marsden--Weinstein reduction is needed.
This exists, and in fact works just the same way.
The statement is the following: let $\mP = \big( \mP, \omega, G\times H, (\mu_G, \mu_H) \big)$ be  a quasi-Hamiltonian space, then the pullback of $\omega$ on the pre-image $\mu_G^{-1}(\E)$ of the identity $\E \in G$ descends to the reduced space%
\be
\mP_\E^\text{red} = \mu_G^{-1}(\E)/G
\ee
in which symmetry-related configurations are identified. 
All the extra structure, in turn, naturally descends to the quasi-Hamiltonian $H$-space $\mP//G$,%
\footnote{Here, $\omega^\text{red}_\E$ and  $\mu_H^\text{red}$ are defined in the obvious way. Denote $\iota: \mu_G^{-1}(\E)\hookrightarrow \mP$ and $\pi:\mu^{-1}_G \to \mP_\E^\text{red} $ the embedding and the projection, respectively. Then, $\omega^\text{red}_\E$ is the only two-form on $\mP_\E^\text{red} $ such that $\iota^\ast\omega = \pi^\ast \omega^\text{red}_\E $, and similarly for $\mu^\text{red}_H$. Notice that the restriction $\iota^\ast \mu_H$ is $H$-equivariant.}
\be
\mP//\mu_G = \big( \mP_\E^\text{red} , \omega^\text{red}_\E, H, \mu_H^\text{red} \big).
\ee
As a consequence, if the group $H$ is trivial, the reduced space is symplectic (cf. the discussion below equation \eqref{eq_Ad+1}).

Application of the reduction theorem leads to the definition of the {\it  SD symplectic phase space of gauge-invariant  configurations on $\Gamma$} as
\be
\boxed{\;\;\phantom{\int}
\mP^\EM_\Gamma = (\mP')^\EM_\Gamma// (\mu_v)_{v=1,\dots,V}.
\quad}
\ee
Both $(\mP')^\EM_\Gamma$ and $\mP^\EM_\Gamma$ are compact spaces. 

Also, I emphasize once more, this construction---in the formal limit of small fluxes---manifestly recovers the usual phase spaces $\mP'_\Gamma = (\mathrm T^\ast G)^{\times E}$ and $\mP_\Gamma = (\mathrm T^\ast G)^{\times E}//(\mu_v)$.

\section{Remarks on self-duality \label{sec_remarks}}

The notion of self-duality employed so far is rather weak. 
It only requires holonomies and fluxes to be valued in the same (group) space. 
In 3+1 spacetime dimensions, however, the space $\mP^\EM_\Gamma$ carries a richer notion of duality, albeit not manifestly. Henceforth, I will restrict to this case.

The statement is that the space $\mP^\EM_\Gamma$ can be obtained by performing the same construction on $\Gamma^\ast$, the graph dual to $\Gamma$, provided that the roles of holonomies and fluxes are exchanged. 

In the undeformed setting, gauge invariance at the vertices of $\Gamma$ is encoded in the {Gau\ss} constraint $ (\mu_v) = 0$ of equation \eqref{eq_mu=sumX}, and is enforced on the phase space by the reduction procedure. 
Physically this constraint states that no charged particle is present on the lattice, or---in a different but equivalent language---that no electric defect (or excitation) is present.

In the continuum, the vanishing of magnetic fluxes out of a 3-dimensional region is automatically encoded in our choice of gauge-connection variables, from which the magnetic field's property of being divergence-free follows algebraically from the Bianchi identity, $\d_A F[A] \equiv 0$.
On $\Gamma$, this is lifted to the discrete setting: connection variables are replaced by holonomies along the edges of $\Gamma$ (equation \eqref{eq_he}), and the magnetic flux through a given face of $\Gamma$ is just the oriented product of the holonomies (equation \eqref{eq_hf}). The total flux out of a closed surface is then trivial for homological reasons essentially dual to the Bianchi identity itself.%
\footnote{This is most easily seen in the Abelian case, where the Bianchi identity simply follows from $d^2=0$, which is the cohomological dual to the homological identity $\pp^2=0$ used in the lattice construction. }
At any rate, although trivial once appropriately constructed out of the edge holonomies, the vanishing of the total flux out of a 3-cell $c$ in $\Gamma$ schematically reads
\be
\overleftarrow{\prod_{f:f\in\pp c}} h^\epsilon_f = \E,
\label{eq_prodh=e}
\ee
where $h^\epsilon_f$ is the oriented circulation around $f$ (possibly parallel transported to a common reference point).
This is nothing but a deformed (and compactified) version of the sum of the magnetic fluxes out of the 3-cell $c$, $\sum_{f:f\in\pp c} \Phi_f(B) = 0$, where the linear flux%
\footnote{This is also a schematic expression: in the non-Abelian case the magnetic field $B=F[A]$ should also be appropriately parallel transported throughout $f$ to a common reference point before being integrated. The same type of parallel transport is at the origin of the Poisson non-commutativity of the discrete fluxes, equation \eqref{eq_T*G}.}
$\Phi_f(B) = \int_f B \in \frakg$ has been replaced by its ``exponentiated version'' $h_f \in G$ and the sum in $\frakg$ by the group multiplication.

{\it Mutatis mutandis}, the previous paragraph can be turned into a description of the construction of the deformed {Gau\ss} constraint for the exponentiated electric fluxes discussed in the previous section.
Therefore, already at this level, it is clear that the group elements $g_e$---the exponentiated electric fluxes---and $h_f$---the compactified magnetic fluxes---play dual roles (notice the subscripts, standing for the edges and faces of $\G$, respectively). 

Of course, the difference is that $g_e$ is treated as a ``fundamental'' object, while $h_f = \overleftarrow\prod h_e^\epsilon$ as a ``composite'' one. 
This difference is, however, lost on the reduced phase space, where the $g_e$ must satisfy constraints that make them not independent from one another and restore the expected symmetry: in vacuum Yang--Mills, the {Gau\ss} law and the Bianchi identity are, within $\mP^\EM_\G$, dual to each-other as they are in the continuum.

Making this duality completely manifest is the aim of the next section.

\section{Flat connections on a Heegaard surface\label{sec_2surface}}

\subsection{Overview}

\begin{figure}
\includegraphics[width=.35\textwidth]{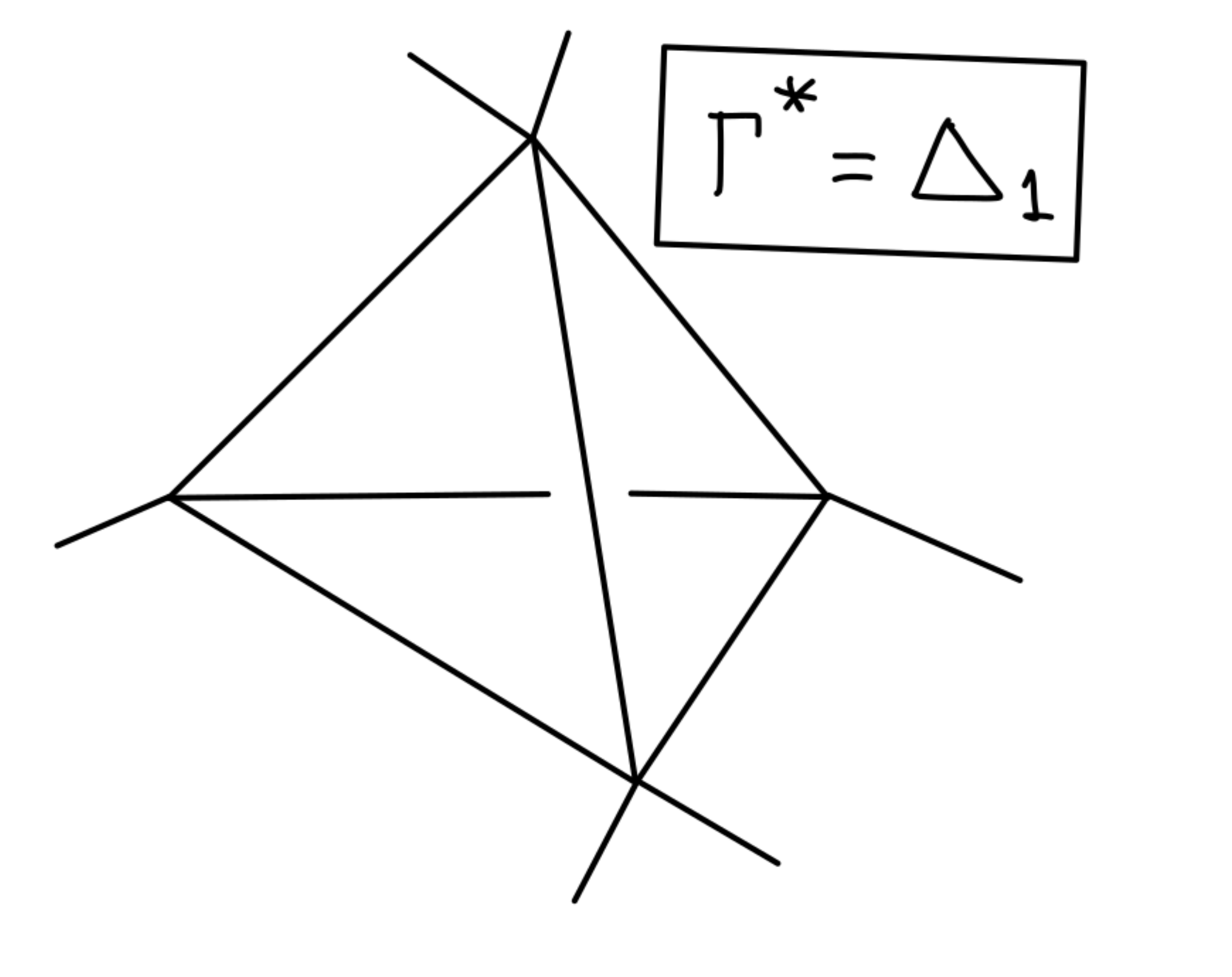}
\quad
\includegraphics[width=.35\textwidth]{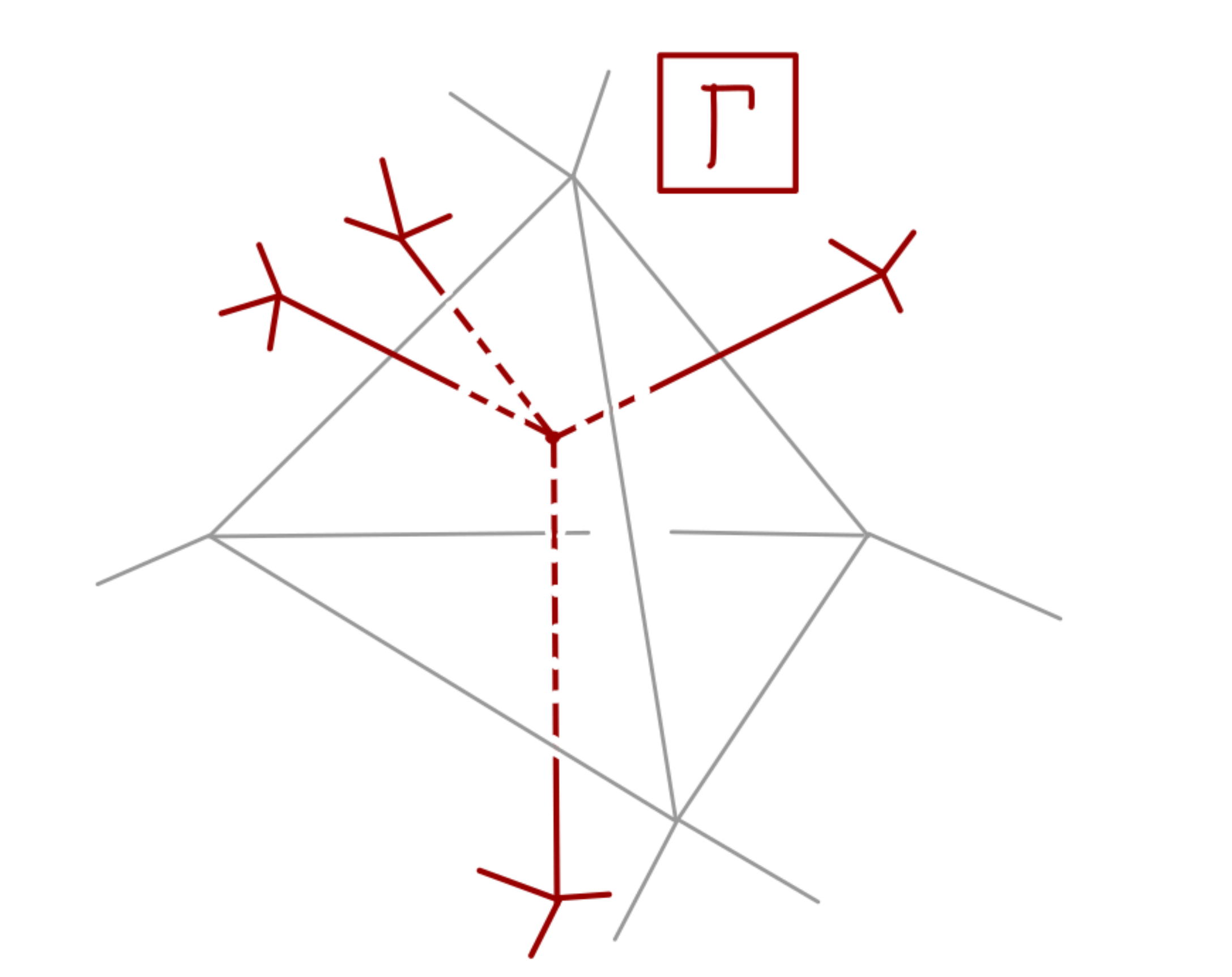}
\caption{Left: A portion of the cellular decomposition $\Delta$, with its 1-skeleton, $\G^\ast=\Delta_1$ highlighted. Right: The corresponding portion of the graph $\G$ dual to $\G^\ast$.}
\label{fig_G_Gstar}
\end{figure}

\begin{figure}
\includegraphics[width=.35\textwidth]{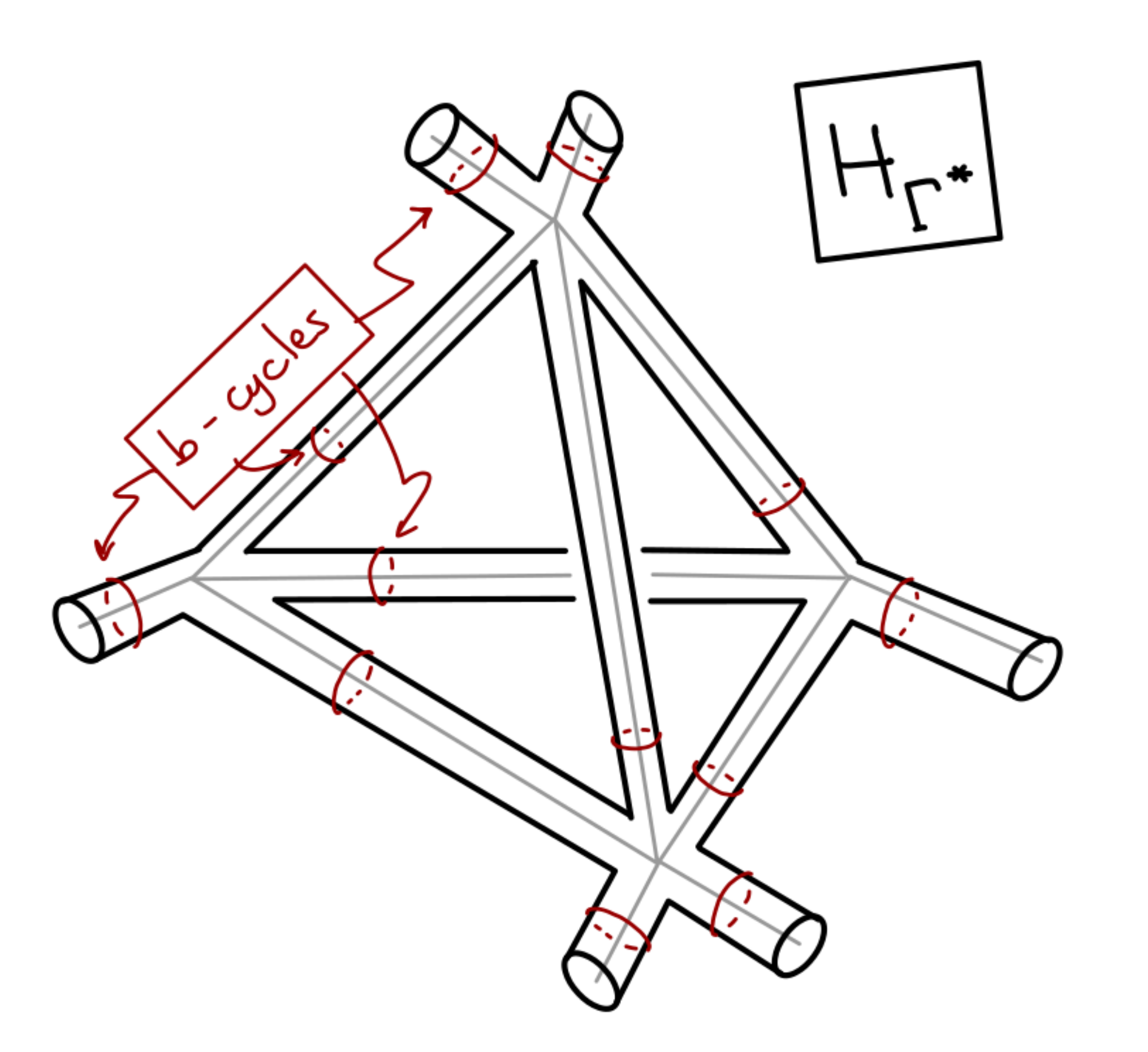}
\quad
\includegraphics[width=.35\textwidth]{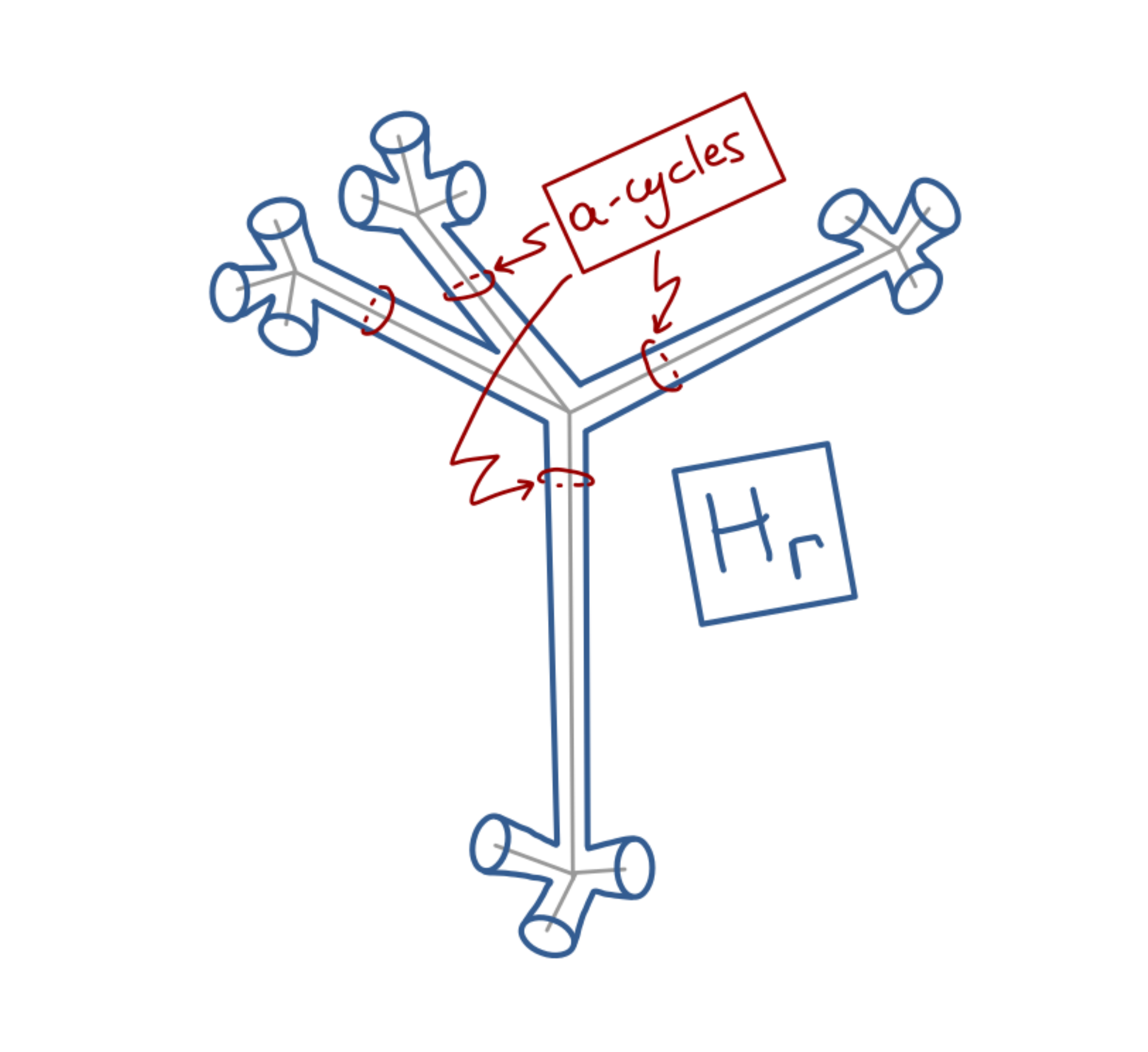}
\caption{Left: The tubular neighborhood $H_{\G^\ast}$ of $\G^\ast$ with the b-cycles highlited. Right: The tubular neighborhood $H_{\G}$ of $\G$ with the a-cycles highlited}
\label{fig_HG_HGstar}
\end{figure}

In this and the following sections, the focus will be on the 3+1 dimensional case.
Thus, $\Sigma$ is a 3-manifold and $\Gamma$ is an embedded graph dual to a cellular decomposition $\Delta$ of $\Sigma$. Name $\Gamma^\ast$ the 1-skeleton of $\Delta$ (figure \ref{fig_G_Gstar}).
Consider now the handlebody $H_\Gamma$ consisting of a tubular neighborhood of $\Gamma$, i.e. the handlebody obtained by ``thickening'' the edges of $\Gamma$ into thin tubes and its vertices into small balls (figure \ref{fig_HG_HGstar}). Schematically, $H_\G$ is the topological manifold defined by
\be
H_\Gamma = \{x\in\Sigma\;:\; \text{distance}(x,\Gamma)\leq\epsilon\},
\ee
for some positive defined metric on $\Sigma$.
Its boundary is a 2-surface, $S_\Gamma=\pp H_\Gamma$.
Up to an orientation reversal, the same topological surface would have been obtained had we started from $\Gamma^\ast$, $S_\Gamma^\text{op} \cong S_{\G^\ast} = \pp H_{\G^\ast}$.
Also, the gluing of the two handlebodies, $H_\Gamma$ and $H_{\G^\ast}$, via an identification of their boundaries gives back the topological manifold $\Sigma$, 
\be
\boxed{\;\;\phantom{\int}
H_\Gamma \cup_{S_\Gamma} H_{\Gamma^\ast} \cong \Sigma.
\quad}
\ee
This is called an Heegaard decomposition of $\Sigma$, and $S_\Gamma=S_{\Gamma^\ast}$ an Heegaard surface.
Non-contractible cycles of $S_\Gamma$ which are contractible in $H_\Gamma$ ($H_{\Gamma^\ast}$) are called $a$-cycles ($b$-cycles, resp.). See figure \ref{fig_HG_HGstar}.

The self-duality described in the remark above follows from the following statement:
the deformed phase space $\mP^\EM_\Gamma$ of a discretized $G$-Yang--Mills theory in the vacuum, i.e. in absence of charged sources, is isomorphic to the (canonical) moduli space of flat $G$-connections on $S_\Gamma$ equipped with the Atyiah--Bott  symplectic structure (see below), 
\be
\boxed{\;\;\phantom{\int}
\mP^\EM_\Gamma \cong \M(S_\Gamma, G),
\quad}
\label{eq_iso}
\ee
where the $g_e$'s ($h_f$'s) are identified%
\footnote{This statement is morally correct, but requires some technical clarification, provided further below.} 
with the holonomies around the $a$-cycles ($b$-cycles, resp.) of $S_\Gamma$. 
Hence, self-duality in the form discussed in the previous remark  immediately follows from the symmetry between $H_{\Gamma}$ and $H_{\Gamma^\ast}$, and $a$- and $b$-cycles.

It is, nonetheless, crucial to have clear the differences between the two spaces appearing in equation \eqref{eq_iso}: $\mP^\EM_\Gamma$ is the (deformed) phase space of a discretized 3+1 gauge theory in which the $G$-holonomies of a Poisson-{\it commutative} connection $A$ are Poisson-conjugated to (deformed) electric fluxes also valued in $G$. 
In turn, $\M(S_\Gamma, G)$ is the reduced phase space of an (auxiliary) (2+1)-dimensional gauge theory of a  {\it flat} Poisson-{\it non}commutative $G$-connection, $\A\in\Omega^1(M)\otimes\frakg$.

The non-trivial curvature of the connection $A$, as well as the non-vanishing of the exponentiated fluxes $g_e$, is hence supported---from the perspective of $\A$---by the {\it non-trivial topology} of $S_\Gamma$, on which the {\it flat} connection $\A$ is defined. In this sense, the excitations of $A$ are mapped onto topological defects carried by $\G$ and $\G^\ast$.

More specifically, $\M(S, G)$ is the finite dimensional symplectic space obtained via symplectic reduction from the infinite dimensional space of $G$-connections on $S$, $\mP'_S(\A)$.
I.e.,
\be
\M(S, G) = \mP'_S(\A)//{\mathscr G},
\ee

This expression needs some clarifications.
Let me start from the group $\mathscr G$. This is the group of gauge transformations on $S$, i.e. the set of $G$-valued functions on $S$ equipped with the point-wise group product. Schematically
\be
\mathscr G = \{ \mathrm{Map}:S\to G\}.
\ee 
Gauge transformations are generated by the momentum map%
\footnote{The inclusion of a non-trivial boundary for $S$ is important e.g. when decomposing $S_\G$ into trinions.}
\be
\tl \mu: \mP'_S(\A) \to \Lie(\mathscr G)^\ast,
\qquad \A \mapsto \tl\mu[\A; \;\cdot \;] = \int_S \la F[\A] , \; \cdot \;\ra + \oint_{\pp S} \la \A , \; \cdot \; \ra ,
\ee
with
\be
F[\A]= \d \A + \tfrac12 [\A \stackrel{\wedge}{,} \A]
\ee
being the $\Lie(\mathscr G)$-valued curvature two-form of the connection $\A$.
That is, for every $\xi\in\Lie(\mathscr{G})\cong \mathcal{C}(S)\otimes \frakg$, to which the infinitesimal gauge-transformation flow $\xi^\sharp\in\mathfrak X^1(\mP'_S(\A))$ is associated---meaning
${\xi^\sharp}\contr\delta \A = \d_\A \xi$---the following Hamiltonian flow equation holds
\be
{\xi^\sharp}\contr\Omega = -\delta\tl\mu[\A; \xi] .
\ee
Here, $\delta$ is the functional%
\footnote{Throughout these notes I neglect functional analytical issues of any sort, see \cite{Atiyah1983,Goldman1984,Jeffrey1994,Alekseev1994,Alekseev1998}.}
deRahm differential on $\mP'_S(\A)$, and $\Omega\in \Omega^2(\mP'_S(\A))$ is the (canonical) Atiyah--Bott symplectic two-form on $\mP'_S(\A)$,
\be
\Omega =\frac12 \int_S \la \delta \A \stackrel{\wedge}{,} \delta \A \ra .
\ee
Observe that $\M(S, G)$ is nothing but the gauge-invariant phase space of Chern--Simons theory on $S\times[0,1]$.

The Atiyah--Bott symplectic form states that the ``longitudinal'' part of $\A$ is conjugated to its ``transverse'' part.
This means that holonomies calculated along transversally crossing paths do not Poisson-commute. In particular, holonomies on $S_\Gamma$ ``parallel'' to the edges of $\Gamma$ are conjugated to those along the corresponding $a$-cycle. This is the first clue towards the identifications of the holonomies $h_e$ and fluxes $g_e$ with longitudinal and transverse holonomies along the ``edge-tubes'' of $S_\G$, respectively. 

To give a precise version of this statement, I will proceed in two steps: ({\it i}) first, I will explicitly show how to read the construction of the previous section in terms of holonomies on $S_\G$, then ({\it ii}) through a natural (albeit highly non-canonical) gauge-fixing procedure I will show that the symplectic form on $\mP'_\G$ matches that on $\M(S_\G, G)$. Beside the (quite simple) gauge fixing, the second step is nothing but the main result of \cite{Alekseev1998}, which in turn constitutes itself the very reason why the quasi-Hamiltonian formalism has been devised in the first place. 

\begin{figure}
\includegraphics[width=.5\textwidth]{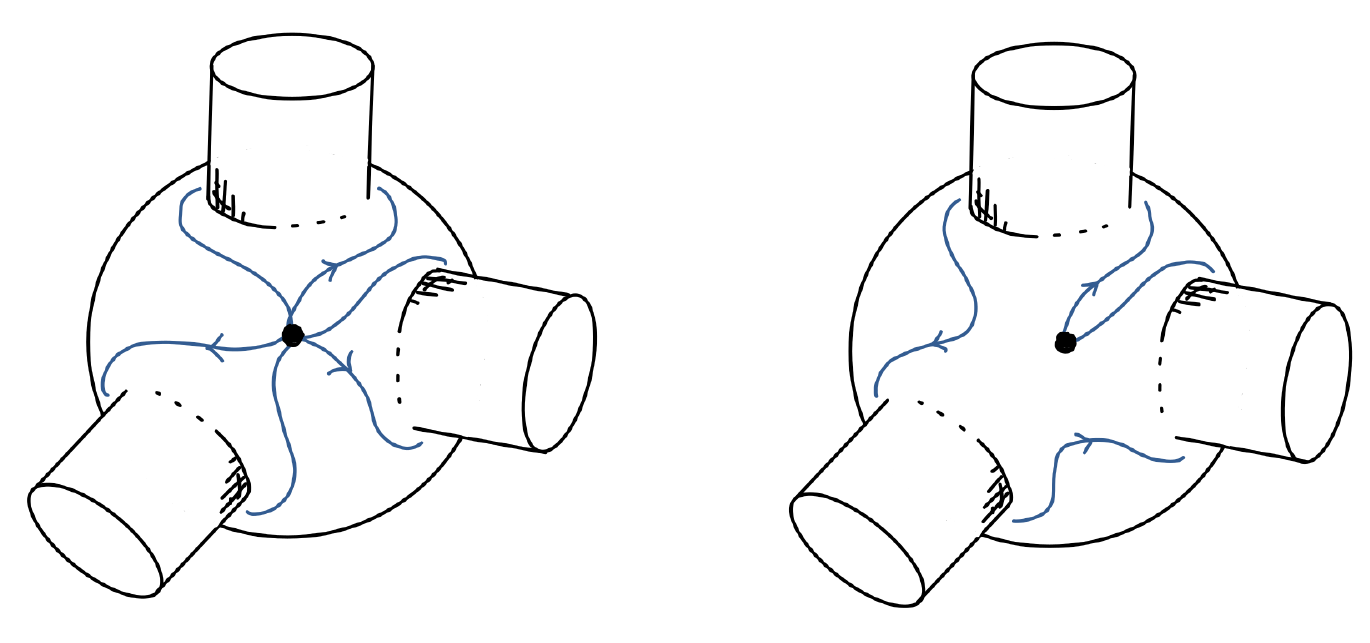}
\caption{Left: The path $\gamma_v = a^v_{e_3}\circ a^v_{e_2} \circ a^v_{e_1}$. Right: The path $\gamma_v$ is homotopic to the trivial path, i.e. it is contractible (via deformation of the path on the ``back'' of the vertex-sphere).}
\label{fig_vertexundone}
\end{figure}

Consider the graph $\G\in\Sigma$, and embed it in $\mathbb R^3$ in such a way that when projected on the plane  $z=0$, only edges intersect (transversally) an the vertices are completely resolved.
This equips each vertex $v\in\G$ with a cyclic orientation. 
By choosing an edge at each vertex (or cilium), the cyclical symmetry can be broken. 
Now, consider the surface $S_\G\subset\mathbb R^3$, the boundary of a tubular neighborhood of $\Gamma$.
Pick on $S_\G$ one of the two graphs homeomorphic to $\G$ such that they have the same projection on the plane $z=0$. Schematically, this graph is obtained by displacing $\G$ in the positive $z$ direction by $\epsilon$. I will denote this graph $\G$, and in order to distinguish it---when necessary---from the original one I rename the latter $\G'$.
Now, $S_\G$ is decorated with a graph: in particular it as $V$ marked points corresponding to the vertices of $\G$, and $E$ lines corresponding to edges of $\G$.
Beside these decorations, introduce at each vertex $v\in\G$ one loop $a_e^v$ for each edge $e$ such that $v\in\pp e$, with the following properties: the loop starts and ends at $v$ and has linking number $+1$ with the corresponding edge $e'\in\G'$ and vanishing linking number with all other links.%
\footnote{More precisely: it has linking number $+1$ with every close path $\ell\subset\G'$ passing through $e'$ once and with the correct orientation, and vanishing linking number with any path $\tl\ell\subset\G'$ which does not pass through $e'$.}
At each vertex, the path obtained by composing the paths $a^v_{e_i}$ according to the edge ordering described above,
\be
\gamma_v = a^v_{e_n}\circ \cdots \circ a^v_{e_2} \circ a^v_{e_1},
\label{eq_gammav}
\ee
is contractible. See figure \ref{fig_vertexundone}.

Define $h_e$ to be the holonomy of $\A$ along the edges $e\in\G\subset S_\G$, and $g_e^v$ to be the holonomy of $\A$ along the loops $a_v^e$.
Using the flatness of the connection $\A$ on $S_\G\setminus\{v\in\G\}$, and the observation above, it is immediate to see that
\be
g^v_{e_n}\cdots g^v_{e_2} g^v_{e_1} = \overleftarrow\Pexp\int_{\gamma_v} \A = \E.
\ee
Furthermore, the same property of $\A$ guarantees that
\be
g_e^{t(e)} = h_e (g_e^{s(e)})^{-1} h_e^{-1}. 
\ee
The two equations above have the same form of the deformed {Gau\ss} constraint and of the parallel transport for the exponentiated fluxes, provided the following notation is introduced
\be
g_e = g_e^{s(e)} 
\qquad\text{and}\qquad
\tl g_e = g_e^{t(e)}.
\ee
Gauge transformations at the marked point also transform $g_e$ and $h_e$ as expected.

Thus, these are coordinates on the space of flat connections on  $S_\G$ which can be readily identified with coordinates on $\mP^\EM_\G$, since they satisfy the same set of constraints.
Of course, this does not guarantee that such an identification also respect the symplectic structure carried  by the two spaces, i.e. that it also defines a symplectomorphism between $\M(S_\G, G)$ and $\mP^\EM_\G$.

The proof of this fact will be provided in the last part of this section. 
Before, I want to present the basic intuitions and one simple example.

\subsection{Examples}

\paragraph*{\bf The double $D(G)$}
The first example I want to consider is the double $D(G)$.
This is the (quasi) phase space associated to a single edge.
As I discussed, it is most naturally described in terms of the $\text{a}$- and $\text{b}$ group coordinates, or, via a change of variables, in terms of holonomies and (exponentiated) fluxes
\be
D(G) = G_\text{a} \times G_\text{b} \cong G_\text{h} \ltimes G_\text{f},
\ee
where
\be
D(G) \ni (a,b)  = (h, hg).
\ee
The intuition I want to provide here regards how to read these variables from the flat connection picture.
To build this one, one needs first to ``explode'' the spin-network edge $e\in\G$ into a two surface with the topology of a cylinder
$S_{e\in\G} \cong \mathbb S^1 \times [0,1]$.
Then, one needs to pick a marked point on each of the two $\mathbb S^1$ boundary components of $S_e$.
These will serve as reference points, at which gauge transformations are ``frozen''.
In the language used later, $S_e$ is a 2-punctured sphere with marked boundaries, denoted $\mathbb S^2_{n=2}$. 
Schematically, 
\be
e\in\G \leadsto S_{e\in\G} \cong \mathbb S_1 \times [0,1] \cong {\mathbb S}^2_{n=2}.
\ee

\begin{figure}[t!]
\includegraphics[width=.5\textwidth]{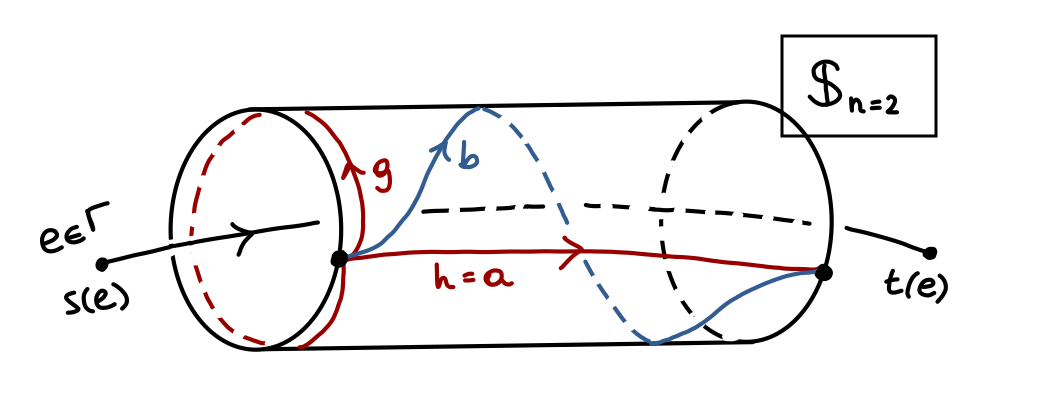}
\caption{The relation between $D_e(G)$ and the moduli space of flat connections on the 2-punctured sphere with marked boundaries, $\mathbb S^2_{n=2}$. For brevity, I have written ($a$, $b$, ...) instead of ($\gamma_\text{a}$, $\gamma_\text{b}$, ...).}
\label{fig_cyl}
\end{figure}

On $S_e$ a flat connection $\A$ is defined.
A complete set of gauge-invariant observables of $\A$ is then given by its two parallel transports defined on paths respectively around and along the tube, which start and end at the two marked points.
In figure \ref{fig_cyl}, I have represented both the paths $\gamma_\text{a}$ ($\gamma_\text{b}$), which correspond to the group elements $a\in G_\text{a}$ ($b\in G_{\text{b}}$, respectively) via
\be
a = \overleftarrow\Pexp \int_{\gamma_\text{a}} \A
\qquad\text{and}\qquad
b = \overleftarrow\Pexp \int_{\gamma_\text{b}} \A,
\ee
and the paths $\gamma_\text{h}$ ($\gamma_\text{f}$), which correspond to the group elements $h\in G_\text{h}$ ($g\in G_\text{f}$, respectively) via analogous equations.
The space $D(G)$, i.e. the space of flat connection on $\mathbb S^2_{n=2}$ with marked boundaries, is a quasi-Hamiltonian $G^{\times 2}$-space, where the $G$ action corresponds to the residual gauge symmetry one can perform at the two marked points.

\begin{figure}[t!]
\includegraphics[width=.58\textwidth]{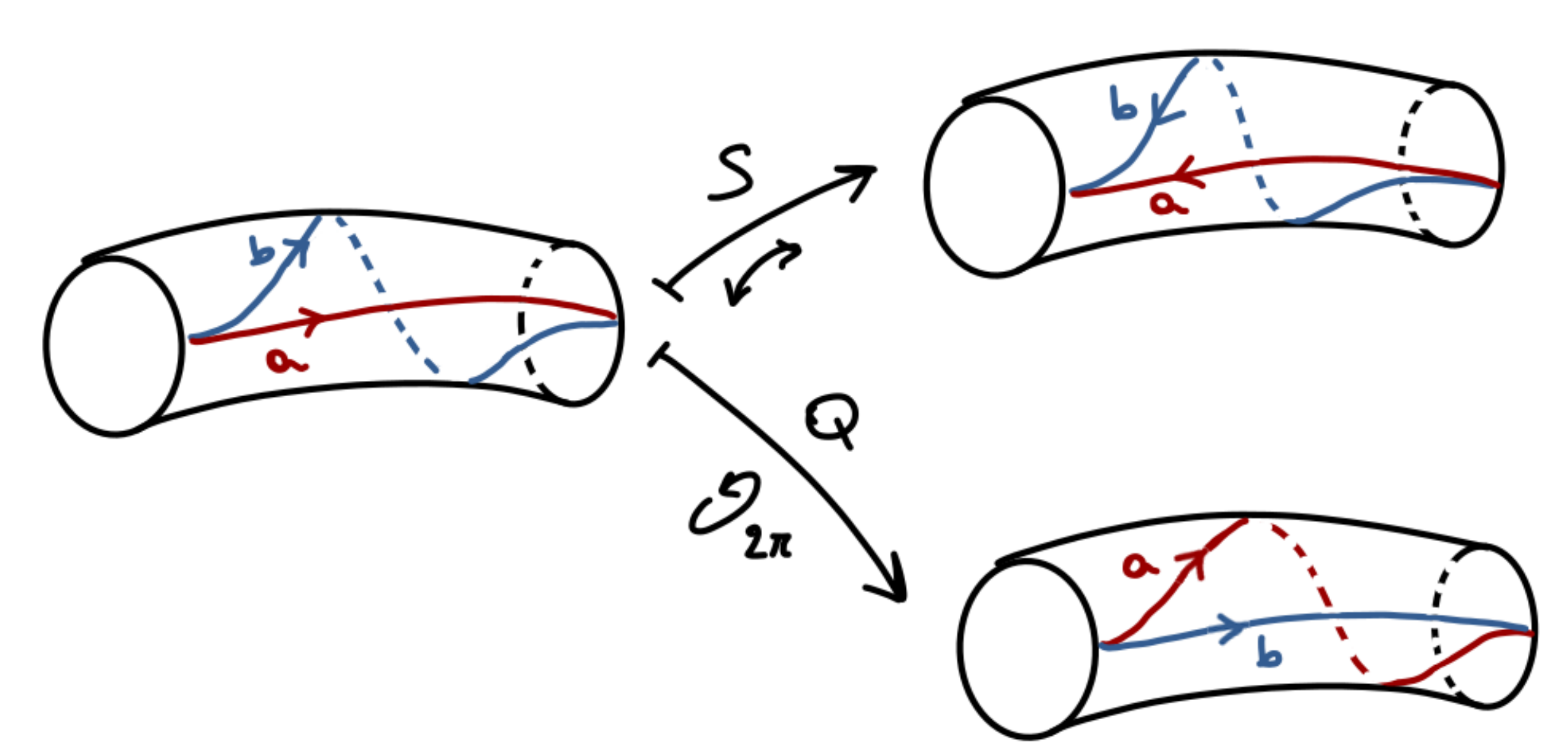}
\caption{The action of the quasi-Hamiltonian-space isomorphisms $S,Q:D(G)\to D(G)$ from the viewpoint of the moduli space of flat connections on the 2-punctured sphere with marked boundaries $\mathbb S^2_{n=2}$. (The $2\pi$ rotation shown in the pictorial representation of $Q$ is meant to act on the ``source end'' of the cylinder.)}
\label{fig_SQ}
\end{figure}

In figure \ref{fig_cyl}, $\gamma_\text{a}$ is represented as a longitudinal holonomy which does not wind around the cylinder, while $\gamma_\text{b}$ winds around it once. Note that this characterization is purely conventional, and only the relative statement ``$\gamma_\text{f} = \gamma^{-1}_\text{a}\circ\gamma_\text{b}$ winds around the cylinder once'' is actually meaningful.
At this purpose, see also \cite{Alekseev1998}[Sect 9.4] on  the action of the mapping class group as an isomorphism of quasi-Hamiltonian spaces between $D(G)$ and itself. In particular, the maps $S,Q:D(G)\to D(G)$ are discussed, which exchange the two boundary components (i.e. reverse the orientation of the edge) and add a $2\pi$ Dhen twist to the cylinder, respectively. See figure \ref{fig_SQ}.


\paragraph*{\bf The torus}
The simplest non-empty $\G$, is the graph $\G_1$ composed by a single edge closed on itself via a 2-valent vertex. 
The momentum map at this vertex is 
\be
\mu_v = \mu_{t(e)}\mu_{s(e)} .
\label{eq_muvG1}
\ee
Thus, the corresponding reduced phase space on the preimage $\mu_v^{-1}(\E)\subset D(G)$ is parametrized by pairs $(h,g)\in G_\text{h}\times G_\text{f}$ such that
\be
\mu_v(h,g) = (\AD_h g^{-1} ) g = h g^{-1} h^{-1} g = \E,
\ee
modulo overall gauge transformations where both $h$ and $g$ transform by conjugation.
One immediately recognizes the description of the moduli space of flat connections on a torus, which is precisely the ``exploded'' version of the graph $\G_1$.
Technically, this can be obtained by fusion of $D(G)$ with itself, and reduction by the constraint $\mu_v=\E$. In the notation of \cite{Alekseev1998}, the quasi-Hamiltonian space obtained by fusion of the two factors of $D(G)$ is denoted by $\mathbf D(G)$.

\begin{figure}[h!]
\includegraphics[width=.5\textwidth]{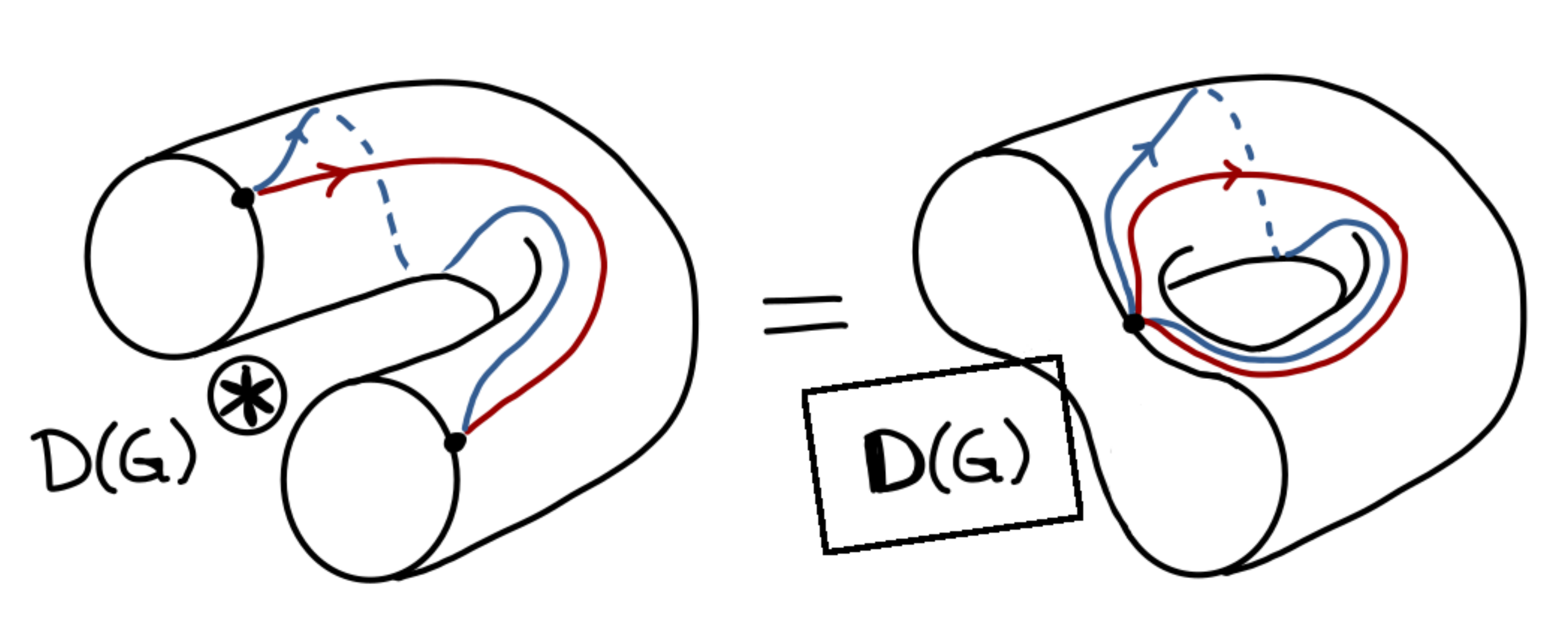}
\caption{The fusion of the two factors of $D(G)$, leading to the quasi-Hamiltonian space $\mathbf D(G)$, as represented from the viewpoint of the moduli space of flat connections on $\mathbb S^2_{n=2}$.}
\label{fig_Dfus}
\includegraphics[width=.95\textwidth]{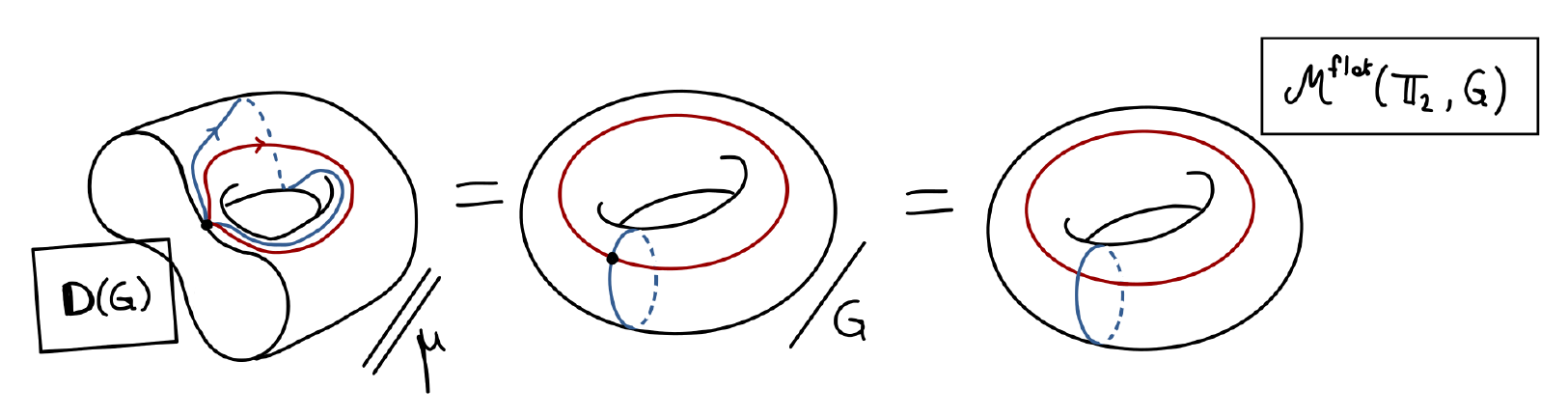}
\caption{The construction of the deformed phase space on $\G_1$, the graph with a single edge, from the perspective of the moduli space of flat connections on $\mathbb S^2_{n=2}$. The figure represents the formula: $\mathbf D(G)//\mu_v = \M(S_{\Gamma_1}=\mathbb T_2,G)$.}
\label{fig_torus}
\end{figure}

From the flat-connection perspective, the fusion product $\mathbf D(G)$ corresponds to the gluing of the two boundary components of the cylinder into a single $\mathbb S^1$ boundary component, together with the identification of the two marked points, see figure \ref{fig_Dfus}. This produces the quasi-Hamiltonian phase space associated to a single handle, that is a torus with a disk removed and a marked point on its single boundary component. It is a quasi-Hamiltonian $G$-space: a single marked point is present where to act with the residual gauge transformations. Going to the preimage $\mu_v^{-1}(\E)$ corresponds to ``filling'' the missing disk, and reducing corresponds to ``erasing'' that last marked point: fully gauge-invariant observables on the torus are the only ones left. See figure \ref{fig_torus}.\\

\subsection{Proof sketch}

To show  that the spaces $\M(S_\G, G)$ and $\mP^\EM_\G$ are symplectomorphic, it is convenient to introduce an extra step in the procedure above, that is it is convenient to split all edges into two ``half-edges'' via the insertion of auxiliary 2-valent vertices. 
The following fact provides the mathematical justification for this step:
the fusion and reduction of two edge spaces $D_{e'}$ and $D_{e''}$, where $v = s(e'')=t(e')$, simply gives another edge space $D_e$, such that $s(e) = s(e')$ and $t(e)=t(e'')$, i.e. 
\be
D_{e'} \fus D_{e''} // G_v = D_{e}.
\label{eq_DD=D}
\ee
This means that one two-valent vertex can be freely inserted on each edge, and thus that $\mP^\EM_\G$ can be built out of ``open-vertex'' spaces fused together at the two-valent vertices.

Another useful observation is the fact that the edge-orientation reversal, as the already-mentioned braid isomorphism, is an isomorphism between quasi-Hamiltonian spaces. 
In the following, I will use these isomorphisms without further mention.

Now, let me recall one of the main results of \cite[Thm 9.3]{Alekseev1998}.
The fusion product of $n$ copies of the double $D(G)$ is isomorphic to the moduli space of flat connections on an $(n+1)$-punctured sphere with one marked point per boundary component.%
\footnote{This is the space of flat connection modulo gauge transformations on a 2-sphere with $n+1$ disks removed and a marked point at which gauge transformations are ``frozen'' on each of the $n+1$ boundary components homeomorphic to the circle $\mathbb S^1$. This space carries an action of the group $G^{\times (n+1)}$.}
Call this space $\mP'_{n+1}$,
\be
\mP'_{n+1} = D_{e_n} \fus \cdots \fus D_{e_2} \fus D_{e_1} .
\label{eq_Pn+1}
\ee
This is a $G^{\times(n+1)}$-quasi-Hamiltonian space, with momentum map 
\be
\big( \mu_{t(e_{n})}, \; \dots \; , \;\mu_{t(e_2)}, \;\mu_{t(e_1)},  \; \mu_v = {\overleftarrow{\scalebox{1.5}{$\Pi$}}}_{i=1}^n \mu_{s(e_{i})}\big): D^{\times n} \to G^{\times(n+1)},
\ee
where I supposed one is gluing $n$ edge spaces all outgoing from a common vertex $v=s(e_i)$.

This is, of course, one of the building blocks of $\mP'_\Gamma$.
By going on-shell of the constraint $\mu_v=\E$ and reducing by the symmetry it generates, i.e. by gauge transformations at the common vertex, one finds the moduli space associated to a single open vertex with $n$ outgoing edges.

\begin{figure}[t!]
\includegraphics[width=.95\textwidth]{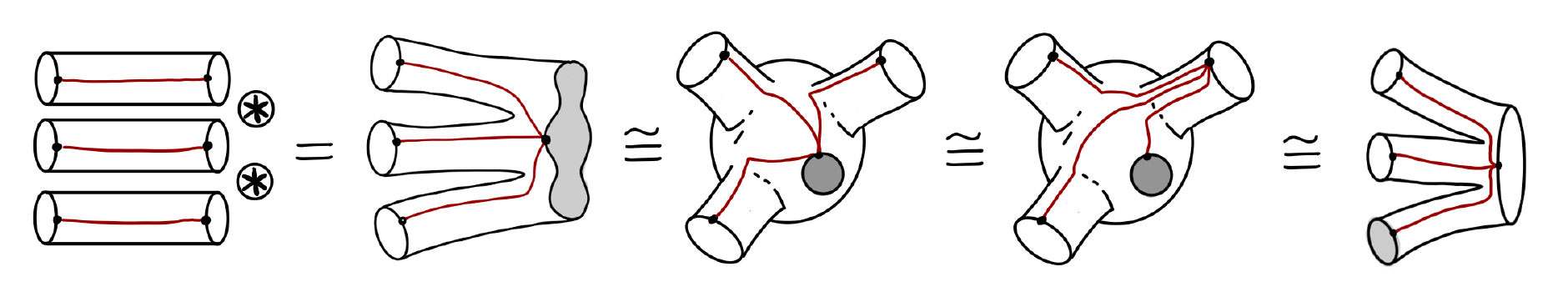}
\caption{The fusion of the 3 half edges at a common vertex of $\G$. The shading signifies that the vertex is left ``open''. This fusion gives a quasi-Hamiltonian space isomorphic to the moduli space of flat connections on $\mathbb S^2_{n=4}$, whose base point can be moved from the open vertex of $\G$ to any other auxiliary two-valent vertex (non-shaded).}
\label{fig_move}
\end{figure}

I prefer, however, to adopt here a slightly different viewpoint. 
The quasi-Hamiltonian space $\mP'_{n+1}$, in spite of having been constructed in an asymmetric fashion with respect to its $n+1$ boundary components, it is actually completely symmetric with respect to their permutation. 
Therefore, I invite the reader to think of $\mP'_{n+1}$ in such a way that vertex $v$ appears just as an open target end of one of the edges appearing in equation \eqref{eq_Pn+1}. 
Pictorially, this corresponds to moving the reference point from the boundary component associated to the ``open'' vertex $v$ (i.e. to the would-be vertex before reduction is taken with respect to the constraint $\mu_v=\E$), to another boundary component associated to one of the two-valent vertices I introduced within the edges of $\G$. See figure \ref{fig_move}.

To emphasize this change in perspective, let me write this space as follows :
\be
\mP'_{n+1} = D^{\fus n},
\qquad
\big(\mu_{t(\tl e_n)},  \;\mu_{t(\tl e_{n-1})},\; \dots\;,\; \mu_{t(\tl e_2)},\; \mu_{t(\tl e_1)},\; \mu_{\tl s}  = {\overleftarrow{\scalebox{1.5}{$\Pi$}}}_{i=1}^n \mu_{s(\tl e_{i})} \big)
\label{eq_Pn+1bis}
\ee
where the correspondence with the previous presentation is the following,
\be
\mu_{t(\tl e_n)} \leftrightarrow \mu_v ,
\qquad
\mu_{t(\tl e_{i<n})} \leftrightarrow \mu_{t(e_{i+1})}
\qquad\text{and}\qquad
\mu_{\tl s} \leftrightarrow \mu_{e_1}
\ee
Let me emphasize once more that these two fusion products are simply two different presentations of the moduli space of flat connections on the $(n+1)$-punctured sphere with marked boundaries.

It is now easy to glue the spaces associated to two neighboring vertices in $\G$. Say the two vertices are $n$- and $m$-valent, respectively. 
First, construct the space $\mP'_n$ and $\mP'_m$ as in equation \eqref{eq_Pn+1}. Then, ``move'' their reference points to the two-valent vertex associated to the edge connecting the two vertices, as in equation \eqref{eq_Pn+1bis}.
Finally, fuse the two spaces together at the two valent vertex with respect to the following composition of momenta at the common vertex:
\be
\mu_{\text{2-valent}} = \mu_{\tl s_n} \mu_{\tl s_{m}},
\ee
with obvious notation. 

\begin{figure}[t!]
\includegraphics[width=.9\textwidth]{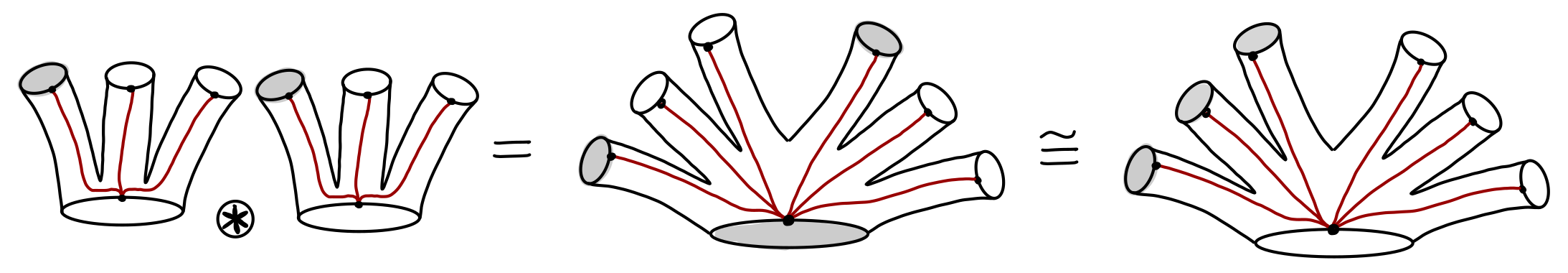}
\caption{The fusion of two 3-valent ``open'' vertices into a 4-valent ``open'' vertex. The shaded disks are not associated to (half) edges of the graph, and rather represent an ``open'' vertex at which gauge invariance has not yet been imposed (or, equivalently, at which reduction has not yet been performed). The two leftmost 3-valent vertices are equal to the rightmost term in figure \ref{fig_move}.}
\label{fig_fusion-halfedges}
\end{figure}

The space so-obtained is isomorphic to the moduli space of the $(n+m+1)$-punctured sphere with marked boundaries:
\be
\mP'_{n+1}\fus \mP'_{m+1} = \mP'_{m+n+1}.
\ee
The $(n+m+1)$ boundaries include the two ``open vertices'' $v_n$ and $v_m$, as well as the ``open'' two-valent vertex which sits within the chosen edge connecting the aforementioned vertices.
The remaining $n+m-2$ boundary components correspond to the $(n+m-2)$ external legs resulting from the gluing of an $n$-valent and an $m$-valent vertices. See figure \ref{fig_fusion-halfedges}.

Clearly, it is now possible to proceed recursively, moving again the reference point to one of the external edges which connect the newly constructed punctured sphere to yet another thickened vertex, and so on. 
This way, one can glue vertices along a spanning tree $\tau\subset \G$. In other words, all vertices can be glued to each other along a selection of edges which do not form closed loops. 
The result of the procedure is the $G^{\times (2E + 1)}$-quasi Hamiltonian space%
\footnote{The counting is the following: there are $V$ ``open'' vertices, plus $(V-1)$ ``open'' auxiliary two-valent vertices along the edges of $\tau$, plus $2(E-V+1)$ other factors due to the fact that every edge in $\Gamma\setminus\tau$ is cut in two and one can act independently on the two halves. Thus $2E+1 = V+(V-1)+2(E-V+1)$.}
$\mP'_{2E+1}$ associated to a $(2E+1)$-punctured sphere with marked boundaries.

At this point, any two half-edges which are fused back together would form a closed loop. 
From the perspective of the $(2E+1)$-punctured sphere above, this corresponds to identifying two boundary components hence forming handles.

This step can be unraveled in the following way.
Observe that the action of $G^{\times (2E + 1)}$ on $\mP'_{2E+1}$ is encoded in the momentum map
\be
\big( \mu_{t(\tl e_{2E})}, \; \dots\;, \; \mu_{t(\tl e_1)}, \; \mu_{\tl s} = {\overleftarrow{\scalebox{1.5}{$\Pi$}}}_{i=1}^{2E} \mu_{s(\tl e_{i})} \big).
\ee
Of the $(2E+1)$ factors above, $(2V-1)$ correspond to ``open'' vertices (here included the auxiliary two-valent vertices), while the remaining $2(E-V +1)$ correspond to open half-edges. Let me also assume that the last factor corresponds to the root vertex of $\tau$, and---up to repeated use of the braid isomorphism---that the edges $(\tl e_{2E}, \dots, \tl e_{2E-2V+3})$ are associated to the other $2(V-1)$ vertices. Eventually, these factors will disappear by fusion and reduction, hence implementing gauge invariance at the vertices of $\G$ (or ``closing'' these auxiliary boundary components from the viewpoint of the punctured sphere).

The remaining $2(E-V+1)$ edges have to be fused pairwise to form the edges in $\G\setminus\tau$.
Again up to repeated use of the braid isomorphism, it is possible to assume that half-edges to be fused together are closed to one another in the listing above. 
That is, pairs of edges $(\tl e_{2k+1}, \tl e_{2k+2})$ are going to be fused to one another to form handles (the global range of the index $k$ is $k\in\{ 0,\dots, E-V\}$). 
Using the orientation reversal isomorphism, edges can be flipped so as to glue the target of $\tl e_{2k+1}$ to the source of $\tl e_{2k+2}$. In this way, at the root vertex, the following ordering of factors will appear:
\be
\mu_\text{root} = {\overleftarrow\prod}_j \mu_{s(\tl e_{j})}  {\overleftarrow\prod}_k \mu_{t(\tl e_{2k})}\mu_{s(\tl e_{2k-1})},
\ee
where the ranges of $j$ and and $k$ are $j\in \{2E-2V+3, \dots, 2E\}$ and $k\in\{ 1,\dots, E-V+1\}$, respectively.

Now, using equation \eqref{eq_DD=D}, half-edges can be glued together into single edges:
\be
D_{\tl e_{2k+1}} \fus D_{\tl e_{2k+2}}//\mu_{v_k} = D_{\tl e_k}.
\ee
where $\mu_k$ stands for the $G$ action at the auxiliary two-valent vertex separating the two half-edges $\tl e_{2k+1}$ and $\tl e_{2k+2}$.
Thus, after fusion and reduction of these pairs of half-edges, the momentum map reads
\be
\big( \mu_{t(\tl e_{2E})}, \; \dots\;, \; \mu_{t(\tl e_{2E-2V+3})}, \; \mu_\text{root} \big).
\ee
where
\be
\mu_\text{root} = {\overleftarrow\prod}_j \mu_{s(\tl e_{j})}  {\overleftarrow\prod}_k \mu_{t(\tl e_k)}\mu_{s(\tl e_k)}.
\ee
The first $(E-V+1)$ factors in $\mu_\text{root}$, which are labeled by $k$, are the momentum maps associated to the handles, $\mu_{\text{handle-}k} = \mu_{t(\tl e_k)}\mu_{s(\tl e_k)}$. This corresponds to the fusion of the two factors of $D_{\tl e_k}(G)$, which simply means that the source and target marked points on the handle transform together, i.e. are attached to the same vertex, as in the ``flower'' representation of gauge-fixed spin-networks (see e.g. \cite{Charles2016}). 
In the notation of \cite{Alekseev1998}, the handle quasi-Hamiltonian space is denoted $\mathbf D(G)$. This is a quasi-Hamiltonian $G$-space.

The remaining $2(V-1)$ factors correspond to the auxiliary edges corresponding to the ``open'' vertices but the root (again, here are included both the vertices of $\G$---except the root vertex---and the auxiliary 2-valent vertices on the edges $e\in\tau$).

As proven in \cite[Thm 9.3]{Alekseev1998}, the quasi-Hamiltonian $G^{\times(2V-1)}$-space
\be
\mP'_{n+1,g} = \mP'_{(2V-1),(E-V+1)} = D(G)^{\fus 2(V-1)} \fus \mathbf{D}(G)^{\fus (E-V+1)}
\label{eq_Pn=1g}
\ee
obtained from the previous construction is (naturally) isomorphic to the moduli space of flat $G$-connections on a $(2V-1)$-punctured Riemann surface with $g=(E-V+1)$ handles and marked boundary components.%
\footnote{Recall that this means that the gauge symmetry has been mod out everywhere on the Riemann surface but one marked point for each of the $2(V-1)$  $\mathbb S^1$ boundary components. Therefore, there is a residual gauge group acting on $\mP'_{n+1,g}$, which is $G^{\times(2V-1)}$.}
The index $g$ stands for ``genus''.

To obtain $\mP^\EM_\G$, it is enough to ``fill-in'' the $(2V-1)$ auxiliary boundary components, i.e. setting the fluxes $(\mu_{t(\tl e_j)}=\E)$. 
This automatically trivializes the $D(G)$-factors in equation \eqref{eq_Pn=1g}.
Accordingly, in the root momentum map only the handle-factors survive
\be
\mu_\text{root}|_{(\mu_{t(\tl e_j)}=\E)} = {\overleftarrow\prod}_k \mu_{t(\tl e_k)}\mu_{s(\tl e_k)}.
\label{eq_rootconstr}
\ee

\begin{figure}[h!]
\flushleft
\hspace{2.1cm}({\it i}) \hspace{2.55cm}({\it ii})\hspace{2.85cm} ({\it iii}) \hspace{2.65cm} ({\it iv}) \hspace{2cm} ({\it v}) \\\vspace{-1em}
\center
\includegraphics[width=.9\textwidth]{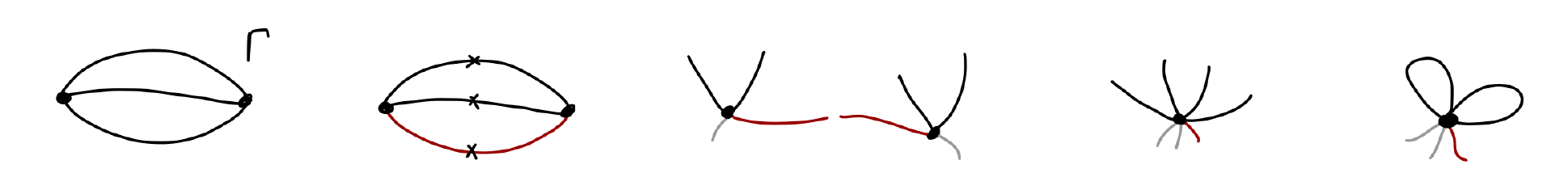}\vspace{5pt}\\
\includegraphics[width=.9\textwidth]{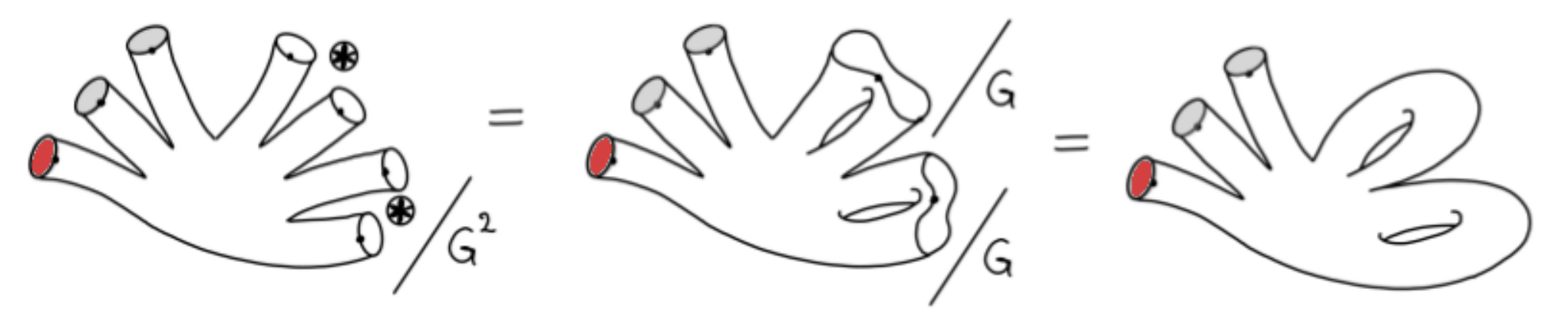}
\caption{Top Row: The deformed phase space of a spin-network graph $\G$---panel ({\it i})---is built as follows: ({\it ii.a}) all the edges of $\G$ are split into half-edges via the insertion of 2-valent vertices (represented by $\times$'s); ({\it ii.b}) a spanning tree $\tau\subset \G$ is chosen (in red); ({\it iii}) the quasi-Hamiltonian space associated to the corresponding ``open'' vertices is constructed (see figure \ref{fig_move}); ({\it iv}) the vertices along $\tau$ are fused together (in this picture the root is represented by a line, cf. figure \ref{fig_fusion-halfedges}); ({\it v}) the remaining edges in $\Gamma\setminus \tau$ are glued back together into ``handles''. Bottom Row: a representation of step ({\it iv}) to ({\it v}) from the perspective of the moduli space of flat connections (one of the shaded holes plays the role of the root, and it is hence shaded in red in the figure). The last step in the construciton (not represented in the figure) is the ``closure'' of the ``open'' vertices: this corresponds to the final reduction of equation \eqref{eq_finalreduction}.}
\label{fig_fusion2}
\end{figure}

What is obtained in this way is the quasi-Hamiltonian $G$-space moduli space of flat $G$-connections on a genus $g=(E-V+1)$ Riemann surface with a marked point (corresponding to the root vertex).
Reduction by the last remaining constraint, $\mu_\text{root}=\E$, gives a quasi-Hamiltonian space with trivial group action. As already observed, this is an actual symplectic space.
As \cite[Thms 9.2 and 9.3]{Alekseev1998} show, such a symplectic space is precisely the moduli-space of flat $G$-connections on a closed genus $g=(E-V+1)$ Riemann surface. I.e.
\be
\mP^\EM_\Gamma \cong \mP'_{(2V-1),(E-V+1)} // (\mu_{t(\tl e_j)}=\E,\mu_\text{root}=\E) \cong \M( S_\Gamma, G),
\label{eq_finalreduction}
\ee
where $S_\Gamma$ is a indeed a closed genus $g=(E-V+1)$ Riemann surface,
\be
S_\Gamma \cong S_{n=0,g=(E-V+1)}.
\ee
The whole procedure is summarized in figure \ref{fig_fusion2}.

\section{Polarizations, excitations, and defects\label{sec_polarization}}

In the previous section, I showed that the deformed phase space $\G$, $\mP^\EM_\G$ acquires in four dimensions a particularly simple interpretation which makes the duality between its electric and magnetic components manifest. 
Indeed, in this case, $\mP^\EM_\G$ is naturally interpreted as the (finite-dimensional) reduced phase space $\M(S_\G , G)$ obtained from the moduli space of flat connections $\A$ on the Riemann surface $S_\G = \pp H_\G$ equipped with the Atyiah--Bott symplectic form. Recall that $S_\G$ is the boundary of a tubular neighborhood of $\G$, denoted $H_\G$.
In this picture, the holonomies and exponentiated fluxes on $\G$ are interpreted as the longitudinal and transverse holonomies along the ``tubes'' of $S_\G$, respectively.

Now, suppose $\G$ is the 1-skeleton of the dual to a cellular decomposition $\Delta$ of the three dimensional Cauchy hypersurface $\Sigma$,  i.e. $\G=\Delta^\ast_1$. Then the magnetic and electric fluxes across the 2-cells of $\Delta$ and $\Delta^\ast$, respectively, correspond to holonomies along the $a$- and $b$-cycles of $S_\G$. 
Since these are readily interchanged by considering the dual cellular decomposition of $\G$, this construction makes completely manifest the duality between its electric and magnetic components.

In particular, the discrete versions of the Bianchi identity%
\footnote{See \cite{Freidel2003} for a detailed treatment of the discretization of the Bianchi identities in 3 dimensions.} 
and of the deformed {Gau\ss} constraint have the same nature.
The flatness of the connection $\A$ relates them to the contractibility of certain compositions of $b$- and $a$-cycles associated to the vertices of $\Delta$ and $\Delta^\ast$, respectively. 

The magnetic and electric variables are conjugated to each other, and therefore offer two different choices of polarizations of the phase space $\mP^\EM_\G$.
This can be used to build a Schr\"odinger representation for the quantum state space associated to $\mP^\EM_\G$.
Not seeking for the moment a precise definition of this quantum state space (more about this later), I will limit myself to generic and qualitative considerations.

First of all, each choice of polarization comes with an associated natural notion of squeezed ``vacuum'' state. 
The term ``vacuum'' is here used with the looser, purely kinematical, acceptation of preferred reference state.
Consider e.g. the electric polarization. The corresponding electric vacuum fixes all electric fluxes to their trivial value:%
\footnote{Notice, however, that the electric fluxes must satisfy non-trivial Gau{\ss} constraints, and consequently cannot be treated as fully independent variables. The previous statement, however, turns out to have a more precise analogue, explained in the next section. In this section, I will henceforth neglect this type of subtleties.
}
\be
g_{e\in\G} = \E \qquad \forall e\in \G.
\label{eq_Evac}
\ee 
Similarly, in the magnetic representation, the corresponding magnetic vacuum is defined by
\be
h_f \equiv{\overleftarrow{\scalebox{1.5}{$\Pi$}}}_{e:e\in\pp f} h_e^\epsilon =\E \qquad \forall f\in \G.
\label{eq_Bvac}
\ee
From the perspective of $\G$ the above equation is readily seen to correspond to the so-called $BF$ vacuum \cite{Dittrich2014,Dittrich2014c,Dittrich2015,DittrichGeillerTQFT,Delcamp2017}, in which the curvature of the (3+1) connection vanishes. The electric and magnetic squeezed vacua represent deformed versions of the high- and low-temperature states of Yang--Mills theory, respectively.

Note that faces $f$ of $\G$ (2-cells of $\Delta^\ast$) are dual to 1-cells of $\Delta$, i.e. to edges of the discretization $\Delta$.
Thus, a choice of polarization corresponds to the choice of one of the two complementary handlebody in the Heegaard splitting induced by $\Delta$: the imposition of the corresponding vacuum equations corresponds indeed to the imposition of the flatness equations in the bulk of either $H_\G$ or $H_{\G^\ast}$. This can be implemented, most naturally, via a Chern--Simons theory for (an extension of) $\A$ within either handlebody. 
The space of flat $G$-connections on a three-manifold $H$, with $S_\G=\pp H$, forms a Lagrangian submanifold of $\M(S_\G, G)$  \cite{Gukov2003}, and hence, quantum mechanically, defines a quantum state of the system (e.g. \cite{Bates1997}, or \cite{Gukov2003,HHKRplb}).%
\footnote{
Very roughly, a Lagrangian submanifold of a phase space, $\mathcal L\subset \mP$, is a half-dimensional submanifold which ``cuts the phase space in two'', i.e. denoting the embedding $\iota_\mathcal L:\mathcal L\hookrightarrow \mP$,  $\iota_\mathcal{L}^\ast\omega \equiv 0$. Or, in adapted Darboux coordinates, $\mathcal L = \{ (q,p)\in\mP \,:\, p=p_\mathcal{L}(q)\}$. Then, in the adapted Schr\"odinger representation, the quantum state $\psi=\E^{i S}$ can be introduced which is associated to $\mathcal L$ by the following: $[\hat p - p_\mathcal{L}(\hat q) ]\psi=0$, i.e. $\pp_q S(q) = p_\mathcal{L}(q)$. Formally, this is solved by $S = \int^q \iota_\mathcal{L}^\ast \vartheta$, with $\vartheta = p\d q\in\Omega^1(\mP)$ the symplectic form potential associated to the Darboux coordinates above.
}
Such states can then be interpreted as either a purely electric or a purely magnetic vacuum for the $(3+1)$ theory, depending on which handlebody, $H_\G$ or $H_{\G^\ast}$, is chosen \cite{Dittrich2017}.

Excitations can be added on the top of these vacua  via the relaxation of equation \eqref{eq_Evac} or \eqref{eq_Bvac}.
These excitations correspond to non-vanishing electric fluxes or curvature. Of course, the curvature excitation I am referring to is a curvature excitation of the (discretized and deformed) $(3+1)$-connection $A$, and {\it not} of the connection $\A$ on the two-surface $S_\G$. 
Indeed, the role of the flatness of $\A$ is that of implementing---automatically and at the same time---both the Gau{\ss} constraint and the Bianchi identities for the connection $A$.
The electric and curvature excitation of $A$, on the other hand, are supported and allowed for by the non-triviality of the topology of $S_\G$: the more refined $\Delta$ is, the more excitations it can support. 

From the Chern--Simons handlebody perspective, unfreezing electric degrees of freedom means making the corresponding $a$-cycles non contractible. This can be achieved via the insertion of Wilson-graph operators in $H_\G$. The support of the Wilson-graph operator is nothing but $\G$ itself, while the representation attached to the Wilson-graph operators correspond to the magnitudes of the electric field. Analogue statements hold for the magnetic excitations.
Thus, the deformation described in this paper can be obtained via a coupling of the original Wilson-graph operators supported by $\G$ to an auxiliary Chern--Simons theory. It would be interesting to understand whether this Chern--Simons theory can be somewhat rigorously understood as coming from a QCD $\theta$-term. A similar idea was at the basis of the construction of \cite{HHKR} in a quantum-gravitational context---see section \ref{sec_gravity}.

So far, all considerations regarded solely the {\it vacuum} sector of YM theory. 
In other words, I have not considered so far the coupling to charged sources.
In standard, i.e. undeformed, YM theory, an electric pointlike source appears as a violation of the Gau{\ss} constraint: $d_A \ast E = 4 \pi \rho$, or in its lattice version---assuming only outgoing edges---$\sum_{e:v\in\pp e} X_e = \Phi_v$.
The deformed analogue of the latter equation is
\be
\overleftarrow{\prod_{e:v\in\pp e}} g_e = G_v.
\ee
As I have stressed already, the Gau{\ss} constraint has a purely topological origin from the viewpoint of the flat connection $\A$ on $S_\G$.
This means that the introduction of a source like in the previous equation requires a modification of the topology of $S_\G$ ``around vertex $v\in\G$''. The simplest such variation is the introduction of a puncture carrying a holonomy $G_v$. 
The deformed phase space on $\G$  in presence of the defect $G_v$ can be constructed by quasi-Hamiltonian reduction with respect to the constraint 
\be
\mu_v = G_v,
\ee
rather than $\mu_v=\E$. The resulting phase space will depend only on the conjugacy class  of $G_v$ $C_v=[G_v]_\text{conj}$.
Of course, the same puncture can carry in precisely the same way a violation of the discretized Bianchi identities.
The simplest way to interpret this type of defects is therefore in terms of electro-magnetic dyons.

A feature of this proposal is that it does {\it not}---at least naively---comply to the Dirac--GNO electro-magentic duality \cite{Dirac1931,Goddard1977,Montonen1977,Witten1979,Olive1997}. It would be interesting %
to study this issue further and possibly see whether the present construction can be modified accordingly. 
One---quite vague---hint comes from the following fact: the gauge group $G_{s(e)}\times G_{t(e)}$ does not act faithfully on $D_e(G)$, but $H=G_{s(e)}\times G_{t(e)}/Z(G)_d$ does.%
\footnote{$Z(G)_d\subset Z(G)\times Z(G)\subset G\times G$ is the diagonal embdedding of the center of $G$, $Z(G)$, in $G\times G$.}
From this definition, provided that $G$ is simply connected, it follows $\pi_1(H) \cong Z(G) \cong Z(H)$, which is a necessary condition for the symmetry group to be self-dual in the sense of GNO, i.e. $H^\lor\cong H$. 
In the elementary case $G=\SU(2)$, it turns out precisely that $H=\SO(4)/(\mathbb Z_2)_d \cong \SU(2) \times \mathrm{SO}(3)$ with $\SU(2) = \mathrm{Sp}(3) = \SO(3)^\lor$.

\section{Some remarks on the quantization of $\mP^\EM_\G$\label{sec_quantum}}

As a manifold, the phase space $\mP^\EM_\G$ is isomorphic to the quotient of $G^{2(E-V+1)}$ by the action of the momentum map of equation \eqref{eq_rootconstr}. Therefore, it is a compact space almost-everywhere homeomorphic to $\mathbb R^{2(E-V)\mathrm{dim}(G)}$.

Being compact, upon quantization, the ensuing Hilbert space is expected to possess only a finite number of states.

A more precise version of this statement can be obtained by an analysis of the symplectic volume of $\mP^\EM_\G = \mP'_\G//\mu_\text{root}$. This can be shown via the construction of the appropriate Duistermaat--Heckman measure \cite{Alekseev2002}.
The starting point is the definition of the analogue of the Liouville form on $\mP'_\G$,
\be
\mathcal L = \frac{1}{n!}\frac{\omega^{\wedge n}}{\sqrt{\left|\det\left(\tfrac{1 + \Ad_\mu}{2}\right)\right|     }} \;,
\ee
where $n=(E-V+1)$.
Notice how the denominator vanishes precisely at the points where $\omega$ is degenerate, see equation \eqref{eq_Ad+1}.
As shown in \cite{Alekseev2002}, the Liouville measure on the double $D(G)$ corresponds to the Riemannian measure on $G\times G$, and the symplectic volume of $\M(S_\G,G)$ is shown to match Witten's formula \cite{Witten1992}:%
\footnote{This version of the formula holds in the vacuum sector only, i.e. in absence of electric or magnetic charges. For the most general case, see \cite{Alekseev2002}.}
\be
\mathrm{Vol}(\mP^\EM_\G ) = \# Z(G) \mathrm{vol}(G)^{2(E-V)}\sum_{J\in\mathrm{Irrep}(G)} \frac{1}{(\mathrm{dim} V_J)^{2(E-V)}},
\ee
where $\mathrm{vol}(G)$ is the Riemannian volume on $G$ for the given inner product on $\frakg$, and $\# Z(G)$ is the cardinality of the center of $G$.

On the basis of the discussion of the previous section, the quantization of $\mP^\EM_\G$ is isomorphic to the Hilbert space of a $G$-Chern--Simons theory on $S_\G$. A construction of such a quantum space in a four-dimensional context appeared in a recent paper by Dittrich \cite{Dittrich2017}, in a setup which can be interpreted as the quantum analogue to the one detailed in this paper.
The comparison with the present paper is, however, not completely straightforward due to the following subtlety: rather than attempting a direct construction of the quantum space of non-commutative flat connections on $S_\G$, Dittrich first builds the quantum space of a fiducial three-dimensional $BF$ theory on $S_\G$, which possesses twice as much degrees of freedom,%
\footnote{These degrees of freedom correspond to holonomies and electric fluxes along {\it both} $a$- and $b$-cycles. Indeed, in $BF$ theory, the canonical variables are the same as in Yang--Mills theory, with a Poisson-commutative gauge connection.}
and then removes half of these degrees of freedom imposing some constraints. Finally, she verifies that the resulting Hilbert space is isomorphic to that of the Witten--Reshetikhin--Turaev model \cite{Reshetikhin1991,Barrett2007}, the model widely expected to correspond to quantum Chern--Simons theory \cite{Andersen2012}.
As a result, in Dittrich's paper, states in the electric (magnetic) polarization are labeled by elements of a finite fusion category $\mathcal C$ associated to the $a$- ($b$-)cycles of $S_\G$.
If  $\mathcal C = \mathrm{Rep}(U_q(\SU(2)))$, with $q=\E^{\mathrm{i}\frac{2\pi}{k}}$, the number of states is controlled by the level $k\in\mathbb Z\setminus\{0\}$.

The classical origin of this parameter has to be looked for in the normalization of the Killing form $\la\cdot,\cdot\ra \leadsto \tfrac{k}{4\pi}\la\cdot,\cdot\ra$. 
As seen above, this regulates the total volume $\mathrm{Vol}(\mP^\EM_\G ) $ and hence the number of states at the quantum level.
It can also be seen as the parameter regulating the entity of the flux exponentiation, $g = \exp(2\pi X/k)$, which in turn originates in the particular coupling to the Chern--Simons theory as in \cite{HHKR,HHKRplb}.
All these ways of introducing $k$ are in the end equivalent. For example, the way $k$ appears in the flux exponentiation formula is dictated by the way one maps the original momentum---which takes value in the {\it dual} of the Lie algebra---in a Lie algebra element that can be eventually exponentiated.
In the next section, I will briefly review this series of works and related $k$ to the cosmological constant.

\section{Gravitational interpretation and the cosmological constant \label{sec_gravity}}

When expressed in terms of Ashtekar variables \cite{Ashtekar1986,Barbero1995,Rovelli2007,Thiemann2004}, the kinematical phase space of general relativity, i.e. before the imposition of the Hamiltonian constraint, is the same as that of a gauge invariant $\SU(2)$ YM theory.

In this setting, the electric field $E=E[q]$ encodes the three-dimensional metric $q_{ab}$ of the Cauchy slice $\Sigma$,%
\footnote{More precisely, this is encoded in a triad field $e^i_a$, such that $q_{ab} = \delta_{ij}e^i_a e^j_b$.}
the $\Lie(\SU(2))\cong\mathbb R^3$ indices being identified with indices in the tangent space,
while the connection $A=\G_\text{LC}[q] +\gamma K$ is a weighed sum of the Levi--Civita connection $\G_\text{LC}[q]$ and the extrinsic curvature tensor $K$. The parameter $\gamma$ is known as the Barbero--Immirzi parameter when real, and gives rise to Ashtekar's original self-dual formulation when equal to the imaginary unity.
To agree with my previous notation, here I consider $(A,E)$ to be a canonical pair, although in the loop quantum gravity (LQG) literature the convention is preferred in which $E_\text{here}=\ell_\text{Pl}^{-2} E_\text{LQG}$, where $\ell_\text{Pl}^2 = 8\pi \mathrm G\gamma$ is the fundamental quantum of area (in units of $\hbar$).

In the gravitational context, a spin-network $\G$ is thus interpreted as a finite-resolution description of the gravitational degrees of freedom $(A,E)$.
In particular, the holonomies $h_e$ encodes at the same time the parallel transport along $\Sigma$ and its extrinsic curvature, while the discrete electric fluxes $X_e$ encode the area of the surface dual to the edge $e\in\G$.
In particular, the spin-network data allow to reconstruct a slightly generalized notion of discrete piecewise flat geometry, known as twisted geometries \cite{Freidel2010,Freidel2010twist2,Livine2012}.

At the core of this reconstruction is the interpretation of the electric fluxes as ``area vectors'',
\be
X_e^i = \ell_\text{Pl}^{-2} a_e n_e^i \in \mathbb R^3,
\ee
where $a_e\in\mathbb R^+$ is the area of the dual surface and $n_e^i$ is its three-dimensional normal unit vector, and of the discrete Gau{\ss} constraint
\be
\sum_{e:v\in\pp e} a_e \vec n_e = 0
\label{eq_closure}
\ee
as the requirement for these area vectors to define a (flat) convex polyhedron. 
Indeed, according to a theorem due to Minkowski \cite{Minkowski1897,Alexandrov2005}, a set of (more than three) vectors $\{a_e \vec n_e\}$ satisfying the ''closure equation'' \eqref{eq_closure} define one (and only one) convex polyhedron whose faces have area $a_e$ and outgoing normals $\vec n_e$.

In \cite{HHKR,HHKRplb,Haggard2016}, it was shown that the deformed Gau{\ss} constraint
\be
\overleftarrow{\prod_{e:v\in\pp e}} g_e  = 0
\ee
can be used to reconstruct in the 4-valent case exactly two homogeneously curved (convex) tetrahedra---only one if the cyclical symmetry of the above equation is broken---provided the group elements are interpreted as exponentiated fluxes defined via the non-Abelian Stokes theorem:
\be
g_e = \exp \big( \tfrac{\ell_\text{Pl}^2\Lambda}{3} X_e^i\tau_i \big) = \overleftarrow\Pexp \int \G_\text{LC}[q_o],
\ee
$q_o$ standing for the homogenous metric of constant curvature $\Lambda$.
By an homogeneously curved tetrahedron, it is meant a geodetic tetrahedron flatly embedded in either $\mathbb S^3$ or $\mathbb H^3$ (the round three-sphere or hyperbolic three-space) of curvature radius $r=\pm\sqrt{|3/\Lambda|}$.
The sign of the curvature is automatically encoded in the group elements.

Thus, in this setting, the quantity $\Lambda\ell^2_\text{Pl}$ plays the same role as the parameter $k$ discussed in the previous section.
As such, in this framework, the introduction of a cosmological constant (of either sign) can be used to induce a compactification of the phase space, with $\Lambda$ providing an IR cutoff for the spectra of geometric operators.

Moreover, the ensuing geometries are described in a self-dual manner, with the electric fluxes being replaced by holonomies of $\G_\text{LC}[q]$, hence conjugate to the holonomies of $A$.

A quantum version of the ensuing self-dual geometries is provided by the construction of Dittrich \cite{Dittrich2017}.
It is therefore of interest to point out the connections between her quantum construction, which she proposes to be the canonical version of the Crane--Yetter TQFT, with the analysis of \cite{HHKR,HHKRplb,Haggard2016}, which studies the deformed spin-network as issued from a semiclassical evaluation of the so-called $\Lambda BF$ theory with a specific graph insertion.%
\footnote{Actually, the analysis of \cite{HHKR,HHKRplb} goes further: it shows how to reconstruct a {\it four}-dimensional homogeneously curved simplex, out of a combination of holonomy data.}
Indeed, the comparison is of interest because it reinforces Dittrich's proposal at the light of the results of this paper and of the expectation of the Crane--Yetter TQFT to be a quantum version of $\Lambda BF$.
\footnote{The latter is a theory defined by the action $S_{\Lambda BF} = \tfrac{1}{2\kappa}\int B\wedge F-\tfrac\Lambda6 B\wedge B$. Integration of the momentum $B$, gives the topological term of the $\theta$QCD action, i.e. $\tfrac{3}{4\kappa\Lambda}\int F\wedge F = \tfrac{3}{4\kappa\Lambda}\oint CS$, with $F$ and $CS$ the curvature and Chern--Simons forms of a connection. This correspondence with Chern--Simons is the simple-minded classical analogue of the correspondence between the Crane--Yetter and the Reshetikhin--Turaev models of \cite{Barrett2007}.}

\section{Conclusions \label{sec_conclusions}}

Using the framework for quasi-Hamiltonian $G$-spaces devised by Alekseev, Malkin, Meinrenken, and Kosmann-Schwarzbach in the late 1990s, I have constructed a deformation of the phase space of lattice Yang--Mills theory.
This deformation has various desirable properties: ({\it i}) the discretized electric and magnetic variables are valued in the same space, i.e. the gauge group $G$; ({\it ii}) in $(3+1)$ dimensions this fact is promoted to a full duality between the electric and magnetic variables which admits a geometric interpretation in terms of an auxiliary, Poisson-non-commuting, flat connection defined on the Heegaard surface associated to an Heegaard splitting of the Cauchy surface $\Sigma$;  ({\it iii}) independently from the spacetime dimension, the resulting phase space is compact and of finite volume, a fact that upon quantization gives rise to a finite Hilbert space. 

After a detailed review of the construction of the above phase space, I discussed, from a classical perspective, the notions of electric and magnetic ``vacua'' in this deformed setting, as well as the nature of the excitations these vacua support. Emphasis was put in the description of such excitations from the Heggaard surface viewpoint.
The notion of defects associated to charged particles has been briefly discussed, as well.

After a few remarks on the quantization of the the deformed phase space, I concluded on a succinct summary of the interpretation of this deformed phase space in terms of piecewise homogeneously-curved geometries. My collaborators and I had, in fact, already studied this very topic in the past years as the basis of a framework for covariant $(3+1)$-dimensional loop quantum gravity in presence of a cosmological constant.

The framework presented here is the classical precursor, and as such the geometric and intuitive analogue, of Dittrich's \cite{Dittrich2017} proposal for studying TQFTs and topological order in 4 spacetime dimensions, via a generalization of the Walker--Wang string-net-like model \cite{Walker2012} baring a solid relationship with the Crane--Yetter topological field theory \cite{Crane:1993if,Barrett2007}. I believe, this classical model---being geometric in nature---can be used to develop intuition about the algebraic quantum models, as well as about the (semi-)classical limit of some of their observables, as in \cite{HHKR,HHKRplb,Haggard2016}.  Moreover, as highlighted in my brief discussion on excitations and defects, it can also guide how to perform the coupling of sources and hence the construction of more refined models of (extended) TQFTs. \\

\paragraph*{Note added}

After completing the writing of these notes, I learnt that Frolov proposed a very similar construction in the mid-1990s \cite{Frolov1995a,Frolov1995b,Frolov1995c} (see also \cite{Bonzom2014a}), where he had also noticed the relation to the moduli space of flat connections on a certain 2-surface as well as some of the ensuing dualities \cite{Frolov1995c}. 
His construction makes use of combinatorial quantization, and applies e.g. to the complex groups $\mathrm{SL}(N,\mathbb{C})$. 
He then proposes to add reality conditions to restrict his construction onto its real forms $\SU(N)$ and $\mathrm{SL}(N, \mathbb{R})$.
This procedure, however, requires a careful analysis, which has not been provided. Indeed, subtleties are bound to arise from the phase space reduction with respect to these new ``reality constraints''. E.g., assuming the reality constraints are first class, it has not been discussed what kind of new gauge transformations---to be eventually mod-out---they generate; similarly, assuming they are second class,  new Dirac brackets would have to be introduced. None of these issues, nor others, have however been discussed.
Here, I avoided all them by employing the quasi-Hamiltonian formalism of Alekseev and collaborators, which directly applies to all the groups considered above. 
As shown in the last section of \cite{Alekseev1998} (with reference to \cite{Alekseev1995}), the two procedures agree whenever both are applicable, e.g. for $G=\mathrm{SL}(N,\mathbb{C})$. 
I thank Prof. Frolov for kindly bringing his work to my attention.

\section{Acknowledgements}
I am indebted with Hal Haggard, who introduced me to (quasi)-Poisson-Lie geometry and with whom I have discussed at lengths many of the ideas presented in this notes. He also provided useful comments on a previous version of this paper.
It is also a great pleasure to thank Bianca Dittrich for her support, for our long discussions on this and related topics, and also for carefully reading and commenting a draft of this paper. 

This work was supported by Perimeter Institute for Theoretical Physics. Research at Perimeter Institute is supported by the Government of Canada through Industry Canada and by the Province of Ontario through the Ministry of Research and Innovation.

\appendix

\section{From the symplectic to the Poisson structure on $G\times \frakg$\label{app_A1}}

As a warm up, let me first show how to translate from the symplectic to the Poisson framework in the flat case.
The symplectic form is
\be
\omega = -\d \la X, \thetaL_h \ra = \la \thetaL_h\stackrel{\wedge}{,}\d X \ra + \tfrac12 \la X , [ \thetaL_h \stackrel{\wedge}{,} \thetaL_h] \ra = \thetaL_h{}^a \wedge \d X^a + \tfrac12 f_{ijk}X^i \thetaL_h{}^j \wedge \thetaL_h{}^k
\ee
where Einstein's summation convention is in order, $\thetaL_h{}^i=\la\tau^i,\thetaL_h\ra$, and $f_{ijk}$ is completely antisymmetric.
Now, using the notation of footnote \ref{fn_Pomega=1},
\begin{subequations}
\begin{align}
\omega_\flat(\pp_{ X^i} ) &= \pp_{X^i} \contr \omega = - \thetaL_h{}^i\\
\omega_\flat( (\hat{\tau^i})^L )& = (\hat{\tau^i})^L\contr \omega = \d X^i + f^i{}_{jk}X^j\thetaL_h{}^k
\end{align}
\end{subequations}
Now, using $P^\sharp\circ\omega_\flat = \mathrm{id}$, we obtain by linearity
\begin{subequations}
\begin{align}
 \pp_{ X^i} & = - P^\sharp( \thetaL_h{}^i) \\
 (\hat{\tau^i})^L & = P^\sharp( \d X^i + f^i{}_{jk}X^j\thetaL_h{}^k) = P^\sharp( \d X^i)  + f^i{}_{jk}X^j P^\sharp(\thetaL_h{}^k)
\end{align}
\end{subequations}
Hence,
\be
P^\sharp( \thetaL_h{}^i) = - \pp_{ X^i} 
\qquad\text{and}\qquad
P^\sharp(\d X^i) = - f_{ijk} X^j \pp_{X^k} +  (\hat{\tau^i})^L ,
\ee
and
\begin{subequations}
\begin{align}
P(\thetaL_h{}^i \otimes \thetaL_h{}^j) & = P^\sharp(\thetaL_h{}^i)\contr \thetaL_h{}^j = 0\\
P(\d X^i \otimes \thetaL_h{}^j) &= P^\sharp(\d X^i)\contr \thetaL_h{}^j = \delta^{ij}\\
P(\d X^i \otimes \d X^j)& = P^\sharp(\d X^i)\contr \d X^j = f^{ij}_{}{k} X^k
\end{align}
\end{subequations}
Multiplying by $\tau^j$ the second equation and recalling that $\{f_1,f_2\} = P(\d f_1 \otimes \d f_2)$ ad that---in coordinates---$\thetaL_h=h^{-1}\d h$, one finds 
\be
\{ h , h \} = 0,
\qquad
\{ X^i, h \} = h \tau^i
\qquad\text{and}\qquad
\{ X^i, X^j\} = f^{ij}{}_k X^k.
\ee

\section{Quasi-Poisson structure on $D(G)$ \label{app_A2}}

I will now move on to the translation between the quasi-symplectic and the quasi-Poisson case. 
Since---in this case---$\omega$ is degenerate, one cannot expect to recover the quasi-Poisson bivector on $D(G)$ by simply inverting $\omega$.
To take into account the non-trivial kernel of $\omega_\flat$, equation \eqref{eq_Ad+1}, the relation between the two gets in fact twisted:
\be
\boxed{\;\;\phantom{\int}
P^\sharp \circ \omega_\flat = \mathrm{id}_{\mathrm T D} - \frac14  (\tau^i)^\sharp \otimes \big(\thetaL_\mu{}^i - \thetaR_\mu{}^i).
\quad}
\ee
Notice, in the limit of small fluxes, this reduces to the usual inverse condition.

Now, contracting this equation with $\mathcal Y^\sharp = (Y, 0)^\sharp = \hat Y^L|_\text{h} + (\hat Y^L + \hat Y^R)|_\text{f}$, and using \eqref{eq_qsymplHam}, one obtains
\begin{align}
P^\sharp\circ\omega_\flat(\mathcal Y^\sharp)%
& = \frac12 P^\sharp\big( \la Y, \thetaL_{\mu_s} + \thetaR_{\mu_s} \ra \big) = \frac12 P^\sharp\big( \la (1 + \Ad_{\mu_s}) Y, \thetaR_{\mu_s}\ra\big)
\notag\\
& = \mathcal Y^\sharp - \frac14(2Y - \Ad_{\mu_s} Y - \Ad_{\mu_s}^{-1} Y)^\sharp_s = \frac14 \big( (1 + \Ad_{\mu_s}^{-1})(1 + \Ad_{\mu_s}) Y \big)^\sharp_s.
\end{align}
An analogous equation can be found for $Y^\sharp = (0, Y)$, and by replacing $ \tfrac12 (1 + \Ad_{\mu} \mathcal) \mathcal Y \leadsto \mathcal Y$, the following condition can be deduced
\be
P^\sharp\big(\la \mathcal Y, \theta^R_\mu \ra\big) = \frac12 \big( (1+\Ad_\mu^{-1}) \mathcal Y \big)^\sharp.
\ee
This is the quasi-Hamiltonian flow equation in the Poisson language, generalizating \eqref{eq_PoissonHam}.

Choosing again $\mathcal Y =(Y,0)$, and a further contracting with $\thetaL_h$, gives
\be
P\big(\la  Y, \theta^R_g \ra \otimes \thetaL_h \big)= \frac12 (1 + \Ad_g^{-1})Y.
\label{eq_thetag_thetah}
\ee
Similarly, contraction with $\thetaR_g$ gives
\be
P\big(\la  Y, \theta^R_g \ra \otimes \thetaR_g \big)= -\frac12 (1-\Ad_g)(1 + \Ad_g^{-1})Y = \frac12( \Ad_g - \Ad_g^{-1}) Y.
\ee
Finally, consider $\mathcal Y =(0,Y)$ in the quasi-Hamiltonian flow equation, and contract with $\thetaR_h$ :
\be
P\big(\la  Y, \thetaR_{\tl g} \ra \otimes \thetaR_h \big)= -\frac12 (1 + \Ad_{\tl g}^{-1})Y.
\ee
Now, using
\be
\thetaR_{\tl g} = (1 - \Ad_{\tl g}) \thetaR_h - \Ad_{hg^{-1}} \thetaR_g,
\ee
and $\Ad_h \thetaL_h = \thetaR_h$, the above equation can be recast in the form
\begin{align}
P\big(\la  (1 - \Ad_{\tl g}^{-1})Y, \thetaR_{h} \ra \otimes \thetaR_h \big)%
&= \Ad_h P\big(\la \Ad_{gh^{-1}} Y, \thetaR_g\ra \otimes \thetaL_h\big) -\frac12 (1 + \Ad_{\tl g}^{-1})Y \notag\\
& = 	\frac12 \Ad_h(1+\Ad_g^{-1})\Ad_{gh^{-1}}Y -\frac12 (1 + \Ad_{\tl g}^{-1})Y = 0
\end{align}
where in the second step I used equation \eqref{eq_thetag_thetah}.

Summarizing:
\begin{subequations}
\begin{align}
P(\thetaR_h{}^i \otimes \thetaR_h{}^j) &= 0\\
P(\thetaR_g{}^i \otimes \thetaL_h{}^j) &= \frac12\delta^{ij} + \frac12\la \tau^i, \Ad_g \tau^j\ra\\
P(\thetaR_g{}^i \otimes \thetaR_g{}^j) &= \frac12\la \Ad_g \tau^i, \tau^j \ra - \frac12\la \tau^i, \Ad_g \tau^j\ra.
\end{align}
\label{eq_A15}
\end{subequations}
Curiously, it is found that the holonomies quasi-Poisson commute, even though the momentum space has been curved. 
This mismatch, however, is not worrying, since this is not yet an actual phase space.
Nevertheless, I will comment again on this point at the end of this section.

Finally, in terms of $(a,b)$, these become
\be
P(\thetaR_a{}^i \otimes \thetaR_a{}^j) = 0= P(\thetaR_b{}^i \otimes \thetaR_b{}^j),
\qquad
P(\thetaR_b{}^i \otimes \thetaL_a{}^j) = \frac12 \la \tau^i, \Ad_b \tau^j\ra + \frac12 \la \Ad_a^{-1}\tau^i, \tau^j\ra,
\ee
that is
\be
P = \frac12  \la \hat \tau^L_b \otimes_A \hat\tau^L_a \ra +  \frac12 \la \hat \tau^R_b \otimes_A \hat\tau^R_a \ra.
\ee
Unsurprisingly, $(a,b)$ are the closest to Darboux coordinates in the quasi-Poisson setting.

For $G=\SU(2)$, 
\be
\tau^k = -\tfrac{i}{2}\sigma^k ,
\qquad
\la\cdot,\cdot\ra = -2\Tr(\cdot\; \cdot),
\qquad\text{and}\qquad
f_{ijk}=\epsilon_{ijk}.
\ee
Also, denoting the tensor product $A\otimes B = A_1 B_2$,
\be
4 \tau^i_1 \tau^i_2 =  1 -2 \eta
\qquad\text{with}\qquad
\eta \cdot (a\otimes b) = b\otimes a.
\ee
Hence, the only non-trivial quasi-Poisson bracket in $D=\SU(2)_\text{a}\times \SU(2)_\text{b}$ can be written as
\begin{align}
a_2^{-1} \{ b_1, a_2\} b_1^{-1} 
& = -\tau^i_1 \tau^j_2\left( \Tr(\tau^j \Ad_b^{-1}\tau^i ) + \Tr(\tau^i\Ad_{a}\tau^j )\right) \notag\\
& = -(\Ad_b \tau^i)_1 \tau^i_2 -  \tau^i_1 (\Ad_a^{-1} \tau^i)_2
\end{align}
and thus
\begin{align}
\{b_1, a_2\} 
& = - b_1 a_2 \tau^i_1  \tau^i_2  - \tau^i_1 \tau^i_2  b_1 a_2\notag\\
& = -\frac14\big(2 b_1a_2 - 2b_1a_2\eta  - 2\eta b_1 a_2 \big)
\end{align}
Or, equivalently
\be
\{a_1, b_2\}  = \frac12\big( a_1 b_2 - a_1 b_2\eta - \eta a_1 b_2 \big).
\ee

Finally, fusion can also be performed in the Poisson framework.
Similarly to the symplectic case, a term has to be added to the sum of the quasi-Poisson bivectors to make it again compatible with the new quasi-Hamiltonian flow equation. See \cite{Alekseev2000b} for the original construction and \cite{Haggard2016} for a worked-out example in a simple case. \\

\section{Remarks on exponentiated fluxes, boosts, and quasi-Poisson commutativity \label{app_remarks}}
As I have emphasized at the end of section \ref{sec_D}, the deformed fluxes are best understood as elements in the coset $G_\text{a}\times G_\text{b}/G_\text{diag}$, which---as a manifold---can be identified with the group $G$ (the momentum map then provides a definite one-to-one mapping between this manifold and the {\it group} $G$). In the case of $G=\SU(2)$, the double is ismorphic to (the universal cover of) the group of rotation of $\mathbb R^4$, $D(\SU(2))\cong\widetilde{\SO(4)}$. Then the $a$- and $b$-factors corresponding to its left and right $\SU(2)$ components, the diagonal $\SU(2)$ subgroup corresponds to spacetime rotations, and the exponentiated fluxes to the Euclidean boosts identified with elements of $G_\text{f}$.
With the construction of section \ref{sec_D}, the group of Euclidean rotations of $\mathbb R^4$, $D$ can be turned into a quasi-Hamiltonian $\SU(2)$-space.

A similar construction can be performed on (the universal cover of) the group of Lorentz transformations of $\mathbb R^{3,1}$, $\SL(2,\mathbb C) \cong \widetilde{\SO(3,1)}$: this group can be turned into a quasi-Hamiltonian $\SU(2)$-space with quasi-symplectic form $\omega_K$ and momentum map $\mu_K$. The latter is valued in the space of boosts identified with $K=\{g\in \SL(2,\mathbb C) : g=g^\dagger\}$, where ${}^\dagger$ stands for transposition followed by complex conjugation. The construction is detailed in \cite{Alekseev1998} and, although it differs from the one reviewed in this notes, it is just an adaptation thereof. On the other hand, it is also well-known \cite{Dupuis2013,Dupuis2014,Bonzom2014a,Bonzom2014,Charles2015,Charles2017} that $\SL(2,\mathbb C)$ is a Poisson--Lie space, which supports an actual symplectic two-form $\omega_\text{PL}$ and is equipped with a momentum map $\mu_\text{PL}$ valued in the Iwasawa subgroup $\mathrm{SB}(2,\mathbb C) = \SL(2,\mathbb C)$. 

It turns out that the two constructions are equivalent for $\SL(2,\mathbb C)$, although only the quasi-Hamiltonian one works for $D(\SU(2))$. This is proved in \cite[Section 10]{Alekseev1998}.
In this proof resides the solution of the puzzle of the commuting holonomies discussed after equation \eqref{eq_A15}:
\be
\omega_{PL} = \omega_{K} + \tau_{\mu_K}
\qquad\text{where}\qquad \d\tau = \chi,
\ee
so that equation \eqref{eq_chi} is satisfied. Notice that when pulled-back to $K$, $\chi$ is exact, since it is closed and the boost space contractible. Therefore the non commutativity of the holonomies, rather than being contained in $\omega_K$ is contained in $\tau_{\mu_K}$, or---more generally---in $\chi$.


\bibliographystyle{bibstyle_aldo}

\bibliography{deformedSN_biblio}    

\begin{thebibliography}{10}
\providecommand{\url}[1]{\texttt{#1}}
\providecommand{\urlprefix}{URL }
\renewcommand{\eprint}[1]{\href{http://arxiv.org/abs/#1}{\tt arxiv:#1}}

\bibitem{Alekseev1998}
{Alekseev}, A.~Y, {Malkin}, A, \protect\BIBand{} {Meinrenken}, E,
\newblock 1998
  \hypersetup{urlcolor=MidnightBlue}\href{http://dx.doi.org/10.4310/jdg/1214460860}{\textit{{Lie
  group valued moment maps}}}\hypersetup{urlcolor=MidnightBlue} .
\newblock Journal of Differential Geometry
\newblock 48(3) 445\hypersetup{urlcolor=BrickRed} [\eprint{dg-ga/9707021}]
  \hypersetup{urlcolor=MidnightBlue}

\bibitem{Alekseev2000}
{Alekseev}, A \protect\BIBand{} {Kosmann-Schwarzbach}, Y,
\newblock 2000
  \hypersetup{urlcolor=MidnightBlue}\href{http://dx.doi.org/10.4310/jdg/1090347528}{\textit{{Manin
  Pairs and Moment Maps}}}\hypersetup{urlcolor=MidnightBlue} .
\newblock Journal of Differential Geometry
\newblock 56(1) 133\hypersetup{urlcolor=BrickRed} [\eprint{math/9909176}]
  \hypersetup{urlcolor=MidnightBlue}

\bibitem{Alekseev2000b}
{Alekseev}, A.~Y, {Kosmann-Schwarzbach}, Y, \protect\BIBand{} {Meinrenken}, E,
  2000.
\newblock \textit{{Quasi-Poisson Manifolds}}.
\newblock Canadian Journal of Mathematics
\newblock 54 26\hypersetup{urlcolor=BrickRed} [\eprint{math.DG/0006168}]
  \hypersetup{urlcolor=MidnightBlue}

\bibitem{Alekseev2002}
{Alekseev}, A.~Y, {Meinrenken}, E, \protect\BIBand{} {Woodward}, C.~T,
\newblock 2002
  \hypersetup{urlcolor=MidnightBlue}\href{http://dx.doi.org/10.1007/s00039-002-8234-z}{\textit{{Duistermaat—Heckman
  measures and moduli spaces of flat bundles over
  surfaces}}}\hypersetup{urlcolor=MidnightBlue} .
\newblock Geometric {\&} Functional Analysis GAFA
\newblock 12(1) 1\hypersetup{urlcolor=BrickRed} [\eprint{math/9903087}]
  \hypersetup{urlcolor=MidnightBlue}

\bibitem{Atiyah1983}
{Atiyah}, M.~F \protect\BIBand{} {Bott}, R, 1983.
\newblock \textit{{The Yang-Mills Equations Riemann Surfaces}}.
\newblock Philosophical Transactions of the Royal Society of London. Series A,
  Mathematical and Physical Sciences
\newblock 308(1505) 523

\bibitem{Goldman1984}
{Goldman}, W.~M, 1984.
\newblock \textit{{The symplectic nature of fundamental groups of surfaces}}.
\newblock Advances in Mathematics
\newblock 54 200

\bibitem{Jeffrey1994}
{Jeffrey}, L.~C, 1994.
\newblock \textit{{Symplectic Forms on Moduli Spaces of Flat Connections on
  2-manifolds}}.
\newblock In W.~Kazez (Ed.) \textit{Proceedings of the Georgia International
  Topology Conference}, p.~12.
\newblock International Press\hypersetup{urlcolor=BrickRed}
  [\eprint{alg-geom/9404013v2}] \hypersetup{urlcolor=MidnightBlue}

\bibitem{Alekseev1994}
{Alekseev}, A.~Y \protect\BIBand{} {Malkin}, A,
\newblock 1994
  \hypersetup{urlcolor=MidnightBlue}\href{http://dx.doi.org/10.1007/BF02105190}{\textit{{Symplectic
  structures associated to Lie-Poisson
  groups}}}\hypersetup{urlcolor=MidnightBlue} .
\newblock Communications in Mathematical Physics
\newblock 162(1) 147\hypersetup{urlcolor=BrickRed} [\eprint{hep-th/9303038}]
  \hypersetup{urlcolor=MidnightBlue}

\bibitem{Fock1999}
{Fock}, V.~V \protect\BIBand{} {Rosly}, A.~A, 1999.
\newblock \textit{{Poisson structure on moduli of flat connections on Riemann
  surfaces and r-matrix}}.
\newblock American Mathematical Society Translations
\newblock 191 67\hypersetup{urlcolor=BrickRed} [\eprint{math/9802054}]
  \hypersetup{urlcolor=MidnightBlue}

\bibitem{Kosmann-Schwarzbach1991}
{Kosmann-Schwarzbach}, Y,
\newblock 1991
  \hypersetup{urlcolor=MidnightBlue}\href{http://dx.doi.org/10.1007/978-1-4615-3696-3_18}{\textit{{From
  “Quantum Groups” to “Quasi-Quantum
  Groups”}}}\hypersetup{urlcolor=MidnightBlue} .
\newblock In B~{Gruber {\{}$\backslash$it et al{\}}} (Ed.) \textit{Symmetries
  in Science V}, pp. 369--393.
\newblock Springer US, Boston, MA

\bibitem{Frolov1995a}
{Frolov}, S~A,
\newblock 1995
\hypersetup{urlcolor=MidnightBlue}\href{http://www.worldscientific.com/doi/abs/10.1142/S0217732395002751}{\textit{{Gauge-invariant Hamiltonian formulation of lattice Yang-Mills theory and the Heisenberg double}}}\hypersetup{urlcolor=MidnightBlue}.
\newblock Modern Physics Letters A
\newblock 10(34)  2619--2632 \hypersetup{urlcolor=BrickRed} [\eprint{hep-th/9501143}]
  \hypersetup{urlcolor=MidnightBlue}

\bibitem{Frolov1995b}
{Frolov}, S~A,
\newblock 1995
\hypersetup{urlcolor=MidnightBlue}\href{https://doi.org/10.1142/S0217732395003021}{\textit{{Hamiltonian lattice gauge models and the Heisenberg double}}}\hypersetup{urlcolor=MidnightBlue}.
\newblock Modern Physics Letters A
\newblock 10(37)   2885 \hypersetup{urlcolor=BrickRed} [\eprint{hep-th/9502121}]
  \hypersetup{urlcolor=MidnightBlue}
  
\bibitem{Frolov1995c}
{Frolov}, S~A,
\newblock 1997
\hypersetup{urlcolor=MidnightBlue}\href{https://doi.org/10.1007/BF02634016}{\textit{{Physical phase space of lattice Yang-Mills theory and the moduli space of flat connections on a Riemann surface}}}\hypersetup{urlcolor=MidnightBlue}.
\newblock Theoretical and Mathematical Physics
\newblock 113(1)  1289--1298 \hypersetup{urlcolor=BrickRed} [\eprint{hep-th/9511018}]
  \hypersetup{urlcolor=MidnightBlue}  

\bibitem{HHKR}
{Haggard}, H.~M, {Han}, M, {Kami{\'{n}}ski}, W, \protect\BIBand{} {Riello}, A,
\newblock 2015
  \hypersetup{urlcolor=MidnightBlue}\href{http://dx.doi.org/10.1016/j.nuclphysb.2015.08.023}{\textit{{SL(2,
  C) Chern–Simons theory, a non-planar graph operator, and 4D quantum gravity
  with a cosmological constant: Semiclassical
  geometry}}}\hypersetup{urlcolor=MidnightBlue} .
\newblock Nuclear Physics B
\newblock 900 1\hypersetup{urlcolor=BrickRed} [\eprint{1412.7546, 2014}]
  \hypersetup{urlcolor=MidnightBlue}

\bibitem{HHKRplb}
{Haggard}, H.~M, {Han}, M, {Kami{\'{n}}ski}, W, \protect\BIBand{} {Riello}, A,
\newblock 2016
  \hypersetup{urlcolor=MidnightBlue}\href{http://dx.doi.org/10.1016/j.physletb.2015.11.058}{\textit{{Four-dimensional
  quantum gravity with a cosmological constant from three-dimensional
  holomorphic blocks}}}\hypersetup{urlcolor=MidnightBlue} .
\newblock Physics Letters B
\newblock 752 258\hypersetup{urlcolor=BrickRed} [\eprint{1509.00458}]
  \hypersetup{urlcolor=MidnightBlue}

\bibitem{Haggard2016}
{Haggard}, H.~M, {Han}, M, \protect\BIBand{} {Riello}, A,
\newblock 2016
  \hypersetup{urlcolor=MidnightBlue}\href{http://dx.doi.org/10.1007/s00023-015-0455-4}{\textit{{Encoding
  Curved Tetrahedra in Face Holonomies: Phase Space of Shapes from Group-Valued
  Moment Maps}}}\hypersetup{urlcolor=MidnightBlue} .
\newblock Annales Henri Poincar{\'{e}}
\newblock 17(8) 2001\hypersetup{urlcolor=BrickRed}
  [\eprint{arXiv:1506.03053v1}] \hypersetup{urlcolor=MidnightBlue}

\bibitem{Haggard2015}
{Haggard}, H.~M, {Han}, M, {Kami{\'{n}}ski}, W, \protect\BIBand{} {Riello}, A,
  2015.
\newblock \textit{{SL(2,C) Chern-Simons Theory, Flat Connections, and
  Four-dimensional Quantum Geometry}} \hypersetup{urlcolor=BrickRed}
  [\eprint{1512.07690}] \hypersetup{urlcolor=MidnightBlue}

\bibitem{Han2017}
{Han}, M \protect\BIBand{} {Huang}, Z,
\newblock 2017
  \hypersetup{urlcolor=MidnightBlue}\href{http://dx.doi.org/10.1103/PhysRevD.95.044018}{\textit{{SU(2)
  flat connection on a Riemann surface and 3D twisted geometry with a
  cosmological constant}}}\hypersetup{urlcolor=MidnightBlue} .
\newblock Physical Review D
\newblock 95(4) 044018\hypersetup{urlcolor=BrickRed} [\eprint{1610.01246}]
  \hypersetup{urlcolor=MidnightBlue}

\bibitem{Delcamp2016}
{Delcamp}, C \protect\BIBand{} {Dittrich}, B, 2016.
\newblock \textit{{From 3D TQFTs to 4D models with defects}}
  \hypersetup{urlcolor=BrickRed} [\eprint{1606.02384}]
  \hypersetup{urlcolor=MidnightBlue}

\bibitem{Dittrich2017}
{Dittrich}, B,
\newblock 2017
  \hypersetup{urlcolor=MidnightBlue}\href{http://dx.doi.org/10.1007/JHEP05(2017)123}{\textit{{(3
  + 1)-dimensional topological phases and self-dual quantum geometries encoded
  on Heegaard surfaces}}}\hypersetup{urlcolor=MidnightBlue} .
\newblock Journal of High Energy Physics
\newblock 2017(5) 123\hypersetup{urlcolor=BrickRed} [\eprint{1701.02037}]
  \hypersetup{urlcolor=MidnightBlue}

\bibitem{Noui2006}
{Noui}, K,
\newblock 2006
  \hypersetup{urlcolor=MidnightBlue}\href{http://dx.doi.org/10.1063/1.2352860}{\textit{{Three-dimensional
  loop quantum gravity: Particles and the quantum
  double}}}\hypersetup{urlcolor=MidnightBlue} .
\newblock Journal of Mathematical Physics
\newblock 47(10) 102501\hypersetup{urlcolor=BrickRed} [\eprint{gr-qc/0612144}]
  \hypersetup{urlcolor=MidnightBlue}

\bibitem{Dupuis2013}
{Dupuis}, M \protect\BIBand{} {Girelli}, F,
\newblock 2014
  \hypersetup{urlcolor=MidnightBlue}\href{http://dx.doi.org/10.1103/PhysRevD.90.104037}{\textit{{Observables
  in loop quantum gravity with a cosmological
  constant}}}\hypersetup{urlcolor=MidnightBlue} .
\newblock Physical Review D
\newblock 90(10) 104037\hypersetup{urlcolor=BrickRed}
  [\eprint{gr-qc/1311.6841}] \hypersetup{urlcolor=MidnightBlue}

\bibitem{Dupuis2014}
{Dupuis}, M, {Girelli}, F, \protect\BIBand{} {Livine}, E.~R,
\newblock 2014
  \hypersetup{urlcolor=MidnightBlue}\href{http://dx.doi.org/10.1007/s10714-014-1802-3}{\textit{{Deformed
  Spinor Networks for Loop Gravity: Towards Hyperbolic Twisted
  Geometries}}}\hypersetup{urlcolor=MidnightBlue} .
\newblock General Relativity and Gravitation
\newblock 46(11) 1802\hypersetup{urlcolor=BrickRed} [\eprint{gr-qc/1403.7482}]
  \hypersetup{urlcolor=MidnightBlue}

\bibitem{Bonzom2014a}
{Bonzom}, V, {Dupuis}, M, {Girelli}, F, \protect\BIBand{} {Livine}, E.~R, 2014.
\newblock \textit{{Deformed phase space for 3d loop gravity and hyperbolic
  discrete geometries}} \hypersetup{urlcolor=BrickRed}
  [\eprint{gr-qc/1402.2323}] \hypersetup{urlcolor=MidnightBlue}

\bibitem{Bonzom2014}
{Bonzom}, V, {Dupuis}, M, \protect\BIBand{} {Girelli}, F,
\newblock 2014
  \hypersetup{urlcolor=MidnightBlue}\href{http://dx.doi.org/10.1103/PhysRevD.90.104038}{\textit{{Towards
  the Turaev-Viro amplitudes from a Hamiltonian
  constraint}}}\hypersetup{urlcolor=MidnightBlue} .
\newblock Physical Review D
\newblock 90(10) 104038\hypersetup{urlcolor=BrickRed}
  [\eprint{gr-qc/1403.7121}] \hypersetup{urlcolor=MidnightBlue}

\bibitem{Charles2015}
{Charles}, C \protect\BIBand{} {Livine}, E.~R,
\newblock 2015
  \hypersetup{urlcolor=MidnightBlue}\href{http://dx.doi.org/10.1088/0264-9381/32/13/135003}{\textit{{Closure
  constraints for hyperbolic tetrahedra}}}\hypersetup{urlcolor=MidnightBlue} .
\newblock Classical and Quantum Gravity
\newblock 32(13) 135003\hypersetup{urlcolor=BrickRed}
  [\eprint{gr-qc/1501.00855}] \hypersetup{urlcolor=MidnightBlue}

\bibitem{Charles2017}
{Charles}, C \protect\BIBand{} {Livine}, E.~R,
\newblock 2017
  \hypersetup{urlcolor=MidnightBlue}\href{http://dx.doi.org/10.1007/s10714-017-2255-2}{\textit{{The
  closure constraint for the hyperbolic tetrahedron as a Bianchi
  identity}}}\hypersetup{urlcolor=MidnightBlue} .
\newblock General Relativity and Gravitation
\newblock 49(7) 92\hypersetup{urlcolor=BrickRed} [\eprint{1607.08359}]
  \hypersetup{urlcolor=MidnightBlue}

\bibitem{Meusburger2016}
{Meusburger}, C,
\newblock 2017
  \hypersetup{urlcolor=MidnightBlue}\href{http://dx.doi.org/10.1007/s00220-017-2860-7}{\textit{{Kitaev
  Lattice Models as a Hopf Algebra Gauge
  Theory}}}\hypersetup{urlcolor=MidnightBlue} .
\newblock Communications in Mathematical Physics
\newblock 353(1) 413\hypersetup{urlcolor=BrickRed}
  [\eprint{arXiv:1607.01144v1}] \hypersetup{urlcolor=MidnightBlue}

\bibitem{Meusburger2016a}
{Meusburger}, C \protect\BIBand{} {Wise}, D.~K, 2016.
\newblock \textit{{Hopf algebra gauge theory on a ribbon graph}}
\newblock pp. 1--68\hypersetup{urlcolor=BrickRed} [\eprint{arXiv:1512.03966v2}]
  \hypersetup{urlcolor=MidnightBlue}

\bibitem{Noui2003a}
{Noui}, K \protect\BIBand{} {Roche}, P,
\newblock 2002
  \hypersetup{urlcolor=MidnightBlue}\href{http://dx.doi.org/10.1088/0264-9381/20/14/318}{\textit{{Cosmological
  Deformation of Lorentzian Spin Foam
  Models}}}\hypersetup{urlcolor=MidnightBlue} .
\newblock Classical and Quantum Gravity
\newblock 20(14) 3175\hypersetup{urlcolor=BrickRed} [\eprint{gr-qc/0211109}]
  \hypersetup{urlcolor=MidnightBlue}

\bibitem{Han2011b}
{Han}, M,
\newblock 2011
  \hypersetup{urlcolor=MidnightBlue}\href{http://dx.doi.org/10.1103/PhysRevD.84.064010}{\textit{{Cosmological
  Constant in Loop Quantum Gravity Vertex
  Amplitude}}}\hypersetup{urlcolor=MidnightBlue} .
\newblock Physical Review D
\newblock 84(6) 064010\hypersetup{urlcolor=BrickRed} [\eprint{1105.2212}]
  \hypersetup{urlcolor=MidnightBlue}

\bibitem{Fairbairn2012}
{Fairbairn}, W.~J \protect\BIBand{} {Meusburger}, C,
\newblock 2012
  \hypersetup{urlcolor=MidnightBlue}\href{http://dx.doi.org/10.1063/1.3675898}{\textit{{Quantum
  deformation of two four-dimensional spin foam
  models}}}\hypersetup{urlcolor=MidnightBlue} .
\newblock Journal of Mathematical Physics
\newblock 53(2) 22501\hypersetup{urlcolor=BrickRed} [\eprint{1012.4784}]
  \hypersetup{urlcolor=MidnightBlue}

\bibitem{Walker2012}
{Walker}, K \protect\BIBand{} {Wang}, Z,
\newblock 2012
  \hypersetup{urlcolor=MidnightBlue}\href{http://dx.doi.org/10.1007/s11467-011-0194-z}{\textit{{(3+1)-TQFTs
  and topological insulators}}}\hypersetup{urlcolor=MidnightBlue} .
\newblock Frontiers of Physics
\newblock 7(2) 150\hypersetup{urlcolor=BrickRed} [\eprint{1104.2632}]
  \hypersetup{urlcolor=MidnightBlue}

\bibitem{Wan2015}
{Wan}, Y, {Wang}, J.~C, \protect\BIBand{} {He}, H,
\newblock 2015
  \hypersetup{urlcolor=MidnightBlue}\href{http://dx.doi.org/10.1103/PhysRevB.92.045101}{\textit{{Twisted
  gauge theory model of topological phases in three
  dimensions}}}\hypersetup{urlcolor=MidnightBlue} .
\newblock Physical Review B
\newblock 92(4) 045101\hypersetup{urlcolor=BrickRed} [\eprint{1409.3216}]
  \hypersetup{urlcolor=MidnightBlue}

\bibitem{Lan2017}
{Lan}, T, {Kong}, L, \protect\BIBand{} {Wen}, X.-G, 2017.
\newblock \textit{{A classification of 3+1D bosonic topological orders (I): the
  case when point-like excitations are all bosons}}
  \hypersetup{urlcolor=BrickRed} [\eprint{1704.04221}]
  \hypersetup{urlcolor=MidnightBlue}

\bibitem{Barenz2016}
{B{\"{a}}renz}, M \protect\BIBand{} {Barrett}, J.~W, 2016.
\newblock \textit{{Dichromatic state sum models for four-manifolds from pivotal
  functors}} \hypersetup{urlcolor=BrickRed} [\eprint{1601.03580}]
  \hypersetup{urlcolor=MidnightBlue}

\bibitem{Crane:1993if}
{Crane}, L \protect\BIBand{} {Yetter}, D.~N, 1993.
\newblock \textit{{A categorical construction of 4D TQFTs}}.
\newblock In Dayton 1992, Proceedings, Quantum topology
  \hypersetup{urlcolor=BrickRed} [\eprint{hep-th/9301062}]
  \hypersetup{urlcolor=MidnightBlue}

\bibitem{Barrett2007}
{Barrett}, J.~W, {{Faria Martins}}, J, \protect\BIBand{} {Garcia-Islas}, J.~M,
\newblock 2007
  \hypersetup{urlcolor=MidnightBlue}\href{http://dx.doi.org/10.1063/1.2759440}{\textit{{Observables
  in the Turaev-Viro and Crane-Yetter
  models}}}\hypersetup{urlcolor=MidnightBlue} .
\newblock Journal of Mathematical Physics
\newblock 48(9) 093508\hypersetup{urlcolor=BrickRed} [\eprint{math/0411281}]
  \hypersetup{urlcolor=MidnightBlue}

\bibitem{Kodama1988}
{Kodama}, H,
\newblock 1988
  \hypersetup{urlcolor=MidnightBlue}\href{http://dx.doi.org/10.1143/PTP.80.1024}{\textit{{Specialization
  of Ashtekar's Formalism to Bianchi
  Cosmology}}}\hypersetup{urlcolor=MidnightBlue} .
\newblock Progress of Theoretical Physics
\newblock 80(6) 1024

\bibitem{Freidel2011}
{Freidel}, L, {Geiller}, M, \protect\BIBand{} {Ziprick}, J,
\newblock 2013
  \hypersetup{urlcolor=MidnightBlue}\href{http://dx.doi.org/10.1088/0264-9381/30/8/085013}{\textit{{Continuous
  formulation of the loop quantum gravity phase
  space}}}\hypersetup{urlcolor=MidnightBlue} .
\newblock Classical and Quantum Gravity
\newblock 30(8) 085013\hypersetup{urlcolor=BrickRed} [\eprint{1110.4833}]
  \hypersetup{urlcolor=MidnightBlue}

\bibitem{Charles2016}
{Charles}, C \protect\BIBand{} {Livine}, E.~R,
\newblock 2016
  \hypersetup{urlcolor=MidnightBlue}\href{http://dx.doi.org/10.1007/s10714-016-2107-5}{\textit{{The
  Fock space of loopy spin networks for quantum
  gravity}}}\hypersetup{urlcolor=MidnightBlue} .
\newblock General Relativity and Gravitation
\newblock 48(8) 113\hypersetup{urlcolor=BrickRed} [\eprint{1603.01117}]
  \hypersetup{urlcolor=MidnightBlue}

\bibitem{Freidel2003}
{Freidel}, L \protect\BIBand{} {Louapre}, D,
\newblock 2003
  \hypersetup{urlcolor=MidnightBlue}\href{http://dx.doi.org/10.1016/S0550-3213(03)00306-7}{\textit{{Diffeomorphisms
  and spin foam models}}}\hypersetup{urlcolor=MidnightBlue} .
\newblock Nuclear Physics B
\newblock 662(1-2) 279\hypersetup{urlcolor=BrickRed}
  [\eprint{arXiv:gr-qc/0212001v2}] \hypersetup{urlcolor=MidnightBlue}

\bibitem{Dittrich2014}
{Dittrich}, B \protect\BIBand{} {Geiller}, M,
\newblock 2014
  \hypersetup{urlcolor=MidnightBlue}\href{http://dx.doi.org/10.1088/0264-9381/32/11/112001}{\textit{{A
  new vacuum for Loop Quantum Gravity}}}\hypersetup{urlcolor=MidnightBlue} .
\newblock Classical and Quantum Gravity
\newblock 32(11) 10\hypersetup{urlcolor=BrickRed} [\eprint{gr-qc/1401.6441}]
  \hypersetup{urlcolor=MidnightBlue}

\bibitem{Dittrich2014c}
{Dittrich}, B \protect\BIBand{} {Geiller}, M,
\newblock 2015
  \hypersetup{urlcolor=MidnightBlue}\href{http://dx.doi.org/10.1088/0264-9381/32/13/135016}{\textit{{Flux
  formulation of loop quantum gravity: classical
  framework}}}\hypersetup{urlcolor=MidnightBlue} .
\newblock Classical and Quantum Gravity
\newblock 32(13) 135016\hypersetup{urlcolor=BrickRed}
  [\eprint{gr-qc/1412.3752}] \hypersetup{urlcolor=MidnightBlue}

\bibitem{Dittrich2015}
{Bahr}, B, {Dittrich}, B, \protect\BIBand{} {Geiller}, M, 2015.
\newblock \textit{{A new realization of quantum geometry}}
  \hypersetup{urlcolor=BrickRed} [\eprint{1506.08571}]
  \hypersetup{urlcolor=MidnightBlue}

\bibitem{DittrichGeillerTQFT}
{Dittrich}, B \protect\BIBand{} {Geiller}, M,
\newblock 2017
  \hypersetup{urlcolor=MidnightBlue}\href{http://dx.doi.org/10.1088/1367-2630/aa54e2}{\textit{{Quantum
  gravity kinematics from extended TQFTs}}}\hypersetup{urlcolor=MidnightBlue} .
\newblock New Journal of Physics
\newblock 19(1) 013003\hypersetup{urlcolor=BrickRed} [\eprint{1604.05195}]
  \hypersetup{urlcolor=MidnightBlue}

\bibitem{Delcamp2017}
{Delcamp}, C, {Dittrich}, B, \protect\BIBand{} {Riello}, A,
\newblock 2016
  \hypersetup{urlcolor=MidnightBlue}\href{http://dx.doi.org/10.1007/JHEP02(2017)061}{\textit{{Fusion
  basis for lattice gauge theory and loop quantum
  gravity}}}\hypersetup{urlcolor=MidnightBlue} .
\newblock Journal of High Energy Physics
\newblock 2017(2) 61\hypersetup{urlcolor=BrickRed} [\eprint{1607.08881}]
  \hypersetup{urlcolor=MidnightBlue}

\bibitem{Gukov2003}
{Gukov}, S,
\newblock 2003
  \hypersetup{urlcolor=MidnightBlue}\href{http://dx.doi.org/10.1007/s00220-005-1312-y}{\textit{{Three-Dimensional
  Quantum Gravity, Chern-Simons Theory, and the
  A-Polynomial}}}\hypersetup{urlcolor=MidnightBlue} .
\newblock Communications in Mathematical Physics
\newblock 255(3) 577\hypersetup{urlcolor=BrickRed} [\eprint{hep-th/0306165}]
  \hypersetup{urlcolor=MidnightBlue}

\bibitem{Bates1997}
{Bates}, S \protect\BIBand{} {Weinstein}, A, 1997.
\newblock \textit{{Lectures on the Geometry of Quantization}}.
\newblock Berkeley Mathematics Lecture Notes
\newblock 8

\bibitem{Dirac1931}
{Dirac}, P.~A.~M, 1931.
\newblock \textit{{Quantised Singularities in the Electromagnetic Field}}.
\newblock Proceedings of the Royal Society A
\newblock 133(60)

\bibitem{Goddard1977}
{Goddard}, P, {Nuyts}, J, \protect\BIBand{} {Olive}, D,
\newblock 1977
  \hypersetup{urlcolor=MidnightBlue}\href{http://dx.doi.org/10.1016/0550-3213(77)90221-8}{\textit{{Gauge
  theories and magnetic charge}}}\hypersetup{urlcolor=MidnightBlue} .
\newblock Nuclear Physics B
\newblock 125(1) 1

\bibitem{Montonen1977}
{Montonen}, C \protect\BIBand{} {Olive}, D,
\newblock 1977
  \hypersetup{urlcolor=MidnightBlue}\href{http://dx.doi.org/10.1016/0370-2693(77)90076-4}{\textit{{Magnetic
  monopoles as gauge particles?}}}\hypersetup{urlcolor=MidnightBlue} .
\newblock Physics Letters B
\newblock 72(1)

\bibitem{Witten1979}
{Witten}, E,
\newblock 1979
  \hypersetup{urlcolor=MidnightBlue}\href{http://dx.doi.org/10.1016/0370-2693(79)90838-4}{\textit{{Dyons
  of charge e$\theta$/2$\pi$}}}\hypersetup{urlcolor=MidnightBlue} .
\newblock Physics Letters B
\newblock 86(3--4) 283

\bibitem{Olive1997}
{Olive}, D,
\newblock 1997
  \hypersetup{urlcolor=MidnightBlue}\href{http://dx.doi.org/10.1016/S0920-5632(97)00412-X}{\textit{{Introduction
  to Electromagnetic Duality}}}\hypersetup{urlcolor=MidnightBlue} .
\newblock Nuclear Physics B (Proceedings Supplements)
\newblock 58 43

\bibitem{Witten1992}
{Witten}, E,
\newblock 1992
  \hypersetup{urlcolor=MidnightBlue}\href{http://dx.doi.org/10.1016/0393-0440(92)90034-X}{\textit{{Two
  dimensional gauge theories revisited}}}\hypersetup{urlcolor=MidnightBlue} .
\newblock Journal of Geometry and Physics
\newblock 9(4) 303\hypersetup{urlcolor=BrickRed} [\eprint{hep-th/9204083}]
  \hypersetup{urlcolor=MidnightBlue}

\bibitem{Reshetikhin1991}
{Reshetikhin}, N \protect\BIBand{} {Turaev}, V.~G,
\newblock 1991
  \hypersetup{urlcolor=MidnightBlue}\href{http://dx.doi.org/10.1007/BF01239527}{\textit{{Invariants
  of 3-manifolds via link polynomials and quantum
  groups}}}\hypersetup{urlcolor=MidnightBlue} .
\newblock Inventiones Mathematicae
\newblock 103(1) 547

\bibitem{Andersen2012}
{Andersen}, J.~E, 2012.
\newblock \textit{{A geometric formula for the Witten-Reshetikhin-Turaev
  Quantum Invariants and some applications}}
\newblock pp. 1--44\hypersetup{urlcolor=BrickRed} [\eprint{1206.2785}]
  \hypersetup{urlcolor=MidnightBlue}

\bibitem{Ashtekar1986}
{Ashtekar}, A,
\newblock 1986
  \hypersetup{urlcolor=MidnightBlue}\href{http://dx.doi.org/10.1103/PhysRevLett.57.2244}{\textit{{New
  Variables for Classical and Quantum
  Gravity}}}\hypersetup{urlcolor=MidnightBlue} .
\newblock Physical Review Letters
\newblock 57(18) 2244

\bibitem{Barbero1995}
{{Barbero G.}}, J.~F,
\newblock 1995
  \hypersetup{urlcolor=MidnightBlue}\href{http://dx.doi.org/10.1103/PhysRevD.51.5507}{\textit{{Real
  Ashtekar variables for Lorentzian signature
  space-times}}}\hypersetup{urlcolor=MidnightBlue} .
\newblock Physical Review D
\newblock 51(10) 5507\hypersetup{urlcolor=BrickRed} [\eprint{gr-qc/9410014}]
  \hypersetup{urlcolor=MidnightBlue}

\bibitem{Rovelli2007}
{Rovelli}, C, 2007.
\newblock \textit{{Quantum Gravity}}.
\newblock Cambridge Monographs on Mathematical Physics. Cambridge University
  Press, Cambridge, England.
\newblock
\newblock ISBN 9780521715966

\bibitem{Thiemann2004}
{Thiemann}, T, 2007.
\newblock \textit{{Modern Canonical General Relativity}}.
\newblock Cambridge Monographs on Mathematical Physics.
\newblock Cambridge University Press, Cambridge, England

\bibitem{Freidel2010}
{Freidel}, L \protect\BIBand{} {Speziale}, S,
\newblock 2010
  \hypersetup{urlcolor=MidnightBlue}\href{http://dx.doi.org/10.1103/PhysRevD.82.084040}{\textit{{Twisted
  Geometries: A Geometric Parametrization of {\$}SU(2){\$} Phase
  Space}}}\hypersetup{urlcolor=MidnightBlue} .
\newblock Physical Review D
\newblock 82(8) 084040\hypersetup{urlcolor=BrickRed} [\eprint{1001.2748}]
  \hypersetup{urlcolor=MidnightBlue}

\bibitem{Freidel2010twist2}
{Freidel}, L \protect\BIBand{} {Speziale}, S,
\newblock 2010
  \hypersetup{urlcolor=MidnightBlue}\href{http://dx.doi.org/10.1103/PhysRevD.82.084041}{\textit{{Twistors
  to twisted geometries}}}\hypersetup{urlcolor=MidnightBlue} .
\newblock Physical Review D
\newblock 82(8) 084041\hypersetup{urlcolor=BrickRed} [\eprint{1006.0199}]
  \hypersetup{urlcolor=MidnightBlue}

\bibitem{Livine2012}
{Livine}, E \protect\BIBand{} {Tambornino}, J,
\newblock 2012
  \hypersetup{urlcolor=MidnightBlue}\href{http://dx.doi.org/10.1063/1.3675465}{\textit{{Spinor
  representation for loop quantum gravity}}}\hypersetup{urlcolor=MidnightBlue}.
\newblock Journal of Mathematical Physics
\newblock 53(1) 012503\hypersetup{urlcolor=BrickRed} [\eprint{1105.3385}]
  \hypersetup{urlcolor=MidnightBlue}

\bibitem{Minkowski1897}
{Minkowski}, H, 1897.
\newblock \textit{{Allgemeine Lehrs{\"{a}}tze {\"{u}}ber die convexen
  Polyeder}}.
\newblock Nachrichten von der Gesellschaft der Wissenschaften zu
  G{\"{o}}ttingen
\newblock pp. 198--219

\bibitem{Alexandrov2005}
{Alexandrov}, A~D, 2005.
\newblock \textit{{Convex Polyhedra}}.
\newblock Springer Monographs in Mathematics.
\newblock
\newblock Springer-Verlag, Berlin/Heidelberg
  \href{http://link.springer.com/10.1007/b137434}{ISBN 978-3-540-23158-5}
  
\bibitem{Alekseev1995}
{Alekseev}, A.~Y, \protect\BIBand{} {Malkin}, A, 
\newblock 1995
  \hypersetup{urlcolor=MidnightBlue}\href{https://projecteuclid.org/euclid.cmp/1104272613}{\textit{{Symplectic Structure of the Moduli Space of Flat Connectionon a Riemann Surface}}}\hypersetup{urlcolor=MidnightBlue} .
\newblock Communications in Mathematical Physics
\newblock 169(1) 99\hypersetup{urlcolor=BrickRed} [\eprint{hep-th/9312004}]
  \hypersetup{urlcolor=MidnightBlue}


\end{thebibliography}

\end{document}